\documentclass{aa}  
\usepackage{longtable}
\usepackage{graphicx}	% Including figure files
\usepackage{amsmath}	% Advanced maths commands
\usepackage{amssymb}	% Extra maths symbols
\usepackage{pdflscape}
\usepackage[T1]{fontenc}
\usepackage[utf8]{inputenc}
\usepackage{ae,aecompl}
\usepackage{float}
\usepackage{natbib}
\usepackage{multirow}

\bibpunct{(}{)}{;}{a}{}{,} % to follow the A&A style

\begin{document}

\title{The host of GRB 171205A in 3D}
\subtitle{A resolved multiwavelength study of a rare grand-design spiral GRB host}
\titlerunning{The host of GRB 171205A spatially resolved}

\author{C. C. Th\"one \inst{1} \and A. de Ugarte Postigo \inst{2,3} \and L. Izzo \inst{4,5}  \and M. J. Michalowski \inst{6} \and A.~J. Levan \inst{7} \and J. K. Leung \inst{8,9,10} \and J.~F.~Ag\"u\'i~Fern\'andez \inst{11} \and T. G\'eron \inst{9} \and R. Friesen \inst{10} \and L.~Christensen \inst{12} \and S. Covino \inst{13} \and V. D'Elia \inst{14,15}\and D.~H.~Hartmann \inst{16} \and P. Jakobsson \inst{17} \and M. De Pasquale \inst{18} \and G.~Pugliese \inst{19} \and A.~Rossi \inst{20} \and P. Schady \inst{21} \and K. Wiersema \inst{22} \and T. Zafar \inst{23,24}
}

\institute{
%1
 Astronomical Institute, Czech Academy of Sciences, Fri\v cova 298, Ond\v rejov, Czech Republic  \email{christina.thoene@gmail.com} \and
%2
Universit\'{e} de la C\^ote d'Azur, Observatoire de la C\^ote d'Azur, Artemis, CNRS, 06304 Nice, France \and 
%3
Aix Marseille Univ, CNRS, CNES, LAM Marseille, France \and
%4
INAF, Osservatorio Astronomico di Capodimonte, Salita Moiariello 16, I-80131 Napoli, Italy
%5
DARK, Niels Bohr Institute, University of Copenhagen, Jagtvej 128, 2200-N Copenhagen, Denmark \and
%6
Astronomical Observatory Institute, Faculty of Physics, Adam Mickiewicz University, ul. S\l{}oneczna 36, 60-286 Pozna\'n, Poland\and
%7
Department of Astrophysics/IMAPP, Radboud University, 6525 AJ Nijmegen, the Netherlands
 \and 
 %8
 David A. Dunlap Department of Astronomy \& Astrophysics, University of Toronto, 50 St. George St., Toronto, Ontario, M5S 3H4, Canada \and
 %9
Dunlap Institute for Astrophysics \& Astrophysics, University of Toronto, 50 St. George Street Toronto, ON M5S 3H4, Canada \and
%10
Racah Institute of Physics, The Hebrew University of Jerusalem, Jerusalem, 91904, Israel \and
%11
Centro Astronómico Hispano-Alemán, Observatorio de Calar Alto, Sierra de los Filabres, 04550, Gérgal, Spain \and
%12
 Cosmic Dawn Center, Niels Bohr Institute, University of Copenhagen, Jagtvej 128, 2200-N Copenhagen, Denmark \and
%13
INAF, Osservatorio Astronomico di Brera, Via Bianchi, Merate, Italy \and
%14
Space Science Data Center (SSDC) - Agenzia Spaziale Italiana (ASI), 00133 Roma, Italy \and
%15
INAF - Osservatorio Astronomico di Roma, Via Frascati 33, 00040 Monte Porzio Catone, Italy  \and
%16
Clemson University, Department of Physics and Astronomy, Clemson, SC 29634-0978, USA  \and
%17
Centre for Astrophysics and Cosmology, Science Institute, University of Iceland, Dunhagi 5, 107 Reykjavík, Iceland \and
%18
University of Messina, MIFT Department, Via F. S. D'Alcontres 31, Papardo, 98166 Messina, Italy. \and
%19
Astronomical Institute Anton Pannekoek, University of Amsterdam, 1090 GE Amsterdam, The Netherlands \and 
%20
INAF - Osservatorio di  Astrofisica e Scienza dello Spazio, via Piero Gobetti 93/3, 40129 Bologna, Italy \and
%21
Department of Physics, University of Bath, Bath BA2 7AY, UK \and
%22
Centre for Astrophysics Research, University of Hertfordshire, Hatfield, AL10 9AB, UK  \and
%23
School of Mathematical and Physical Sciences, Macquarie University, NSW 2109, Australia \and
%24
ARC Centre of Excellence for All Sky Astrophysics in 3 Dimensions (ASTRO-3D), Australia
}

\date{Received Oct. 3; accepted xxx}
\authorrunning{Th\"one et al.}
\titlerunning{The host of GRB\,171205A in 3D}

\abstract{Long GRB hosts at $z<1$ are usually low-mass, low metallicity star-forming galaxies. Here we present the until now most detailed, spatially resolved study of the host of GRB 171205A, a grand-design barred spiral galaxy at $z=0.036$. Our analysis includes MUSE integral field spectroscopy, complemented by high spatial resolution UV/VIS HST imaging and CO(1--0) and H~{\sc i} 21cm data. The GRB is located in a small star-forming region in a spiral arm of the galaxy at a deprojected distance of $\sim$ 8\,kpc from the center. The galaxy shows a smooth negative metallicity gradient and the metallicity at the GRB site is half solar, slightly below the mean metallicity at the corresponding distance from the center. Star formation in this galaxy is concentrated in a few H~{\sc ii} regions between 5--7\,kpc from the center and at the end of the bar, inwards of the GRB region, however, the H~{\sc ii} region hosting the GRB is in the top 10\% of regions with highest specific star-formation rate. The stellar population at the GRB site has a very young component ($<$ 5\,Myr) contributing a significant part of the light. Ionized and molecular gas show only minor deviations at the end of the bar. A parallel study found an asymmetric H~{\sc i} distribution and some additional gas near the position of the GRB, which might explain the star-forming region of the GRB site. Our study shows that long GRBs can occur in many types of star-forming galaxies, however, the actual GRB sites consistently have low metallicity, high star formation and a young population. Furthermore, gas inflow or interactions triggering the  star formation producing the GRB progenitor might not be evident in ionized or even molecular gas but only in H~{\sc i}.}
\keywords{stars: gamma-ray bursts, galaxies: kinematics, galaxies: spiral, galaxies: ISM}

\maketitle

%%%%%%%%%%%%%%%%% BODY OF PAPER %%%%%%%%%%%%%%%%%%

\section{Introduction}
Long-duration Gamma-Ray bursts (GRBs) are the most luminous stellar explosions in the Universe. They originate from the collapse of a massive star, possibly a Wolf-Rayet star, stripped of its hydrogen and helium envelope while retaining a high angular momentum to support an accretion disk and launch a jet. In addition to synchrotron emission from the jet created in the actual GRB, the star explodes as a broad-line (BL) Ic supernova (SN) \citep{Hjorth03,Modjaz16, CanoRev}. In recent years, a few exceptions have been found that might point to different progenitors, or are actually compact binary mergers accompanied by a kilonova instead of a BL-Ic SN \citep{Fynbo06, Michalowski15, Rastinejad22} and would hence fall in the class of short GRBs. The peak absolute magnitudes of GRB-SNe are in the range of $M_R = -18.5$ to $-20$ mag \citep{CanoRev}, making their detection at redshifts above $\sim$1 challenging. Furthermore, the average GRB redshift is $z\sim2.2$ and events at $z<0.5$, where we can obtain good S/N spectra and spectral series, are rare. 

GRB host studies suffer from the same redshift problem as do GRB-SNe: Spatially resolving them is only possible at low redshifts with current ground-based instrumentation and is still challenging at higher redshift in the NIR with JWST \citep{Schady24}. Various studies of global host properties have shown that long GRB hosts are all actively star-forming and have predominantly low metallicities \citep{Kruehler15}, although there are exceptions \citep{Savaglio12, Schady15, Michalowski18, Heintz18}. Long GRB hosts at low redshifts furthermore have low average masses of $<$1/10 L* but tend to higher masses at redshifts beyond 2  \citep{PerleySHOALS2, Palmerio19}. L* hosts at redshifts $z\lesssim1$ are extremely rare where most hosts are dwarf compact or irregular galaxies. The host of GRB 171205A is the first grand design spiral long GRB host at $z<0.5$, only recently followed by the even larger spiral host galaxy of GRB\,190829A \citep{Dichiara19_GCN}.

Studying the global properties of a host galaxy might not necessarily reflect the conditions at the GRB site. Due to the low number of hosts at $z<0.4$ where all important emission lines are still in the visible part of the spectrum, the faintness and small size of the hosts and the lack of very sensitive integral field units (IFUs) needed, there are still only a few long GRB hosts studied at high angular resolution: GRBs 980425 \citep{Christensen08, Kruehler17}, 060505 \citep{Thoene14}, 111005A \citep{Tanga17, Michalowski18} and 100316D \citep{Izzo17} and only three short GRB hosts, GRBs 170817 \citep{Levan17}, 050709 \citep{Guelbenzu21a} and 080905A \citep{Guelbenzu21b}. A few other GRB hosts have been observed at low resolution with one or several long slit positions: GRB 120422A \citep{Schulze14}, GRB 161219B \citep{Cano17} and the short GRB 130603B \citep{deUgarte13}. These studies indicate that long GRBs tend to occur at the metal-poorer and higher star-forming places of their galaxies, although in most galaxies, the GRB site is not the site with the most extreme properties. However, all GRB sites consistently show a subsolar metallicity, believed to be necessary to produce a long GRB by theoretical models \citep{Yoon06}, although binary progenitors might change the picture \citep{Chrimes20}. GRB-SNe as well as BL-Ic SNe without GRB also seem to have lower metallicity progenitors than normal Ic SNe, pointing to a similar progenitor and possibly a choked jet for BL-Ics without GRB \citep{Modjaz20}. In a recent study on resolved kinematics in six GRB hosts \citep{Thoene21}, all hosts show evidence for large-scale outflows or winds, confirming the highly star-forming nature of GRB hosts. Models have proposed that GRB progenitors are Wolf-Rayet (WR) stars \citep{WoosleyHeger06, Detmers08}, stripped stars originating from stars at $>$20--30 M$_\odot$ with a lifetime of only a few Myrs, which strengthens the connection to recent, massive star-formation, generally observed in long GRB hosts. 

One of the few, very nearby, GRBs was GRB 171205A, detected by the Swift satellite on Dec.\,5,\,2017, with a duration in $\gamma$-rays of T$_{90}=$189\,s and at a distance of only 163 Mpc or $z=0.037$ \citep{LucaNature}. It was a low-luminosity GRB, which is the predominant class of long GRBs, but due to their faintness they can only be observed at low redshift \citep{Liang07, Patel23}. We initiated an early observing campaign and detected the first traces of the SN only one day after explosion \citep{deUgarte_171205A_cocoon_GCN}. We observed two different components in the SN: an early component with fast ejecta and an inverted composition, which was interpreted as emission from a cocoon around the emerging GRB jet \citep{LucaNature}, and a later component with rather standard GRB-SN features. The proximity allowed us to observe this SN into the nebular phase, yielding one of the longest and high cadence follow-up datasets of a GRB-SN (de Ugarte Postigo et al. in prep.). The host is a high-mass, grand-design spiral ($\log_{10}(M/M_{\odot})$ 10.29$_{-0.05}^{+0.06}$), an unusual type for a long GRB host galaxy \citep[see e.g.][]{Fruchter06, Schneider22}.

In this paper we present extensive multiwavelength data of the host galaxy of GRB\,171205A, focusing in particular on the IFU dataset obtained with the Multi-Unit Spectroscopic Explorer (MUSE) at the Very Large Telescope (VLT), complemented by CO(1-0) observations with the Atacama Large Millimetre Array (ALMA), H~{\sc i} observations of the 21~cm line with the Giant Meterwave Radio Telescope (GMRT) \citep[both datasets are presented in][]{deUgarte24} and the Very Large Array (VLA) as well as imaging with the HST in UV and visible bands. In Sect. \ref{sect:observations} and \ref{sect:analysis} we present the observations and data analysis of the MUSE datacube as well as the auxiliary data Sect. \ref{sect:results} presents the results from the spatially resolved study, Sect. \ref{sect:kinematics} the resolved kinematics of the host using ionized, molecular and neutral gas. Finally, in Sect. \ref{sect:discussion} we discuss the results and compare the host and the GRB site to other GRB hosts at low redshift. Throughout the paper we use a flat lambda CDM cosmology as constrained by Planck with $\Omega_m$$=$0.315, $\Omega_\Lambda$$=$0.685 and H$_0$$=$67.3 km~s$^{-1}$~Mpc$^{-1}$ \citep{planck18}.

\begin{figure*}[hbt!]
    \begin{center}
    \includegraphics[width=18cm]{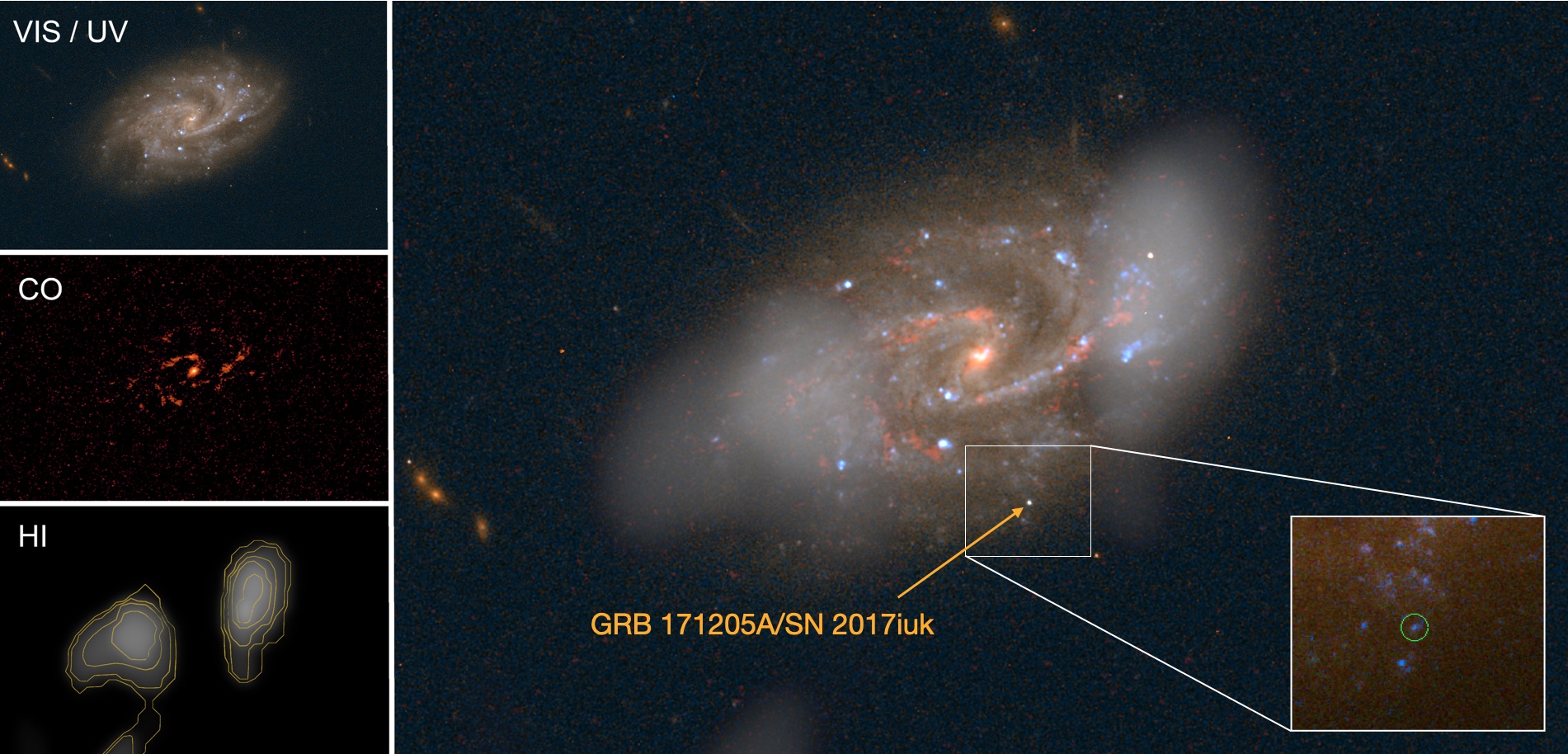}
    \caption{Small panels, top left: False-color image based on HST images in filters F606W (red), F473W (green) and F300X (blue) taken in July 2018 with the GRB-SN still present. Centre left: CO(1-0) emission from ALMA. Bottom left: H~{\sc i} emission from JVLA, including contours. Large panel: False-color image combining the HST data together with the CO(1-0) emission from ALMA (dark red) and H~{\sc i} emission from JVLA (white clouds). The inset shows data from December 2019 when the SN had faded and revealed the underlying star-forming region at the GRB site, marked by a green circle. In all images North is up and East is left.}
    \label{fig:HST}
    \end{center}
\end{figure*}

\section{Observations and data analysis}\label{sect:observations}
\subsection{IFU data and additional spectroscopy}
 
We observed the host galaxy of GRB 171205A \citep[6dFGS gJ110939.7-123512, ][]{Jones2009} with the integral-field unit MUSE instrument mounted on the UT4 at the European Southern Observatory in Paranal (Chile). Observations started on March 20, 2018, i.e. 105 days after the GRB detection, and consisted of four exposure of 900s each in wide field mode.  These observations were obtained within the Stargate program \footnote{PI: N. Tanvir, Program ID 0100.D-0649}. The raw data were reduced using a set of \textsc{esorex} scripts (v. 3.13.7) that provide a fully-combined and flux-calibrated datacube, which was used for the analysis presented in this paper. MUSE spaxels (in IFU data each ``pixel'' in the spatial direction also contains spectral information and is hence called a ``spaxel'') have a size of 0\farcs{2}$\times$0\farcs{2}, however, due to the seeing conditions, the actual resolution is 0\farcs{7}, corresponding to a physical resolution of 0.5\,kpc. The observing log of this and all other observations used in this paper is displayed in Table~\ref{table:obs}.

Additional data of the environment of SN 2017iuk were obtained 14 months after the SN explosion with X-shooter at the VLT. A set of eight exposures of 300s were obtained in the three arms of X-shooter on February 14, 2019, using the ABBA nod-on-slit method, with a nod throw of 6 arcsec along the slit. The data were not reduced using the X-shooter nodding mode pipeline, instead we reduced each single spectrum as if it had been obtained without any offset along the slit (stare mode). Given that the full 11 arcsec length of the X-shooter slit was entirely within the host galaxy, a conventional background subtraction was not possible. To correct the sky background in these spectra we used the Cerro Paranal Advanced Sky Model provided by ESO \citep{Noll2012,Jones2013}. In spite of the quality of this model it is never as precise as an observed correction and this could be the reason for the small wiggles seen in the resulting spectra (see Fig. \ref{fig:integratedspecs}). The spectral series have finally been stacked to result in a spectrum of the immediate environment of SN 2017iuk, but contaminated by sky emission lines and continuum.  

We also obtained IFU data of a companion galaxy at a distance of 186\,kpc discovered originally in \ion{H}{i} to have the same redshift and which could have interacted with the host (see Sect. \ref{sect:discussion}). The data have been presented already in \cite{deUgarte24}, but they only show the integrated spectrum and do not list emission line fluxes. The galaxy was observed with the Potsdam Multi-Aperture Spectrophotometer (PMAS) mounted on the 3.5~m telescope of Calar Alto Observatory \citep[Spain;][]{Roth05} using the 1\farcs0 lens array, which gives a field-of-view of 16$^{\prime\prime}\times16^{\prime\prime}$ \footnote{CAHA project no. 19A-3.5-029, PI: L. Izzo}. We used the V500 grism, providing a wavelength coverage from 3700--7700 \AA{} at an average spectral resolution of $\sim4.3~$\AA{} at a rotation angle of 143.5 deg. Four pointings were needed to cover the galaxy. Data reduction was done with the P3D reduction tools \citep{Sandin10} and the different pointings combined to a single cube.

\subsection{Imaging data from HST}
We obtained observations of GRB 171205A and its host galaxy with the {\em Hubble Space Telescope (HST)}. Observations were obtained at 
three epochs on 25 December 2017 using filter F300X, on 02 July 2018 in filters F300X, F475W and F606W, and a final epoch on 02 December 2019 using F300X, F475W and F606W (1083\,s). The last images were taken two years after the GRB, hence we expected the contribution from the afterglow or supernova to be minimal. 

The imaging data were retrieved from the {\em HST} archive after standard processing (debiasing, flat-fielding and charge transfer efficiency corrections). The data were subsequently drizzled via {\tt astrodrizzle} to a plate scale of 0.025 arcseconds per pixel. A color composite image of the second epoch images is shown in Fig.~\ref{fig:HST}. The last epoch at two years post GRB reveals an H~{\sc ii} region right at the location of the GRB, making an association of this region to the progenitor site of the event very likely.

\subsection{Long wavelength data}
Our study is complemented with data from ALMA and GMRT already presented in \cite{deUgarte24} as well as data from the JVLA \citep{Arabsalmani22}. ALMA Band~3 observations covering CO(1--0) at the redshift of the host were taken on Dec. 7 and 8, 2017 with a total integration time of 4.9\,h. The data have a spectral resolution of 0.977~MHz or $\sim2.6$ km~s$^{-1}$ and a spatial resolution of $0\farcs31\times0\farcs24$. Data were calibrated with Common Astronomy Software Applications (CASA) \citep{McMullin07,CASA2022}, smoothed to 10\,km~s$^{-1}$, which yields an r.m.s. of 0.2\,mJy and continuum subtracted. GMRT observed the field with a total time of 12 hr distributed in two epochs at an observing frequency of 1.362 GHz, a total bandwidth of 16.7 MHz, and a channel width of 32.6 kHz, equivalent to 7 km~s$^{-1}$ \citep{deUgarte24}. The data were reduced with the in the Astronomical Image Processing software \citep{vanMoorsel96} and combined to a single cube. The final cubes have beam sizes of 13'' $\times$ 19'' and 21'' $\times$ 27''. 

We redid the analysis of the data taken by JVLA, which observed the field for 10.5\,hr over two epochs on 9 Mar 2019 and 4 May 2019 (proposal ID: VLA/19A-394; \citealt{Arabsalmani22}). The observing frequency of 1.368\,GHz, with a channel width of 3.9\,kHz and a 16\,MHz bandwidth, which is equivalent to a velocity resolution of 0.8\,km\,s$^{-1}$ and a total velocity coverage of 3500\,km\,s$^{-1}$, and have a beam size of 7\farcs1 $\times$ 6\farcs3.

The standard pipeline-calibrated data \citep{Kent20}\footnote{\url{https://science.nrao.edu/facilities/vla/data-processing/pipeline}} were processed through standard procedures in CASA. Corrupted visibility was  flagged before the datasets from the two epochs were combined together. On the combined dataset, we used the line-free data to produce a standard continuum image, using channels averaged to 0.5\,MHz widths and a robustness of 0.5. The image has a restoring beam of $6.0\arcsec \times 4.3\arcsec$ and provides a model of the continuum emission, which was used to apply one round of phase-only self-calibration to the data. Our self-calibrated continuum image shows bright extended emission from a source North-East of the GRB as well as emission from the GRB itself, with a measured flux density of $2.76\pm0.11$\,mJy, consistent with previous studies at a similar epoch \citep{Arabsalmani22, Leung21}. The continuum emission, interpolated from line-free channels on each side of the \ion{H}{i} emission line, was subtracted from the self-calibrated visibility, leaving only the line emission data. A spectral cube was imaged, centered on $1369.85\,$MHz (corresponding to a redshift of $z=0.0369$), with a robustness of 0.5, 15 channels, and a velocity resolution of 34\,km\,s$^{-1}$, which had been suggested in \cite{Arabsalmani22} as the optimal velocity resolution for this dataset. The cube channels were spatially smoothed to a common resolution of $12\arcsec \times 8\arcsec$ and then smoothed in the velocity axis using the Hanning smoothing function to ensure that only emission components correlated along the spatial and velocity axes are identified. This resulting cube was then masked to a threshold of $0.55\,$mJy, below which local noise peaks start to contaminate the line emission. Using the masked cube we finally produced the moment 0 (total intensity) and moment 1 (velocity field) maps. A similar procedure was followed to obtain the total intensity and velocity field maps for the companion galaxy. The main difference is a cube that is phase-rotated to the companion coordinates in the imaging step and a velocity resolution of 20\,km\,s$^{-1}$, instead of 34\,km\,s$^{-1}$.

To find and detect sources in the JVLA data we used version 2 of the \ion{H}{i} Source Finding Application \citet[SoFIA2]{Serra15, Westmeier21}. SoFIA2 is a pipeline for identifying extragalactic \ion{H}{i} galaxies in 3D spectral line data in \ion{H}{i} surveys such as the Widefield ASKAP L-band Legacy All-sky Blind surveY \citet[WALLABY]{Koribalski20}. It uses the \textit{smooth+clip} algorithm \citep{Serra12} to search for sources of \ion{H}{i} emission at various angular and velocity resolutions by convolving the calibrated spectral cube with a user-specified 3D kernel. Sources above a user-specified threshold with spatial and spectral correlations in the cube are identified and automatically characterised, with the output for each identified source being their moment 0 and 1 maps as well as their line profile. In addition, moment 0 and 1 maps are  output for the entire field. When we applied SoFIA2 to our data cube (smoothed to a velocity resolution of 10\,km~s$^{-1}$), we optimised our parameters\footnote{For a complete description of the parameters, see: \url{https://gitlab.com/SoFiA-Admin/SoFiA-2/-/wikis/SoFiA-2-Control-Parameters}} for source finding (increase recall, decrease precision) and subsequently checked the properties of the sources output by the pipeline individually to check if they are real or artifacts. The key parameters were: \texttt{flag.threshold=7.0} (in multiples of standard deviation - used for to discard data affected by interference or artifacts), \texttt{reliability.threshold=0.8}  (reliability of a detection - between 0 and 1 - evaluated by comparing the total positive flux of the detection with the density of positive and negative detections in 3D space), \texttt{scfind.threshold=3.8} (lower source-finding threshold score - which is in units of the measured noise level - the higher the recall/lower the precision),  \texttt{scfind.kernelsXY=0,6,12,18,24} (Gaussian kernels applied to the data cube in units of the FWHM of the Gaussian used to smooth the data in the spatial axes), and \texttt{scfind.kernelsZ=0,3,7,15,31} (full width of the Boxcar kernels used to smooth the data cube along the velocity axis).

\begin{table*}[ht!]
\caption{Log of the observations used in this paper. The last column refers to the presence or absence of the underlying GRB-SN.}            
\label{table:obs}      
\centering                       
\begin{tabular}{c c c c c c c}       
\hline\hline                
Date            & Instrument    & Band          & Exposure    & Spatial res. & Spectral res. & GRB/SN? \\    
            %[Galactic extinction EBV=0.138]
\hline  
20 March 2018   &   MUSE/VLT    &4650-9300\AA{} & 4$\times$900\,s & 0\farcs7 & 171-80 km\,s$^{-1}$ & yes\\
3 April 2019 & PMAS/CAHA & 4700-7700 & 4$\times$900\,s & 1\farcs0 & $\sim$200 km\,s$^{-1}$  & companion!\\
14 Feb. 2019    & X-shooter/VLT & 3000-20000    & 8$\times$300\,s & (longslit)  & 55-34 km\,s$^{-1}$ &no\\
\hline 
25 Dec. 2017    &   HST         & F300X         &  1044\,s  & 0\farcs10 &--- &yes   \\
2 July 2018     &   HST         & F300X         &  2400\,s  & 0\farcs10 & ---&yes     \\
2 July 2018     &   HST         & F475W         &  1200\,s  & 0\farcs10 & ---&yes     \\
2 July 2018     &   HST         & F606W         &  1050\,s  & 0\farcs10 &--- &yes      \\
2 Dec. 2019     &   HST         & F300X         & 2400\,s   & 0\farcs10 &--- &no   \\
2 Dec. 2019     &   HST         & F475W         & 1083\,s   & 0\farcs10 & ---&no   \\
2 Dec. 2019     &   HST         & F606W         & 1083\,s   & 0\farcs10 &--- &no   \\
\hline 
7/8 Dec. 2017   &  ALMA         & CO(1-0)       & 4.9\,h    & 0\farcs31$\times$0\farcs24& 10 km\,s$^{-1}$ & yes$^1$    \\
11 Feb./15 Mar. 2018 & GMRT     & H~{\sc i}            &  12\,h & $13^{\prime\prime}\times19^{\prime\prime}$ & 7 km\,s$^{-1}$& yes$^1$         \\
9 Mar/4 May 2019 & JVLA & H~{\sc i} & 10.5h &  7\farcs1 $\times$ 6\farcs3 & 0.8 km\,s$^{-1}$& no         \\
\hline                                
\end{tabular}
\tablefoot{$^1$GRB not visible after continuum subtraction.}
\end{table*}

\section{Analysis of the different datasets}\label{sect:analysis}
\subsection{Analysis of individual spaxels}
\subsubsection{Emission line maps}
For the MUSE data we extract emission line maps of all lines detected in individual spaxels (see figures in the Appendix). We obtain the line fluxes by summing the flux in a window corresponding to 12.5 \AA{} (10 wavelength steps or 400--750 km\,s$^{-1}$ depending on the wavelength) and subtracting the mean of a background window in a region free of other emission lines 40--60 wavelength steps away from the emission line, unless this wavelength region was contaminated by another line or atmospheric feature, in which case we adopt a different distance of the window from the line. Since the galaxy has a considerable velocity field we have to adjust the window to the position of the line in different parts of the galaxy. We do this by fixing the center of the window to the center of a Gaussian fit to the brightest line, H$\alpha$, which also gives us the velocity field which will be discussed further in Sect.~\ref{sect:kinematics}. With this method of choosing a narrow window we avoid getting dominated by noise from the continuum of the galaxy, which is crucial for obtaining fluxes from faint lines. Finally, we apply a cut of S/N$=$5 for the bright lines (H$\alpha$, H$\beta$ and [O~{\sc iii}]\, $\lambda$5008) and S/N$=$3 for the rest of the lines. 

H$\alpha$ and H$\beta$ maps were derived separately after correcting for underlying stellar absorption. The stellar absorption was fitted in the process of the stellar population fitting described in Sect.~\ref{subsect:SPanalysis} in Voronoi binned maps. The corrections derived for each Voronoi bin were then interpolated to derive H$\alpha$ and H$\beta$ maps for each MUSE spaxel. We furthermore correct the final emission line fluxes for a Galactic extinction value of E(B--V)$=$0.138 \citep{Schlegel1998} and the intrinsic extinction in the host using the E(B--V) map derived in Sect.~\ref{subsect:extinction} smoothed with a Gaussian kernel of 3 spaxels. 

Using the data from PMAS/CAHA we also produce emission line maps of H$\alpha$ and [O~{\sc iii}] of the companion galaxy (see Fig.~\ref{fig:companion}). [N~{\sc ii}]~$\lambda$6585 and H$\beta$ were not detected with sufficient S/N in most of the individual spaxels to make maps of those lines.

\subsubsection{Pixel counting statistics}
\label{Sect:pixcounting}
Several previous works have established a method to determine the fraction of light at the GRB site compared to the light distribution in the rest of the galaxy. The reason behind this is that bright regions are usually associated with star-formation. This work was pioneered by \cite{Fruchter06} and has subsequently been updated including larger sample of GRB hosts \citep{Kelly08, Svensson10, Blanchard16, Lyman17}, almost exclusively using {\em HST} data due to the higher spatial resolution. However, there is quite some diversity in which filters have been used in individual analyses. 

We first establish a mask containing the values where there is flux from the galaxy and furthermore filter out values that are infinite or below zero. Then we take the flux value of all remaining spaxels in the F606W and F300X filters and sort them by increasing flux. For each ranked pixel we then determine the cumulative flux, which is the sum of all fluxes from the faintest to the respective pixel, and normalize it by the total flux. Hence each pixel now has a value between 0 (faintest pixel) and 1 (brightest pixel), the fractional flux. Finally, we determine the fractional flux at the GRB position, which can then be compared to the fractional flux of other GRB sites in the samples mentioned above (see Sect.~\ref{subsect:sfr}).

\subsection{Voronoi binned maps}

\subsubsection{Voronoi tessellation procedure}\label{sect:Voronoi}
To use the continuum emission for stellar population fits or absorption lines, the S/N per spaxel is not high enough and rebinning is needed. To this end we perform Voronoi tessellation as described by \citet{Cappellari2003} which bins the spaxels within an adaptive region size to achieve a constant S/N for a given property, in this case the continuum emission.

As reference to determine the S/N we use a region of the continuum near the NaD doublet at the host redshift, between 6130 and 6200 {\AA}. We first create a mask for the galaxy by taking only the spaxels that have an average S/N per dispersion element $\geq$3 in that spectral range. In this masked region, we run the Voronoi tessellation procedure requiring a S/N of 40 for the specified spectral region. The resulting bins have a fractional S/N scatter of less than 7\% around the goal of 40. The resulting map is composed of 232 spatial bins (see Fig.~\ref{fig:voronoi_segmap}).

\subsubsection{Stellar population fitting}\label{subsect:SPanalysis}
\label{starlight}
For each of the Voronoi bins we extract a single spectrum and error spectrum, which we correct for Galactic extinction assuming a value of $A_V=0.138$, as derived from the \citet{Schlegel1998} maps updated with the recalibration by \citet{Schlafly11}. The spectra are transformed to rest frame taking as reference the central wavelength of a Gaussian fit to the Balmer H$\alpha$ line. Strong emission lines and artefacts remaining from the reduction, mostly due to residuals of atmospheric emission line subtraction, were masked out before proceeding to the stellar population fit. 

Each of these spectra is fit using the STARLIGHT stellar population  code \citep{Cid2005,Cid2009}. STARLIGHT performs a fit of the observed spectral continuum using a combination of the synthesis spectra of different single stellar populations (SSPs). As reference sample, we use the SSP spectra from \citet{Bruzual2003}. To run STARLIGHT we use a set of models of 6 different metallicities (Z = 0.0001, 0.0004, 0.0040, 0.0080, 0.0200 and 0.0500, corresponding to 1/500 to 1.5 solar metallicty) and 25 ages for each metallicity ranging from 1 Myr to 18 Gyr. For the fit we used a Calzetti reddening law \citep{Calzetti2000}. STARLIGHT simultaneously fits the ages and relative contributions of the different SSPs, as well as the average reddening, providing a star formation history (SFH) for each binned spectrum of the datacube. We also made a fit for the integrated spectrum of the galaxy using all the spaxels in the mask described in Sect.~\ref{sect:Voronoi}

\subsubsection{NaD absorption}

The Voronoi binned spectra used for stellar population fit also allow for a resolved study of the absorption of the NaD doublet throughout the galaxy, which is an independent measure for extinction in the sightline. For each spectrum we fit a double Gaussian to the absorption of the doublet. The spacing of these two components is fixed to the known spacing of the NaD doublet (NaD\,$\lambda\lambda$5892,5898). 

Furthermore we determine the equivalent width (EW) of the doublet and its error by measuring the absorption of the features in a window that covers the two lines plus a range of 6\,{\AA} on either side. This measurement is thus model independent and gives us a more reliable value of the EW than the Gaussian fit.

\subsection{Integrated H~{\sc ii} regions and global properties}
Studying the properties of integrated H~{\sc ii} regions improves the S/N of the data. The values should otherwise follow similar trends to individual spaxels, except for changes within a single H~{\sc ii} region. However the physical resolution of the seeing-limited MUSE dataset does not allow to trace deviations within a single H~{\sc ii} region. We use the program ``H{\sc ii}explorer'' \citep{SanchezHII} written in python, which searches for peaks in an H$\alpha$ map and grows the region starting from the peaks down to a maximum radius defined beforehand, and which should match the typical size of an H~{\sc ii} region at the resolution of the data. For our purpose we use a peak value of 2.5$\times$10$^{-18}$~erg~cm$^{-2}$~s$^{-1}$\AA$^{-1}$, a threshold of 0.6$\times$10$^{-18}$~erg~cm$^{-2}$~s$^{-1}$\AA$^{-1}$ and a maximum radius of 4 pixels.

This yields a final number of 60 individual H~{\sc ii} regions (see Fig.~\ref{fig:voronoi_segmap}). The spectra of all spaxels in each of the 60 regions are integrated to 1D spectra which are then analyzed in the same way as we do for individual spaxels. Region 60 does not have enough S/N to measure individual lines and was therefore discarded. Region 36 was affected by a foreground star and is therefore also discarded. Because of the underlying stellar absorption in the Balmer lines (see Sect.~\ref{sect:Voronoi}) we determined the fluxes of H$\alpha$ and H$\beta$ separately from the previously corrected maps by extracting the integrated flux using the same mask as for the extraction of the spectra. We also determine the physical distance of the center of each region from the center of the galaxy by deprojection as described in Sect. \ref{sect:deprojection} and include the derived properties in the deprojected plots in Sect.~\ref{sect:results}. The integrated properties for all H~{\sc ii} regions are listed in Tab.~\ref{tab:HIIregions}.

\begin{figure*}[h!]
\centering
	\includegraphics[width=15cm]{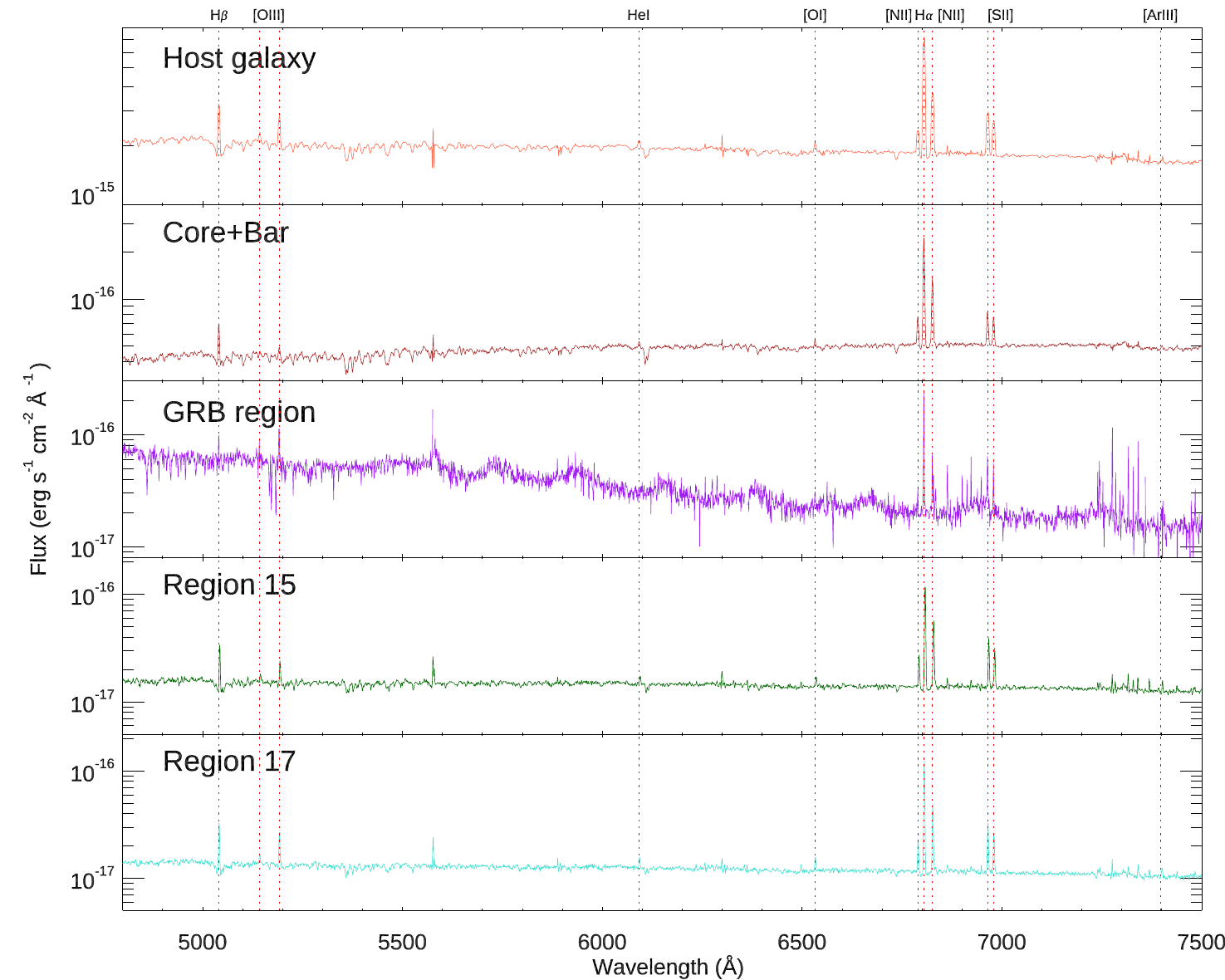}
 \caption{Integrated spectra and detected lines in (top to bottom): the integrated galaxy spectrum, the GRB region, the galaxy core and H~{\sc ii} regions 15 and 17 which show a very young stellar population.}
    \label{fig:integratedspecs}
\end{figure*}

Last, we extract integrated spectra of the entire host galaxy from the MUSE cube taking the spaxels included in the mask created in the Voronoi tessellation (see Fig.\ref{fig:voronoi_segmap} and Sect.~\ref{sect:Voronoi}), which excludes some low S/N spaxels. We also exclude the GRB region due the presence of the SN as well as a foreground star, which falls on top of the host in the N-W part of the galaxy. To study the integrated properties of the companion galaxy we use the integrated spectrum described in \cite{deUgarte24}. The integrated spectra of the host, the GRB region, the galaxy core and two young H~{\sc ii} regions of interest are shown in Fig.~\ref{fig:integratedspecs}, integrated fluxes and properties are presented in Tab.~\ref{tab:integratedspecs}.

\subsection{Deprojection}\label{sect:deprojection}
To study the different properties as function of physical distance from the center, we geometrically correct for the inclination of the galaxy, which is called ``deprojection''. This method can only be applied to disk galaxies, since the orientation of irregular galaxies is very difficult to derive. To calculate the deprojection needed for this galaxy we measure the position angle (PA) and inclination by determining the major/minor axis and the rotation angle of the projected ellipse of the galaxy on the sky (assuming that the galaxy in reality is a circular disk). The PA is the angle of the major axis measured North through East, for which we obtain 110 deg. This is well in agreement with the PA found in the kinemetry analysis with a median of 106\,$\pm$\,5.3 deg (see Sect. \ref{sect:kinematics}). The inclination $i$ is obtained via the ratio between the major ($a$) and minor axis ($b$) as $\mathrm{cos(i)=b/a}$, resulting in an inclination of 50 deg.

\begin{figure}[h!]
  \includegraphics[width=8.3cm]{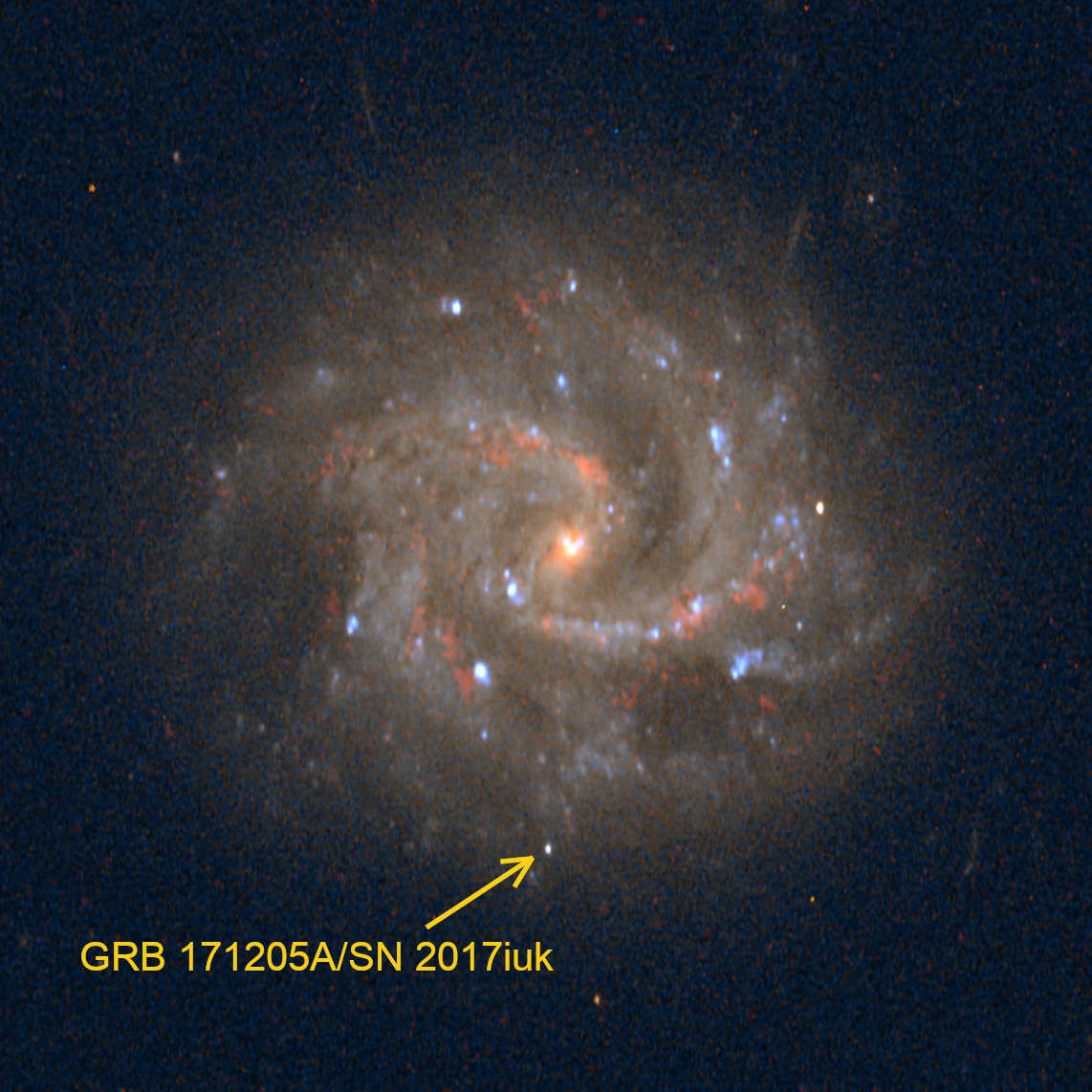}\\
  \includegraphics[width=8.4cm]{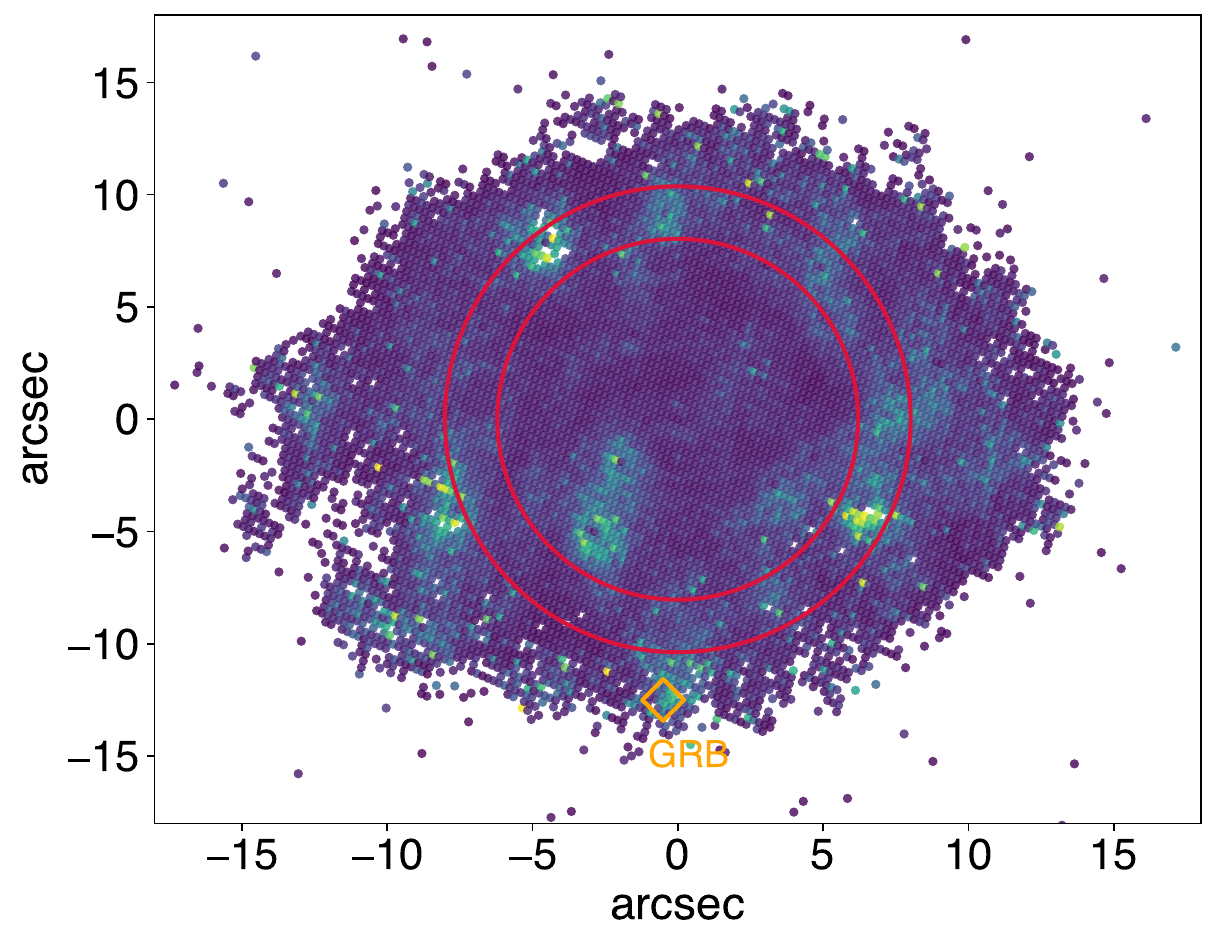}
    \caption{Top: Deprojected false-color image of the galaxy using the HST and ALMA data  using a PA of 110\,deg and an inclination of 50~deg. The whitish dot at the outskirts of the lower spiral arm is the GRB-SN also indicated in Fig. \ref{fig:HST}. Bottom: Deprojected specific SFR, obtained from H$\alpha$ and the HST UV continuum. Overplotted is the ring which contains most of the SF regions in the host, with exception of the GRB site (diamond) and some SF regions at the southern end of the bar (see also Sect.\ref{sect:discussion_alldata}).
    \label{fig:deprojectedhost}}
\end{figure}

We then use these values for $i$ and PA to correct each position in the galaxy for its ``real'' offset compared to the galaxy center using the \texttt{ im\_hiiregion\_deproject.pro} procedure\footnote{https://github.com/moustakas/moustakas-projects/} in IDL. The code outputs a radial distance of each pixel from the center from which we then derive a distance in kpc using the cosmology mentioned in the introduction. The same deprojection has been applied to the H~{\sc ii} regions, taking the center of each H~{\sc ii} region as the original position. The code also outputs the new position of each spaxel or H~{\sc ii} region, which can then be used to recreate a deprojected image of the galaxy or one of its properties. This is shown in Fig.~\ref{fig:deprojectedhost} where we plot a deprojected color image of the host and the specific SFR. In Sect.~\ref{sect:results} we plot the derived properties for each spaxel or H~{\sc ii} region as physical distance from the center derived via the deprojection algorithm. 

\section{2D properties of the host}\label{sect:results}

\begin{figure*}
\centering
	\includegraphics[width=8.5cm]{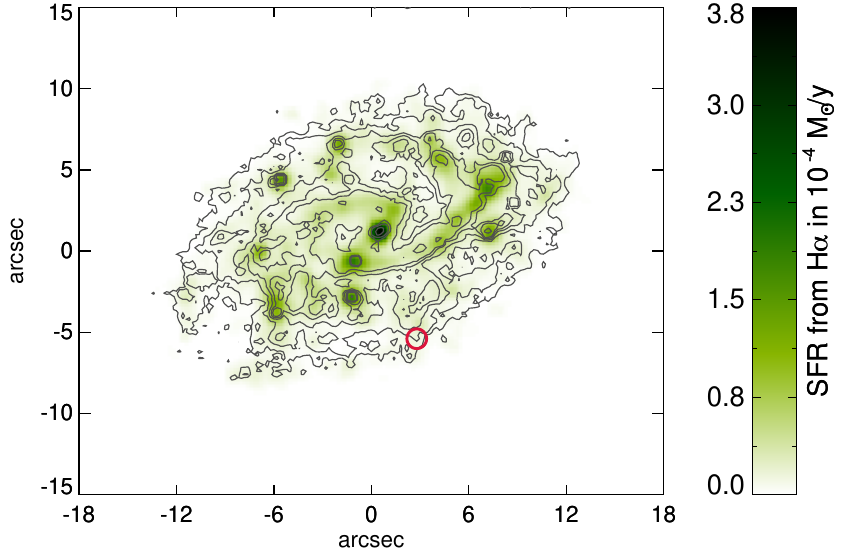}
	\includegraphics[width=7.cm]{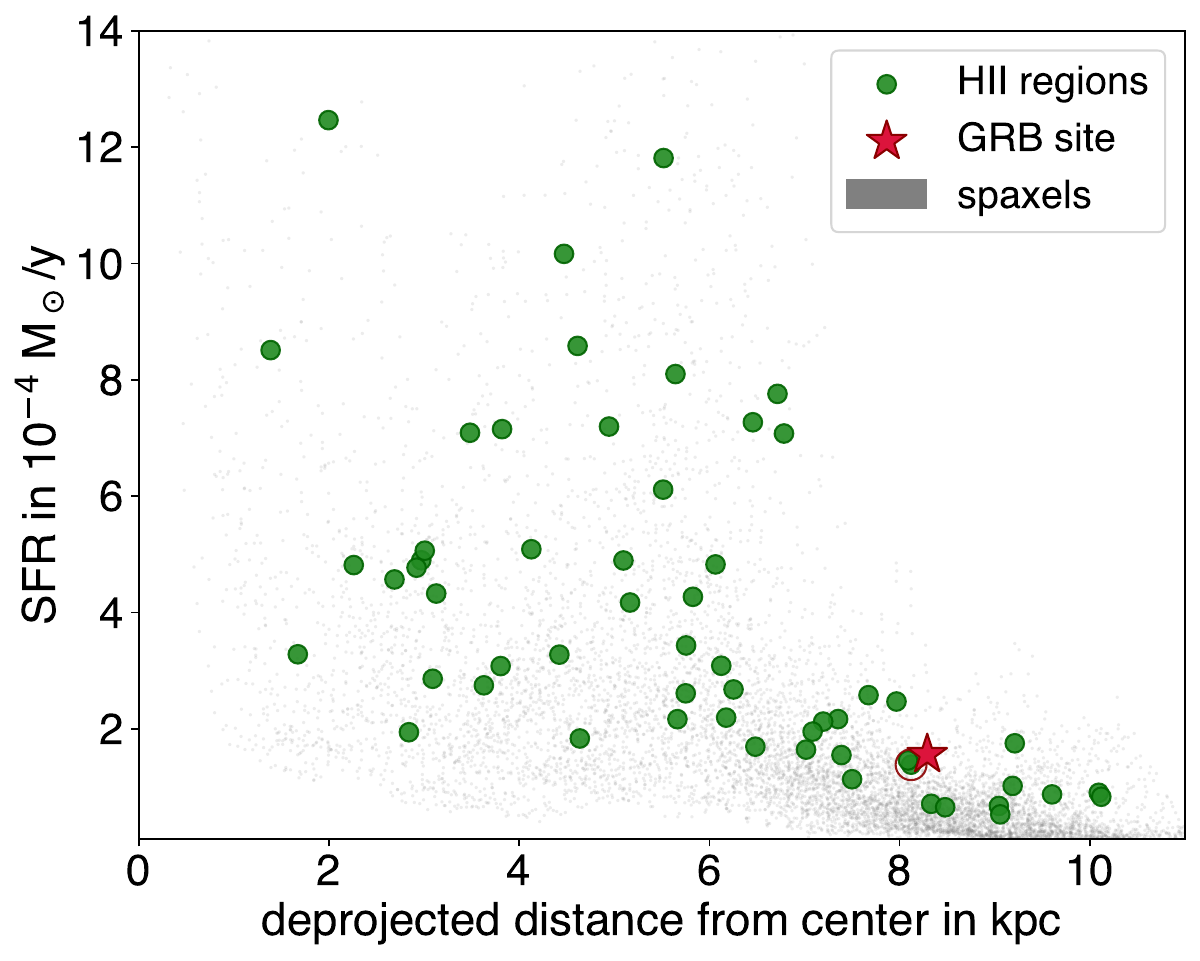}\\
		\includegraphics[width=8.5cm]{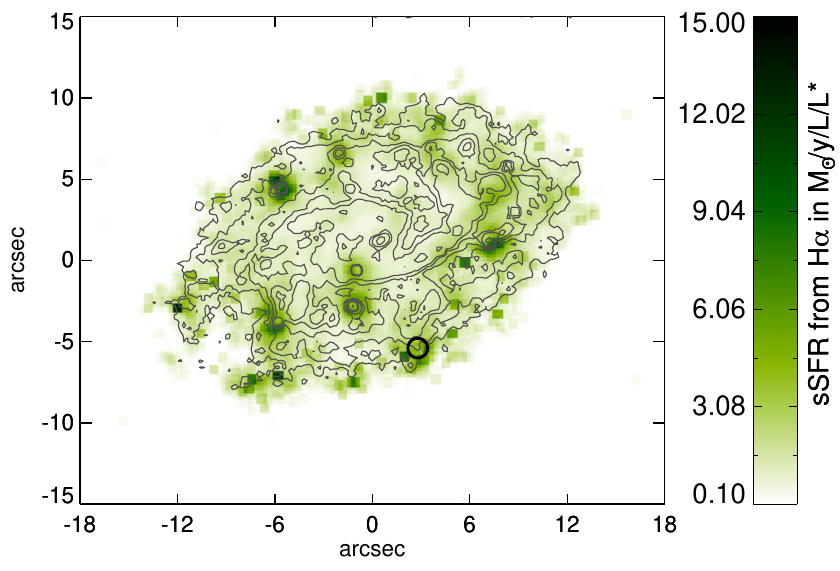}
	\includegraphics[width=7.cm]{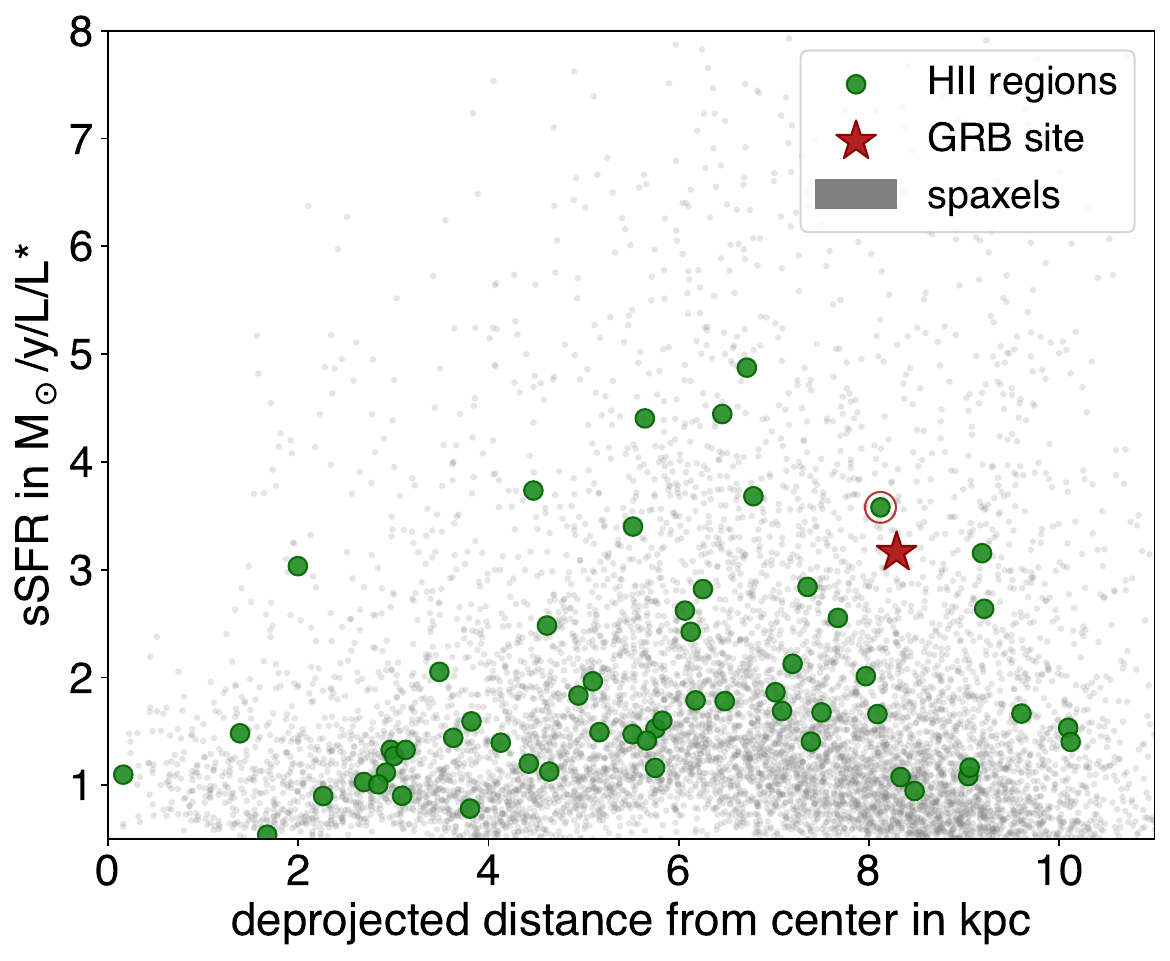}\\
	\includegraphics[width=8.5cm]{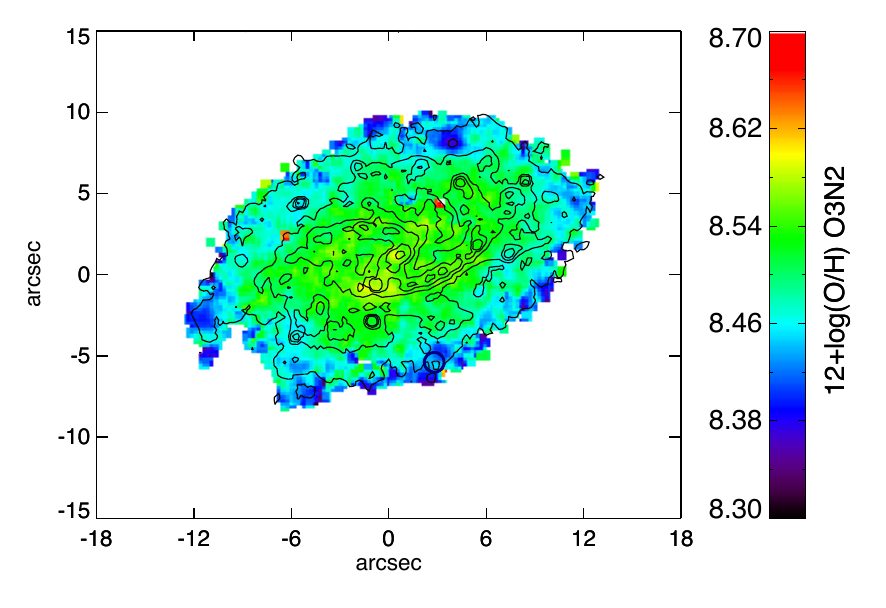}
	\includegraphics[width=7.cm]{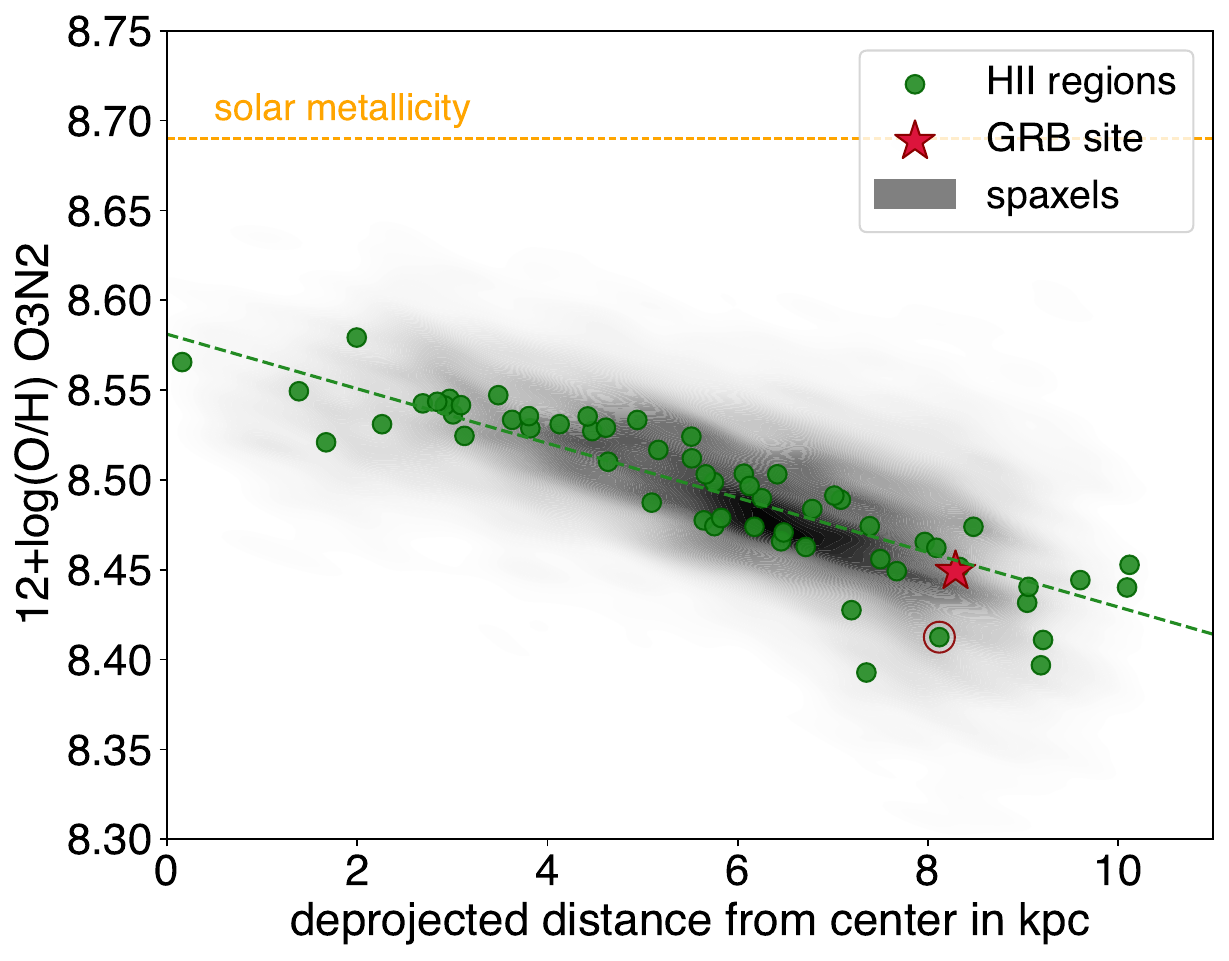}\\
 \includegraphics[width=8.5cm]{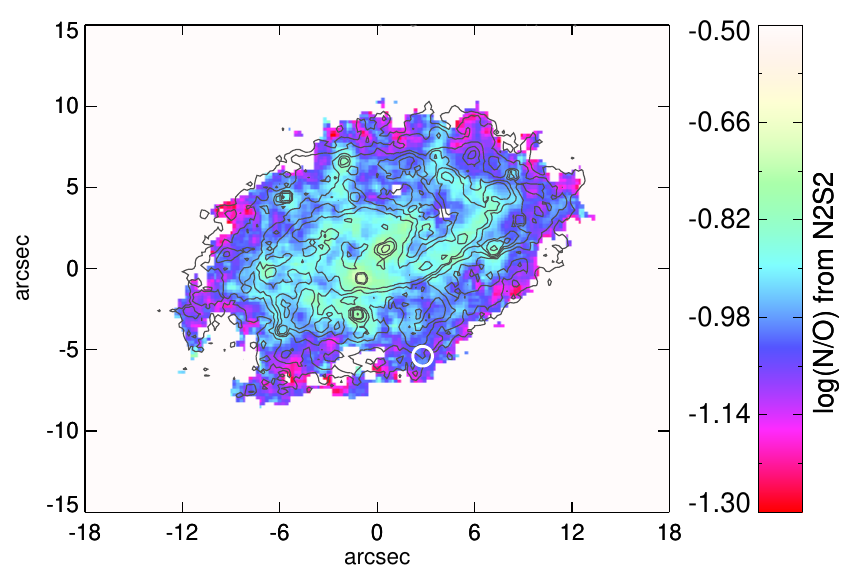}
 \includegraphics[width=7.cm]{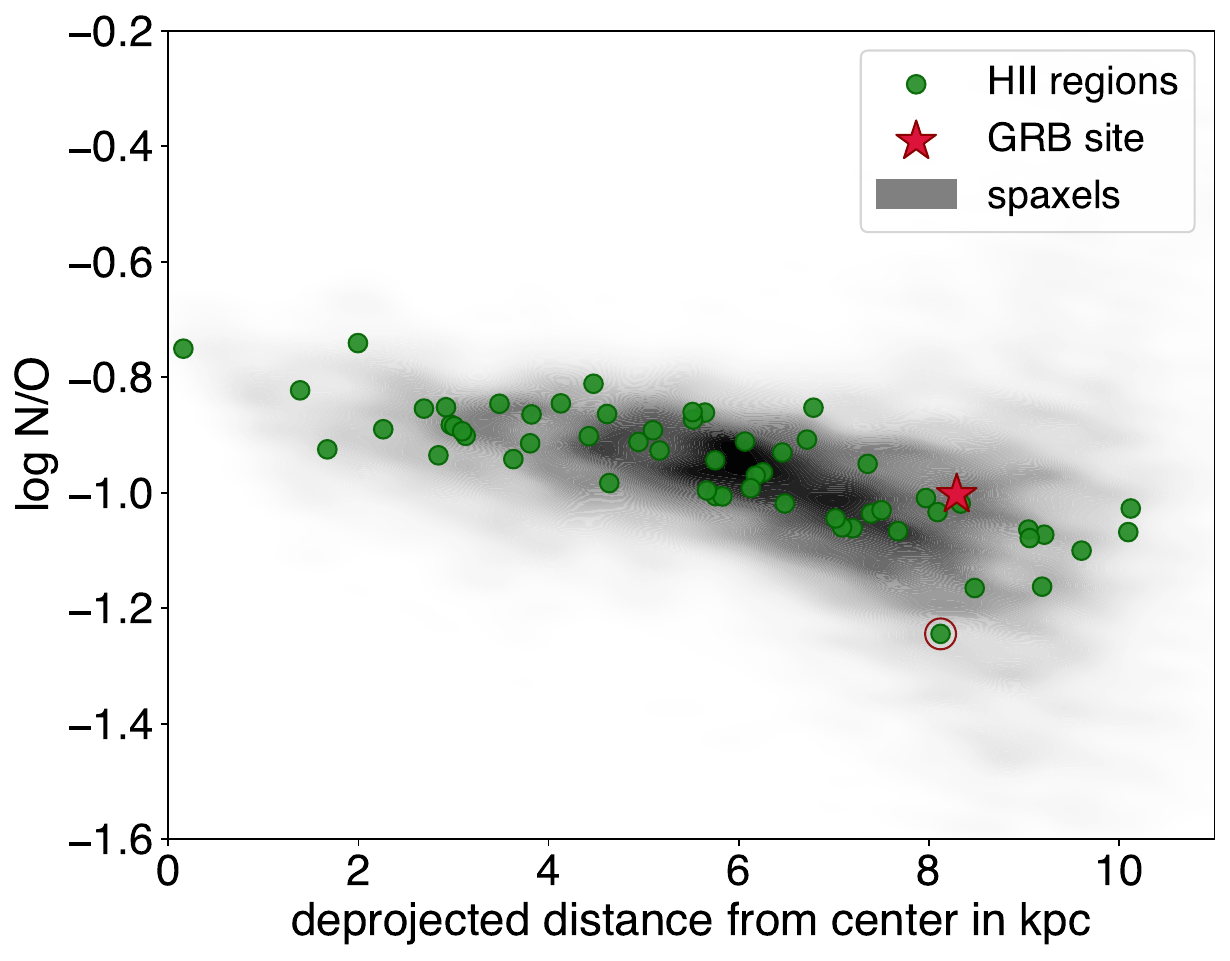}
	    \caption{1. row: SFR map and deprojected SFR from H$\alpha$, 2. row: Luminosity weighted SFR map and deprojected specific SFR from H$\alpha$. 3. row: Metallicity using the O3N2 parameter \citep{MarinoZ} and deprojected values, fitted with a linear gradient of --0.009\,dex/kpc. 4. row: Map and deprojected plot of the N/O abundance derived from the N2S2 parameter. The star marks the value for a region of 2$\times$2 spaxels at the GRB site, the red encircled H~{\sc ii} region marks the one of the GRB.}
    \label{fig:maps1}
\end{figure*}

\subsection{Star formation rate}\label{subsect:sfr}
We determine the current star-formation rate (SFR) from the H$\alpha$ flux according to \cite{Kennicutt92} assuming a Salpeter initial mass function (IMF) leading to a conversion of $\mathrm{SFR}=7.2\times$10$^{-42}$L$_{H\alpha}$. The result is plotted in Fig. \ref{fig:maps1} showing the 2D map and the values at the deprojected distance from the center. We also obtain a map of the luminosity weighted specific SFR (SSFR/L), where the SFR is multiplied with the ratio between the absolute $B$-band magnitude and the magnitude of an L$^\star$ galaxy M$_B\,=\,-21$\,mag \citep{Christensen04}: 
\begin{equation}
\mathrm{SSFR}_L=\mathrm{SFR} \times 10^{0.4\,(\mathrm{M}_B+20.1)}
\end{equation} . As $B$-band magnitude we use the HST F475W filter image from the observations in Dec. 2019, which is similar in wavelength coverage to a $B$-band filter (see Fig. \ref{fig:maps1}). The HST images from December 2019, which likely do not have any contamination from the GRB-SN any more, clearly show a small underlying SF region at the exact location of the GRB. Another, somewhat brighter SF region, lies to the South of it (see Fig.~\ref{fig:HST}).  

Ongoing star-formation is present in most of the host galaxies, however, the highest values of both absolute and specific SFR are found in several H~{\sc ii} regions distributed in a ring-like shape around the galaxy center, between $\sim$\,5\,kpc and $\sim$\,7\,kpc from the center, slightly closer to the center than the region of the GRB (see Fig.~\ref{fig:deprojectedhost} and Fig.~\ref{fig:maps1}). The absolute SFR at the GRB site is slightly above the average value at that deprojected distance, however, the size of its SF region is small ($\sim$ 300\,pc in diameter), compared to other SF regions in the host. The total SFR of the galaxy derived from H$\alpha$ is 2.28\,$\pm$\,0.05\,M$_\odot$~yr$^{-1}$. The SSFR/L at the  exact GRB location has a value of 3.1\,M$_\odot$~yr$^{-1}$ (L/L*)$^{-1}$, the SSFR/L of the total H~{\sc ii} region in which the GRB is located is slightly higher with 3.5\,M$_\odot$~yr$^{-1}$~(L/L*)$^{-1}$, which puts it among the highest 10\% in of all H~{\sc ii} regions in the galaxy. We will elaborate further on this in the discussion. 

The high SSFR does not reflect in the global luminosity of the GRB H~{\sc ii} region. Taking the results of the fractional flux analysis detailed in Sect. \ref{Sect:pixcounting}, the GRB region is among the 45\% brightest pixel in UV (F300X), but in the 80\% brightest pixels in F606W, which includes the H$\alpha$ emission line (see Fig.~\ref{fig:cumulative}). Most studies on the fractional flux use HST data that are at restframe UV wavelengths, due to the higher average redshift of GRB hosts, corresponding to our observations in the F300X filter. Compared to previous studies \citep{Kelly08, Svensson10, Blanchard16, Lyman17} the GRB site here is not a particularly bright region at wavelengths dominated by emission from massive stars. This might be explained by the unusual type of host galaxy for this GRB, where the GRB site was only one of several highly star-forming regions, while in the more common irregular hosts the GRB is more likely to be found in one of the few luminous star-forming regions.

\subsection{Metallicity and N/O abundance}
We determine the metallicity from emission lines using different strong emission line ratios. The direct $T_e$ method \citep{Peimbert67} cannot be applied here as the temperature sensitive [O~{\sc iii}] $\lambda$4363 line is only detected at a low S/N in a few regions of the host. Instead we use the O3N2 and N2 parameters in the re-parametrisation of \cite{MarinoZ} based on the CALIFA sample of low redshift galaxies. 

The O3N2 parameter shows a smooth negative metallicity gradient with no particular features. We fit the metallicity vs. distance with a simple linear fit and obtain a gradient of --0.015\,dex~kpc$^{-1}$, which is shallow compared to other local disk galaxies \citep[see e.g.][]{Sanchez14,Bresolin19} and does not show any flattening as observed in some of them. The metallicity in the GRB H~{\sc ii} region is slightly lower than the average metallicity at that distance from the center (the typical metallicity error of the integrated H~{\sc ii} regions is $\sim$0.01\,dex), the GRB spaxel itself is consistent with the metallicity grandient. The N2 metallicity map shows a lower metallicity in the strong SF regions on top of the metallicity gradient (see Fig.~\ref{figapp2}). In the past it has been pointed out that the N2 parameter depends on the ionization, which can affect the results \citep{Schady15} and, in fact, the regions of low metallicity clearly correlate with regions of high ionization (see next Section). The O3N2 parameter handles the ionization issue somewhat better and, as shown in Fig.~\ref{fig:maps1}, the resulting metallicity map is smooth and shows no low metallicity regions specifically in the bright SF regions. While the absolute value of the metallicity can depend on the metallicity calibrator used, the relative calibrations between different regions and spaxels, however, are reliable.

\begin{figure}
	\includegraphics[width=9cm]{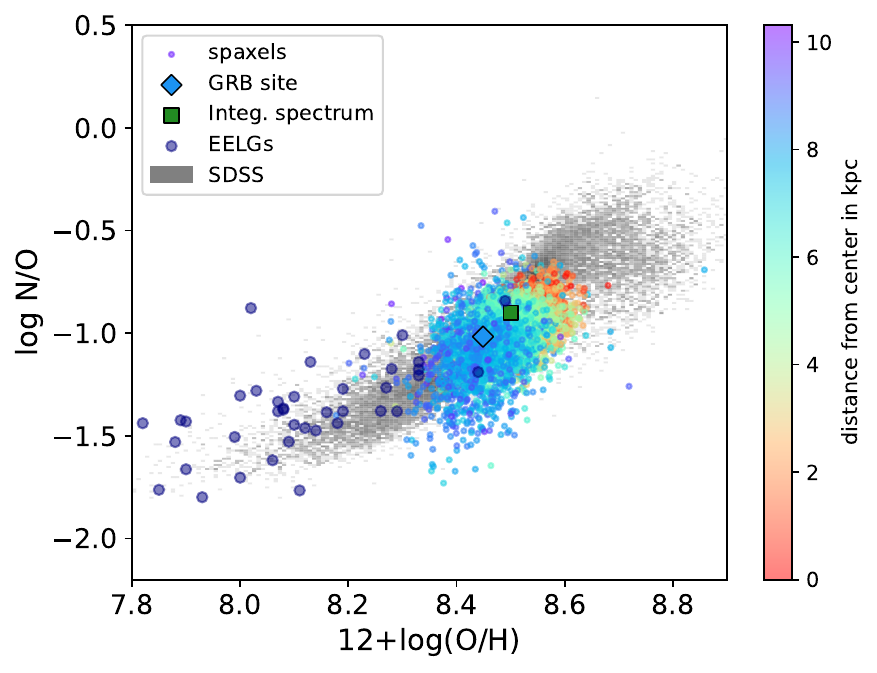}
	    \caption{N/O abundance vs. metallicity determined by the O3N2 parameter for different spaxels in the host of GRB171205A color coded by distance from the center compared to star-forming galaxies in the SDSS (grey grid) and a sample of EELGs from \cite{Amorin15}. EELG metallicities have been determined largely by the T$_e$ method \cite{Amorin15} while for the SDSS galaxies we use the same O3N2 calibration as for the spaxels in the host.}
    \label{fig:NO}
\end{figure}

To further investigate the abundances in the host we plot the ratio of N/O (see Fig. \ref{fig:maps1}), derived from the N2S2 parameter (N2S2$=$log$_{10}$([N~{\sc ii}] $\lambda$6585 / [S~{\sc ii}] $\lambda$6714,6731) as originally introduced by \cite{Pilyugin04, Molla06}. We cannot determine the oxygen abundance directly since [O~{\sc ii}]~$\lambda\lambda$3727,3729 is not in the range of the MUSE spectrum. Metal-poor, highly star-forming galaxies such as extreme emission line galaxies (EELGs) show an overabundance of nitrogen vs. oxygen compared to the general population of star-forming galaxies in the SDSS \citep{Amorin15}. The reason for this could be WR stars, low-metallicity intermediate stars or in-/outflows, but is likely linked to young stellar populations. The host of GRB171205A does not show a high ratio of N/O like EELGs (see Fig. \ref{fig:NO}). The different spaxels in the host, in fact, follow very closely the N/O vs. metallicity correlation of SDSS SF galaxies. The regions of higher N/O ratio follow the inner spiral arms, and the ratio is highest in the bar, in particular at the end of the bar S-E of the center. The N/O  abundance ratio at the GRB site is very similar to the one expected at this metallicity, looking at the entire H~{\sc ii} region of the GRB, the value, however, is one of the lowest in the host (see Fig. \ref{fig:maps1}).

\subsection{Extinction}\label{subsect:extinction}
\begin{figure*}
	\includegraphics[width=\columnwidth]{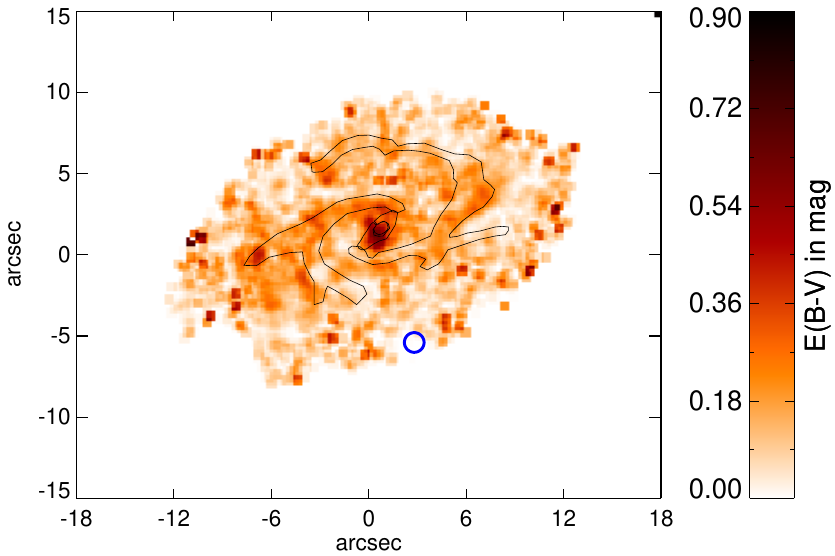}
	\includegraphics[width=7.8cm]{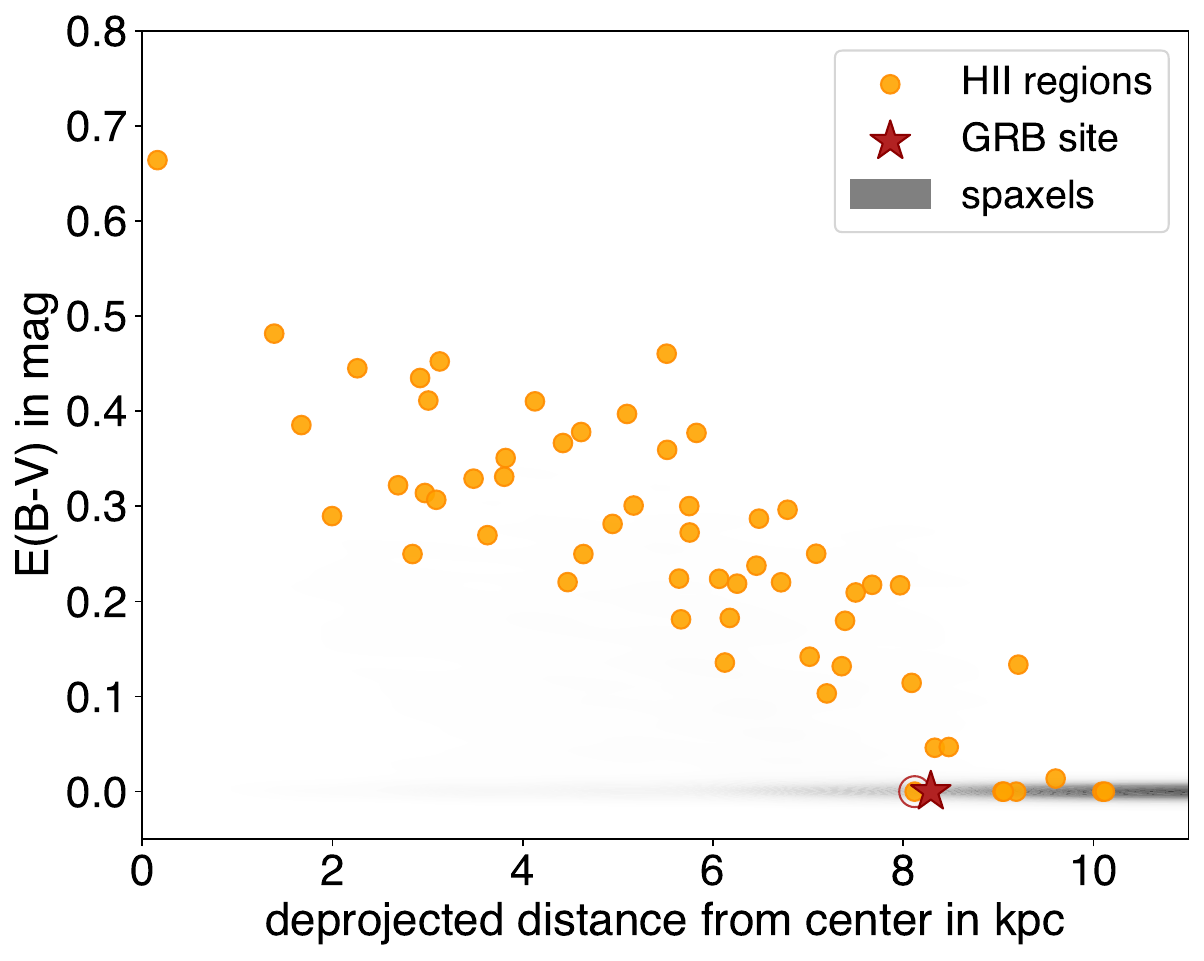}\\
	\includegraphics[width=\columnwidth]{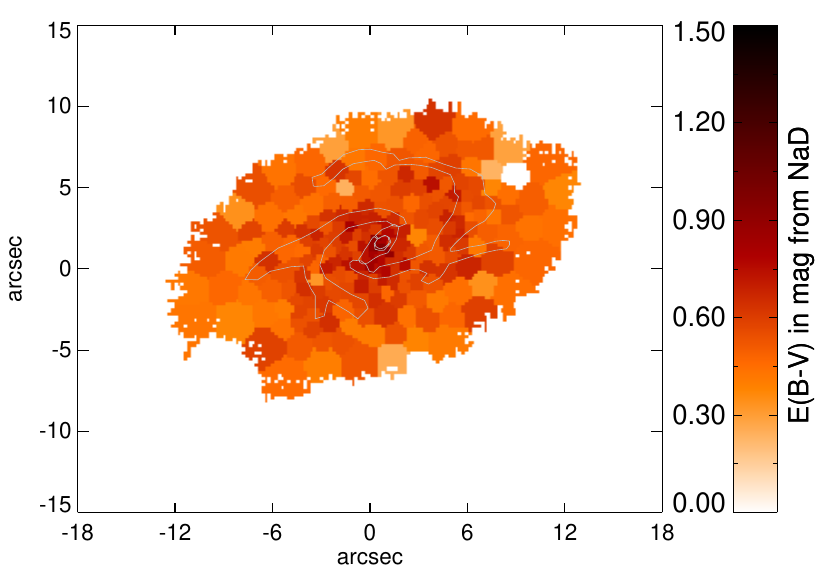}
 \includegraphics[width=\columnwidth]{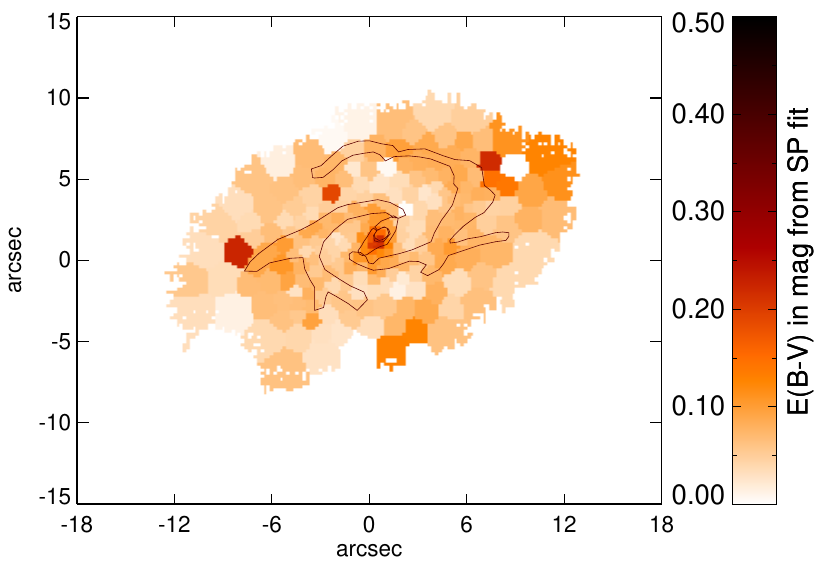}
	
	    \caption{Top: E(B--V) map from the Balmer decrement, after correcting H$\alpha$ and H$\beta$ for stellar absorption (top) and deprojected values. Bottom left: Extinction map using the correlation between the NaD EW and extinction from \cite{Poznanski12}. Bottom right: Extinction map from the stellar population fitting. The maps at the bottom have been obtained from the Voronoi binned cube, the white region is a bin affected by a foreground star and has been omitted. Overplotted contours are derived from the ALMA CO(1-0) map (see \citealt{deUgarte24}).}
    \label{fig:ext}
\end{figure*}

In Fig. \ref{fig:ext} we plot the 2D extinction map and the radial distribution of the extinction, derived from the Balmer decrement using H$\alpha$ and H$\beta$, which, in the Case B recombination and zero extinction have an intrinsic ratio of 2.76 \citep{osterbrock}. The colour excess in most regions is low or even zero with a mean of 0.20\,$\pm$\,0.17\,mag. The highest extinction is observed in the center of the galaxy, which has $\mathrm{E(B-V)}=$1.3\,mag. The location and H~{\sc ii} region of the GRB does not show any extinction in the MUSE cube but shows some extinction in the late X-Shooter spectrum (see Tab.~\ref{tab:integratedprops}). The regions of highest extinction correlate very well with the CO emitting regions, which again correlate with the inter-arm regions and dust along the spiral arm (see Fig.~\ref{fig:deprojectedhost}). 

We furthermore derive an extinction from NaD absorption using the Voronoi binned cube, according to the relation between the NaD EW and E(B--V) established in \cite{Poznanski12}. The distribution follows a similar pattern with higher extinction in the center, but with overall higher values, however the relation between NaD EW and extinction has larger uncertainties than the Balmer decrement. The correlation with CO emission is less clear due to the lower spatial resolution used for this analysis. Again, the GRB region shows zero extinction. Finally, we derive an extinction map using the stellar population fitting described in the next Section (see Fig.\ref{fig:ext}). Here we get again higher values in the core and zero extinction at the GRB site. Some of the young SF regions (see Sect.~\ref{sect:SP}) in the spiral arms also show high extinction.

\subsection{Ionization}
\begin{figure*}
\centering
		\includegraphics[width=8.5cm]{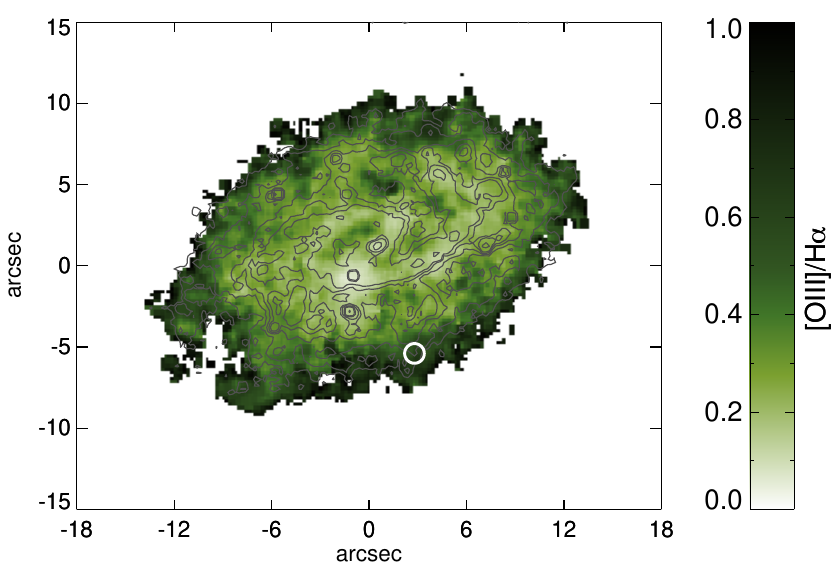}~~~
	\includegraphics[width=7.4cm]{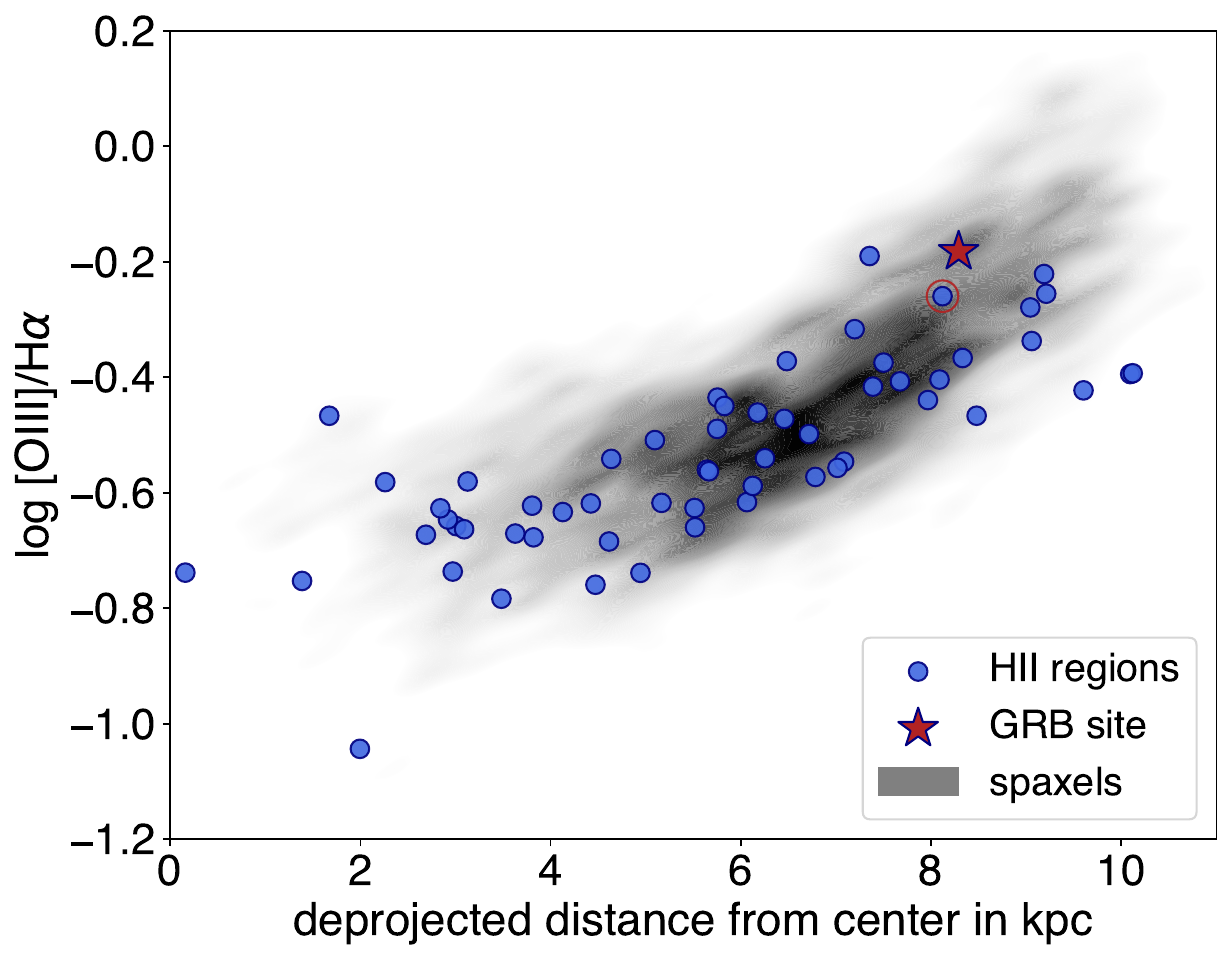}\\
		\includegraphics[width=8.5cm]{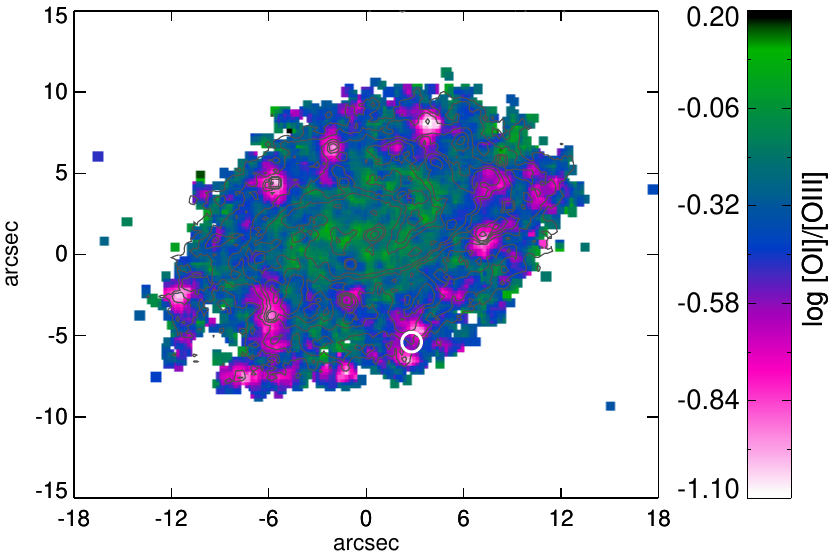}~~~
	\includegraphics[width=7.4cm]{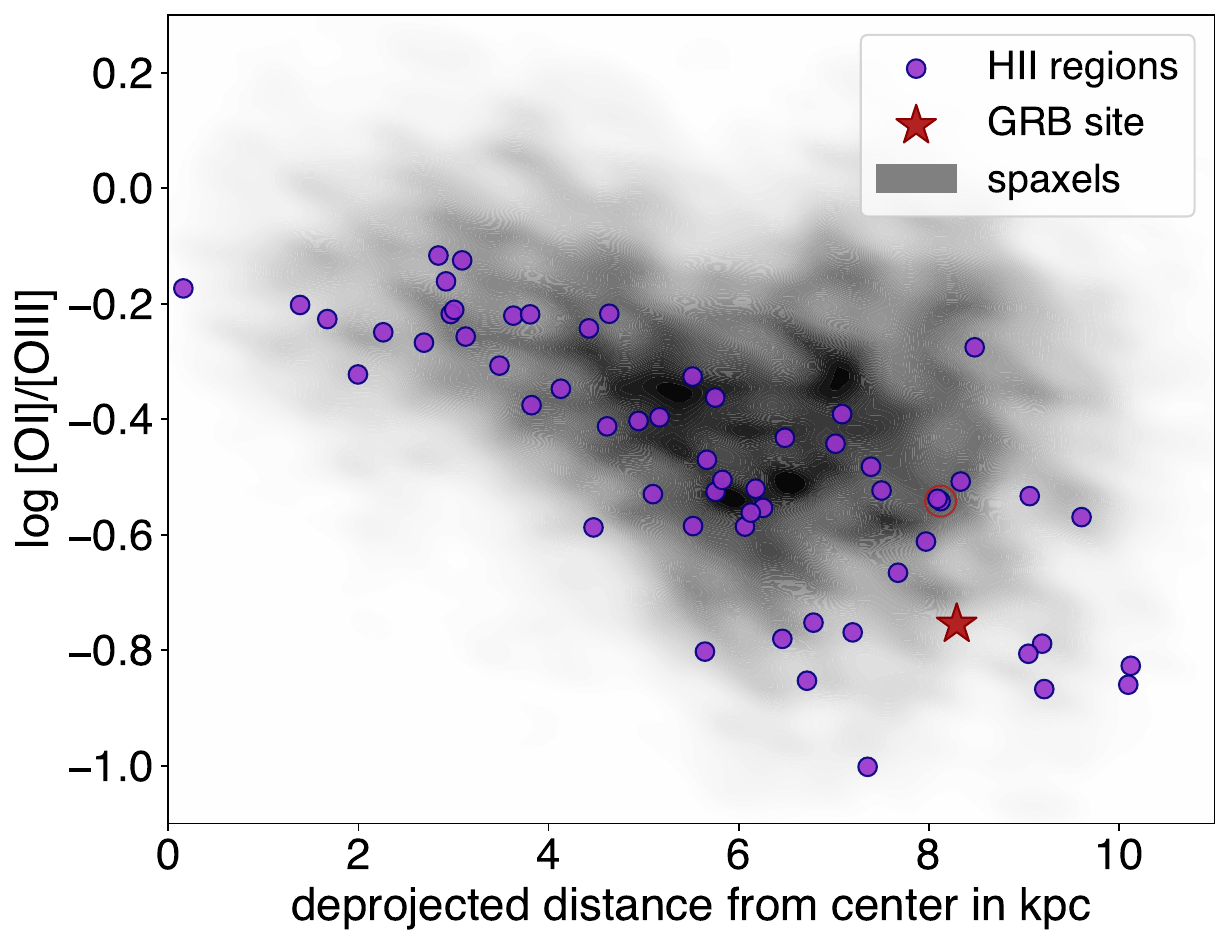}\\
 	\includegraphics[width=8.5cm]{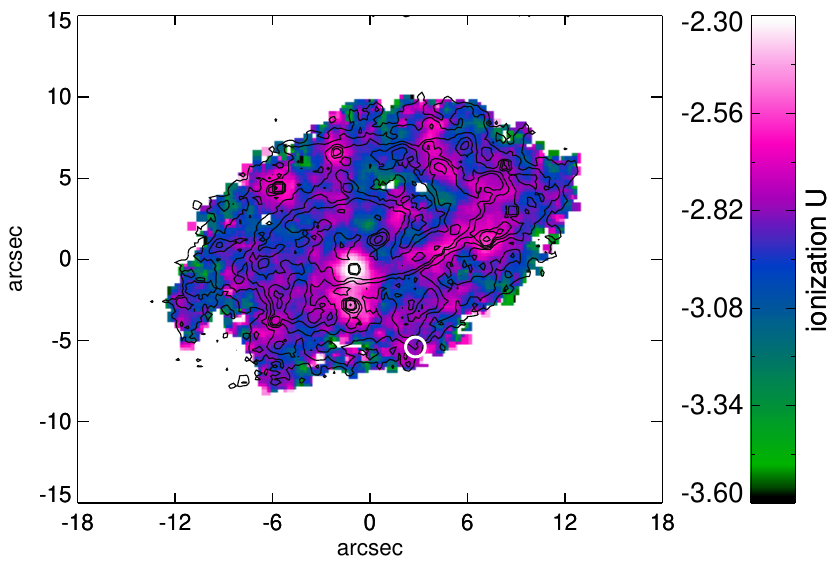}~~~
	\includegraphics[width=7.4cm]{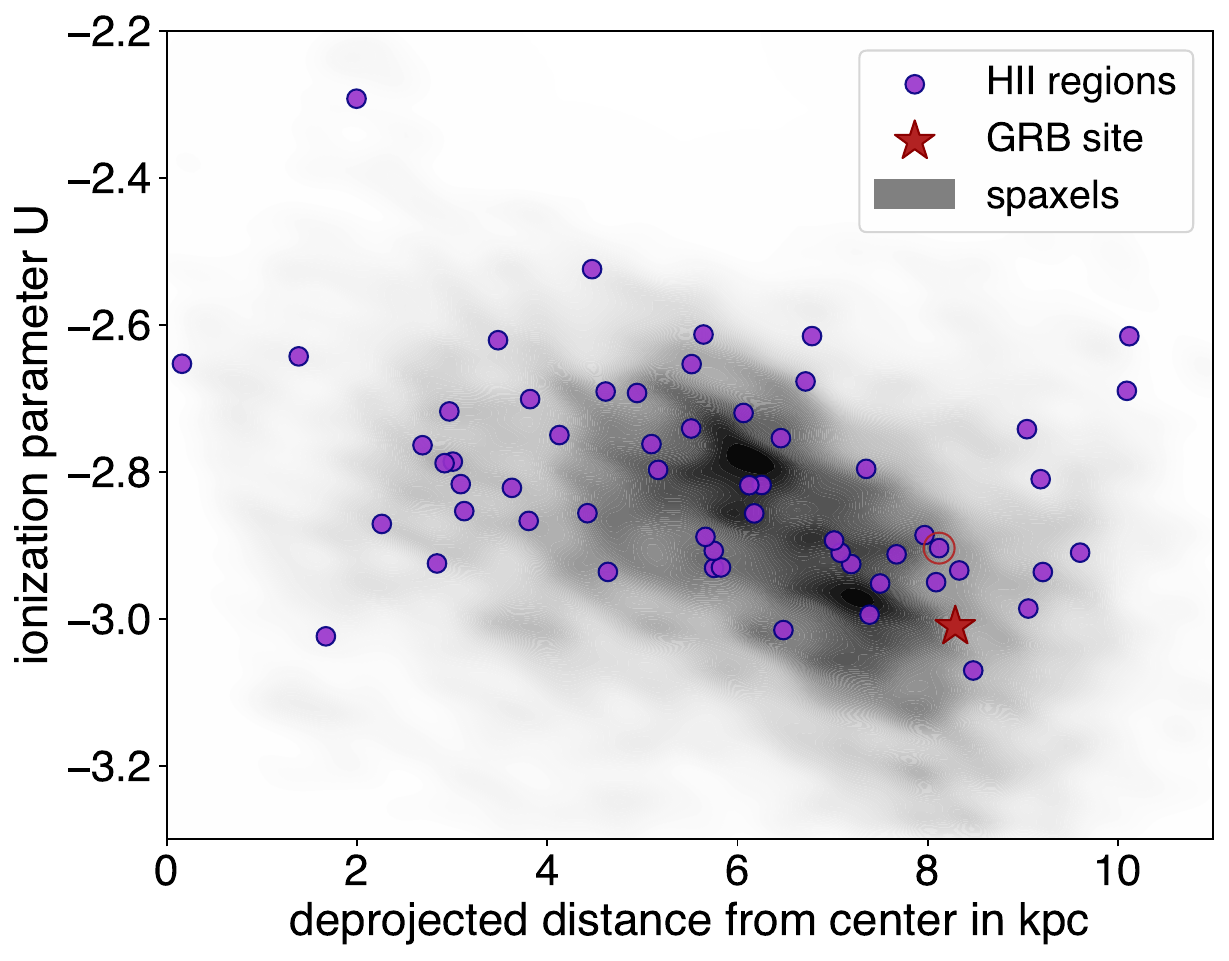}\\
  \includegraphics[width=8.5cm]{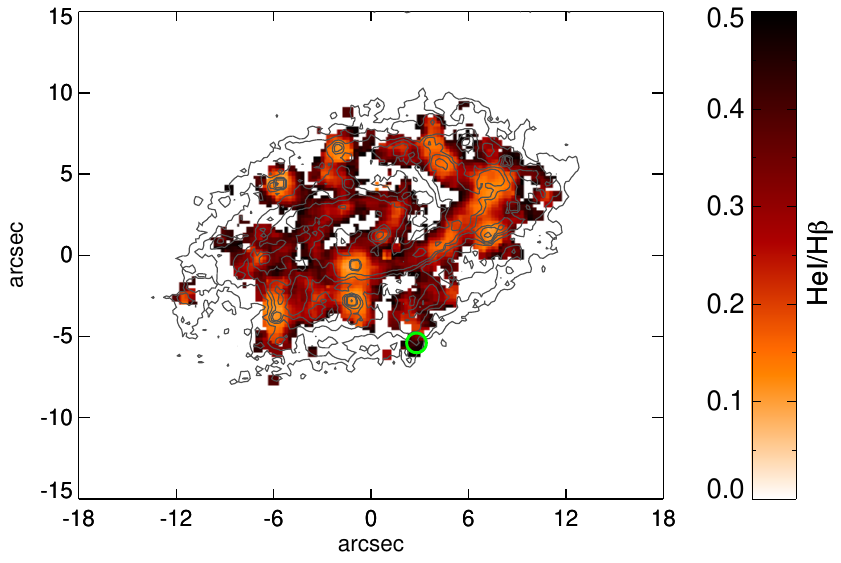}~~~
	\includegraphics[width=7.4cm]{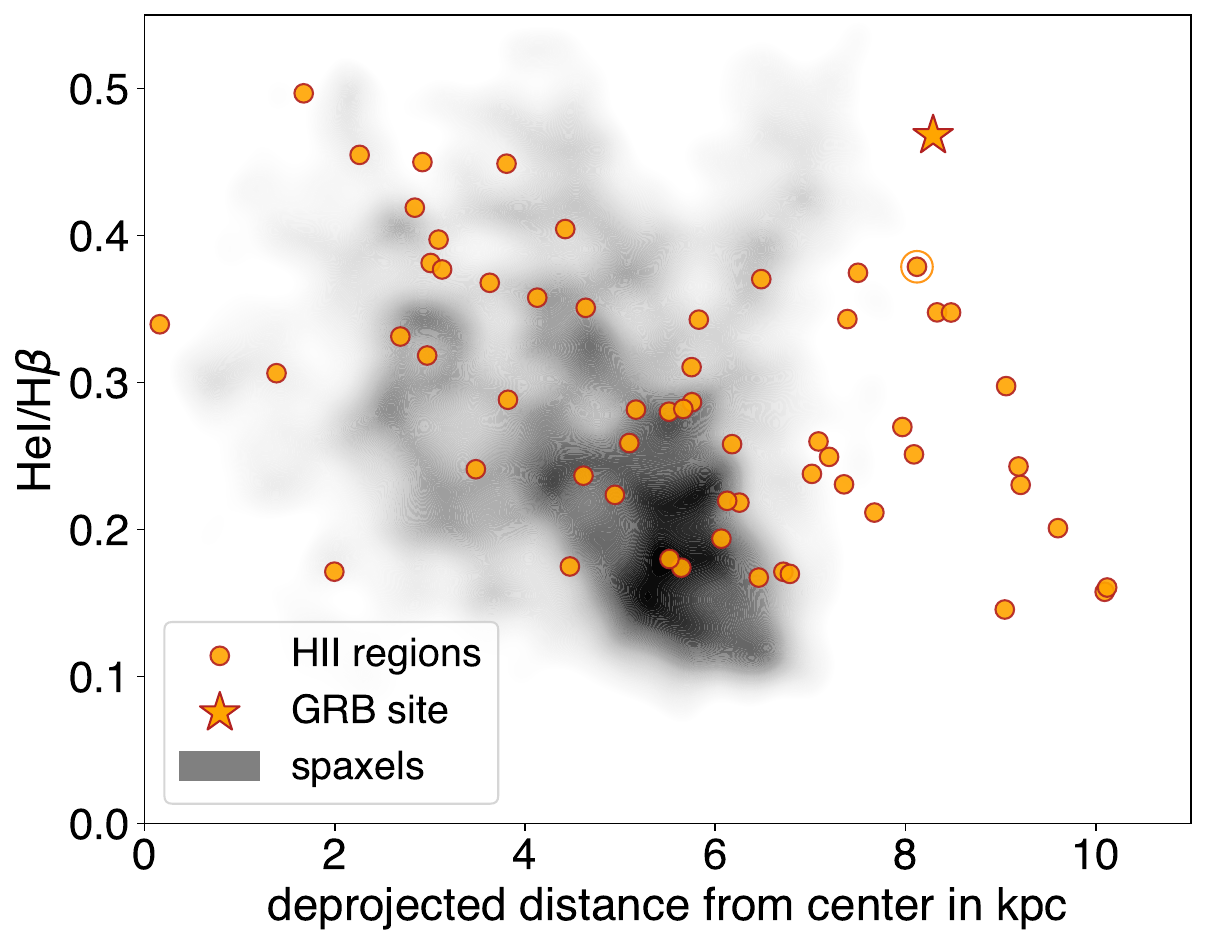}
	    \caption{1. row: 2D map of [O~{\sc iii}] $\lambda$ 5008 \AA{} vs. H$\alpha$ and deprojected plot. 2. row: Ionization of the ISM probed by [O~{\sc i}]/[O~{\sc iii}]. 3rd row: Ionization parameter U derived from [S~{\sc ii}]/H$\beta$ and the O3N2 metallicity according to \cite{Diaz00}, 4th row: He~{\sc i} $\lambda$ 5876 vs. H$\beta$.}
    \label{fig:maps2}
\end{figure*}

SF regions usually also show a high ionization level due to the strong UV radiation field from massive stars. The ratio of [O~{\sc iii}] $\lambda$ 5008 vs. H$\alpha$ shows a clear slope of increasing [O~{\sc iii}] towards the outer regions of the host, the GRB region and site are among the highest in the galaxy (see Fig. \ref{fig:maps2}). However, [O~{\sc iii}] is almost always weaker than H$\alpha$, in contrast to more extreme galaxies such as EELGs \citep[e.g.][]{Amorin15}.

We cannot determine directly the ionization of oxygen using [O~{\sc ii}] $\lambda\lambda$\,3727,29 since those doublet lines are outside the wavelength range of MUSE. The ratio [O~{\sc i}]/[O~{\sc iii}] decreases with distance from the center of the galaxy (meaning that ionization increases) and the GRB H~{\sc ii} region is in line with this trend. Here, however, the outer \ion{H}{ii} regions particularly stand out as more ionized regions, while the [O~{\sc iii}]/H$\alpha$ distribution is more smooth.

We also determine the ionization from the dimensionless parameter U based on the emission lines of [S~{\sc ii}], H$\beta$ and the metallicity \citep{Diaz00}. The ionization parameter follows a similar pattern towards increasing ionization (lower U parameter) towards the outer regions of the host, but somewhat less pronounced as from the [O~{\sc i}]/[O~{\sc iii}] ratio. The least ionized region is the H~{\sc ii} region at the southern end of the bar, which also shows up as outlier in the [O~{\sc iii}]/H$\alpha$ map. Again, the H~{\sc ii} regions in the outer part of the galaxy show a higher ionization and there is higher ionization along the Western spiral arm extending from the end of the bar at the highly ionized region. The GRB region does not stand out as a particularly ionized region using this parameter. 

He~{\sc i} $\lambda$5876/H$\beta$ can be used as indication for the hardness of the radiation field caused by either massive stars or another ionizing source. Usual values for galaxies such as the Magellanic Clouds show ratios of $\sim$0.03 \citep{Martin97}. He~{\sc i} is also an indication of a young stellar population \citep{Delgado99}. Photoionization models give values of $\sim$0.13 during the first 5\,Myr of a starburst. Our 2D map shows ratios around 0.15 in the inner regions of the regions with higher SF and especially along the spiral arm extending to the West, similar to what is observed in the U-parameter. He~{\sc i} around the GRB site shows a few higher value pixels, but here the extraction is complicated by the underlying supernova. The values are generally higher at the edges of the spiral arms, which could indicate some mildly shocked regions from the ISM driven away from the star-forming regions along the spiral arms.

\subsection{Shocks}\label{sect:shocks}

\begin{figure*}[h!]
	\includegraphics[width=8.5cm]{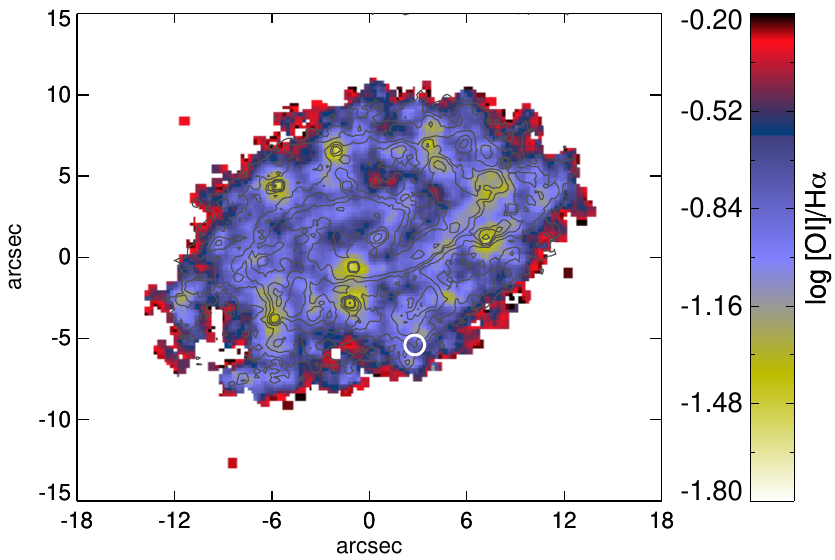}~~~
	\includegraphics[width=7.4cm]{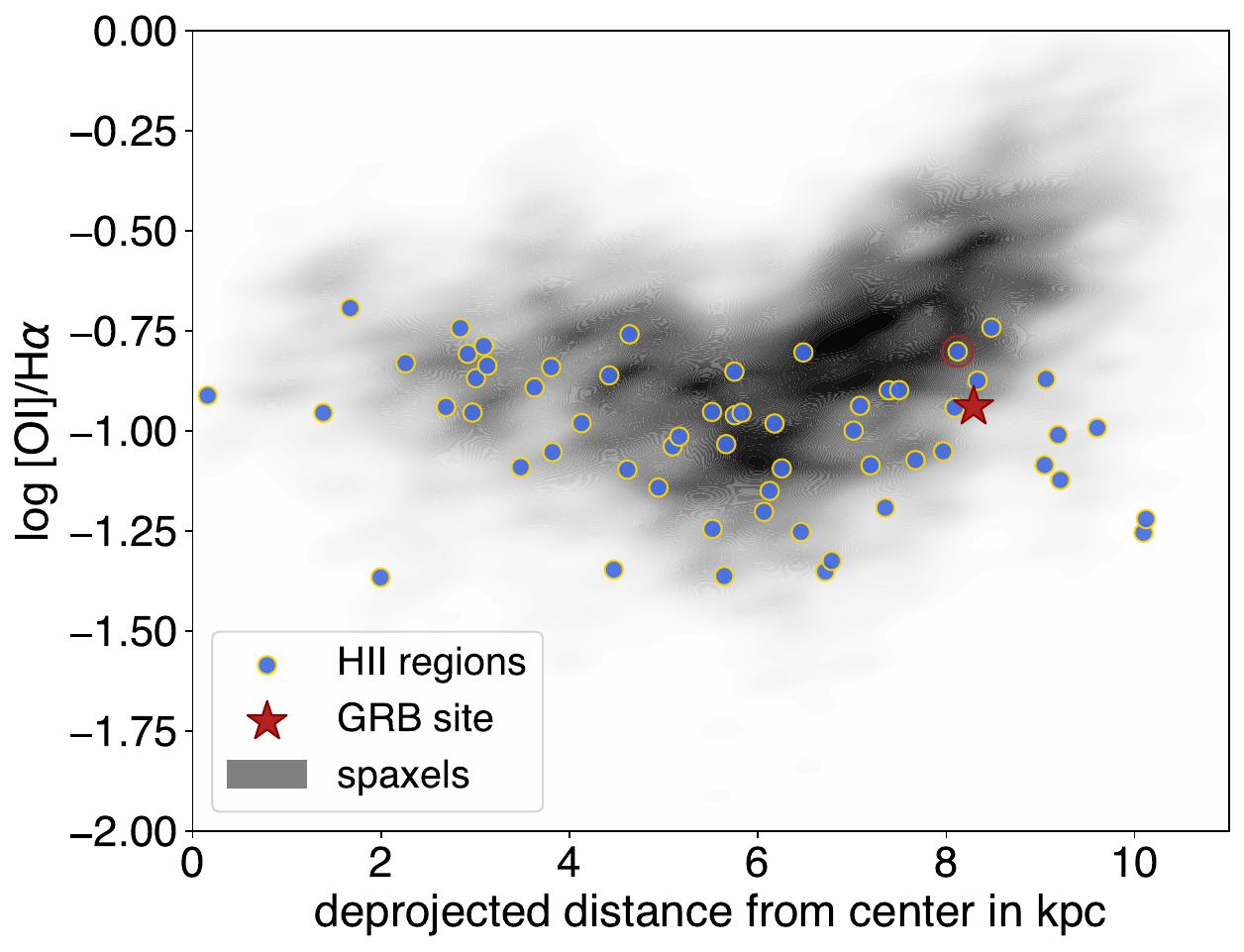}\\
	\includegraphics[width=8.5cm]{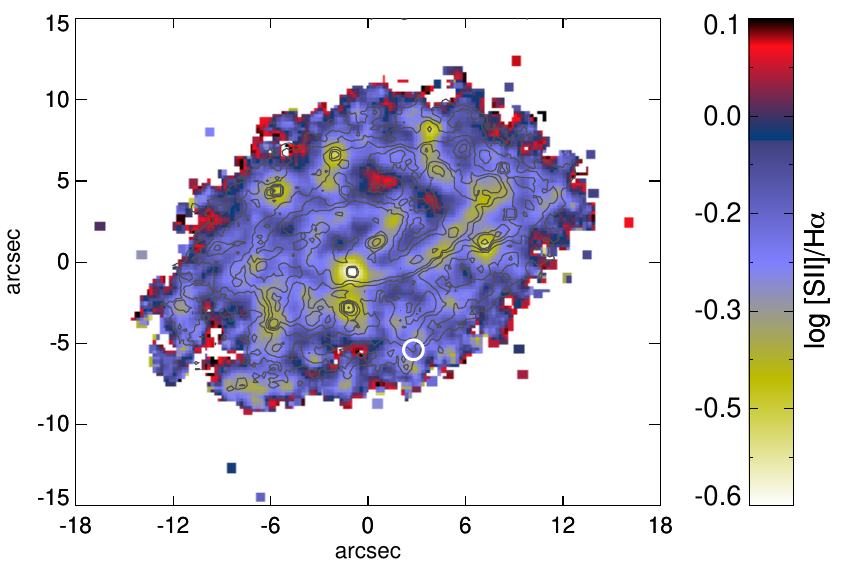}~~~
	\includegraphics[width=7.4cm]{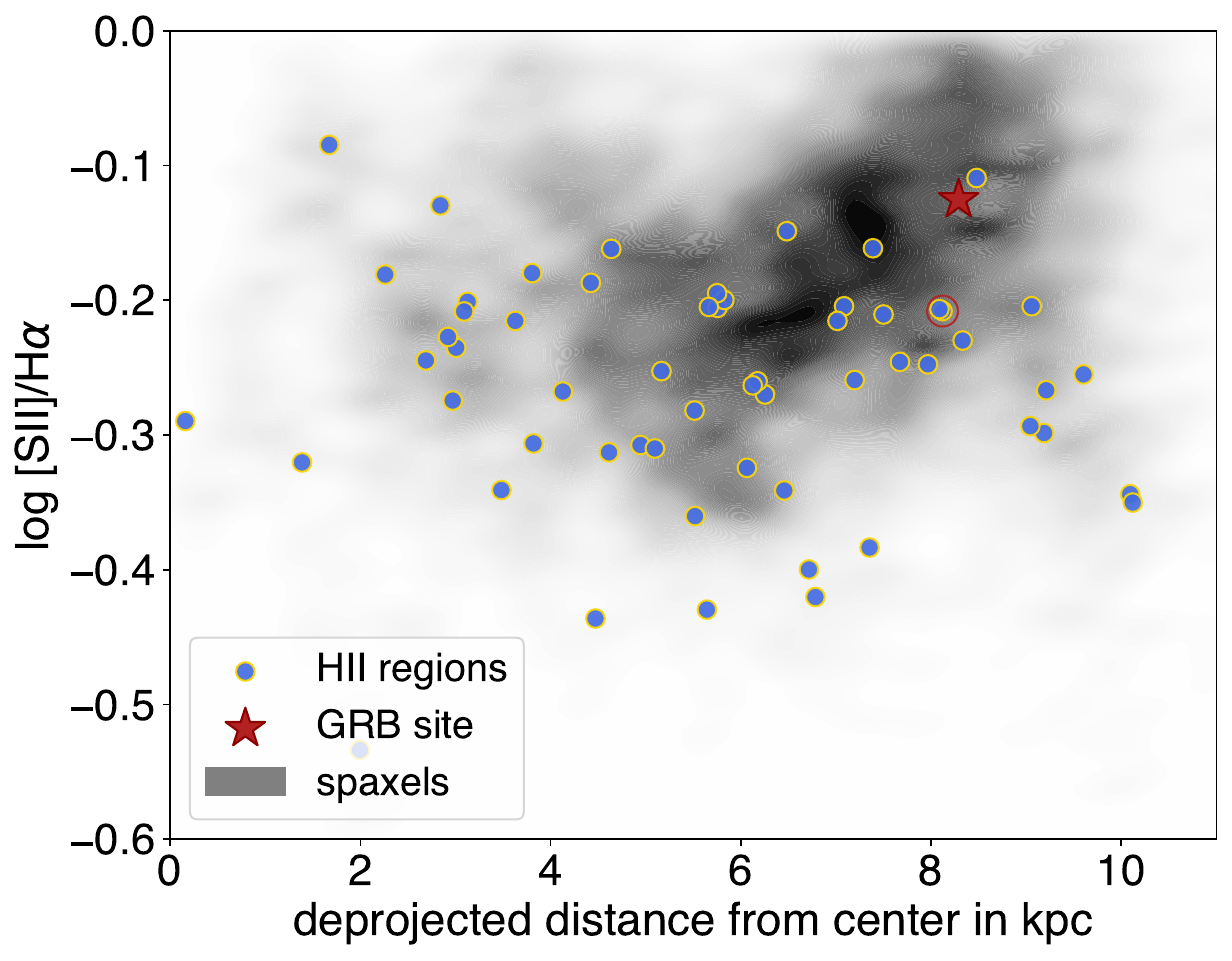}\\
   \includegraphics[width=8.5cm]{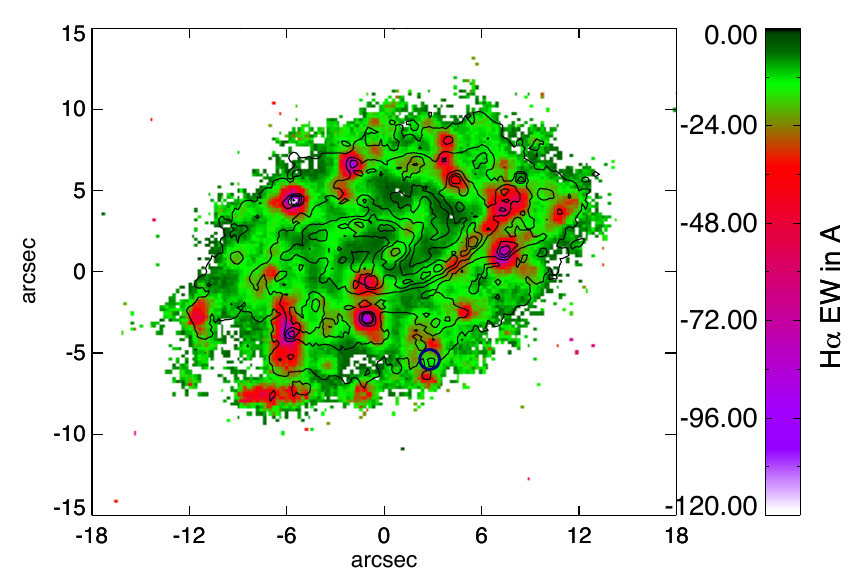}
	\includegraphics[width=7.5cm]{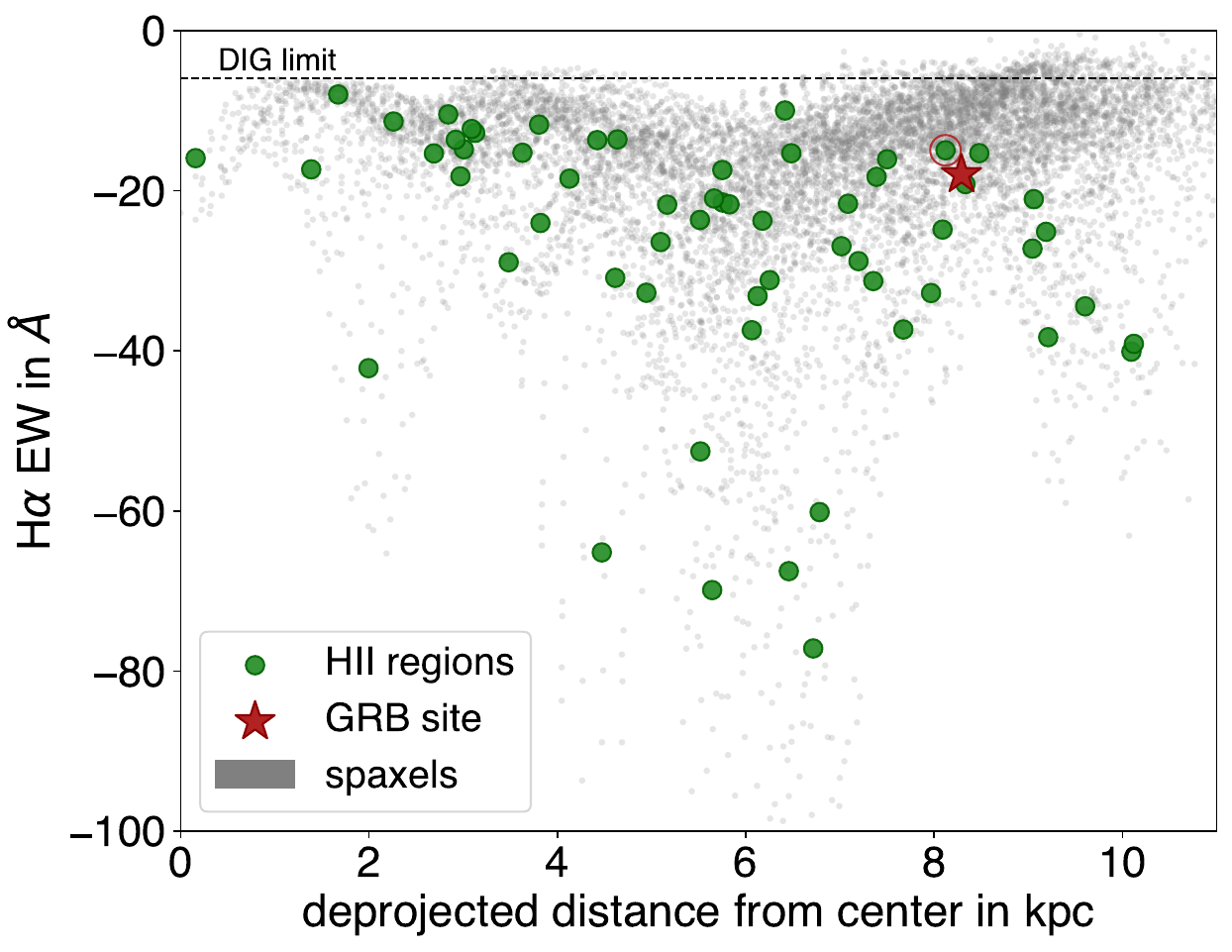}
	    \caption{First row: Map of [O~{\sc i}]/H$\alpha$ and deprojected values. Second row: log$_{10}$ [S~{\sc ii}]/H$\alpha$, indicative of shocked regions. Third row: Map of the H$\alpha$ EW and deprojected values, the EW at the GRB site is only a lower limit due to the contamination of the continuum by the GRB-SN. The dashed line indicates the cut in EW applied to account for possible emission from diffuse ionised gas (see Sect. \ref{sect:discussion_local}).}
    \label{fig:shocksEW}
\end{figure*}

As a proxy for shocked regions we use the ratio of [S~{\sc ii}]/H$\alpha$, plotted in Fig.~\ref{fig:shocksEW}, where strong [S~{\sc ii}] indicate possible shocks. The high values at the edge of the galaxy are likely artefacts. Most of the strong star-forming regions show low values of [S~{\sc ii}]/H$\alpha$ while moderate values can be observed in the inter-arm regions. The highest values are observed in a region at the North-Western end of the bar and might actually be a signature of shocked gas induced by movements along the bar.

No similar signature is observed at the S-E end of the bar, which contains the low ionized region mentioned in the previous subsection. The GRB region shows average values except for a few patches of higher values in the last inter-arm region East of the GRB site, while the spaxels around the GRB site show relatively high values, but not a larger pattern indicative of shocks. These observations, together with the kinematic analysis (see Sect.~\ref{sect:kinematics}) do not point to an ongoing, violent shock or interaction scenario as an origin for the star-formation in the GRB region.

A similar patter is observed for the ratio of [O~{\sc i}]/H$\alpha$, which is also indicative of shocks \citep{Bik18}, with a median around --1.0, a scatter $\sim$ 0.5 dex and a slight increase towards the outer regions of the host. Regions with log([O~{\sc i}]/H$\alpha$) larger than $\sim$--1.0 could be regions excited by AGN activity, however, this seems unlikely in this galaxy and even the central H~{\sc ii} region is below --1.0. A BPT diagram (Baldwin, Philips \& Terlevich diagram, \citealt{BPT}) using [N~{\sc ii}]/H$\alpha$ (see Fig.~\ref{fig:BPT}) does not show any indication for AGN excitation anywhere in the galaxy.

\subsection{Star formation history and stellar population age}\label{sect:SP}

\begin{figure}[h!]
	\includegraphics[width=\columnwidth]{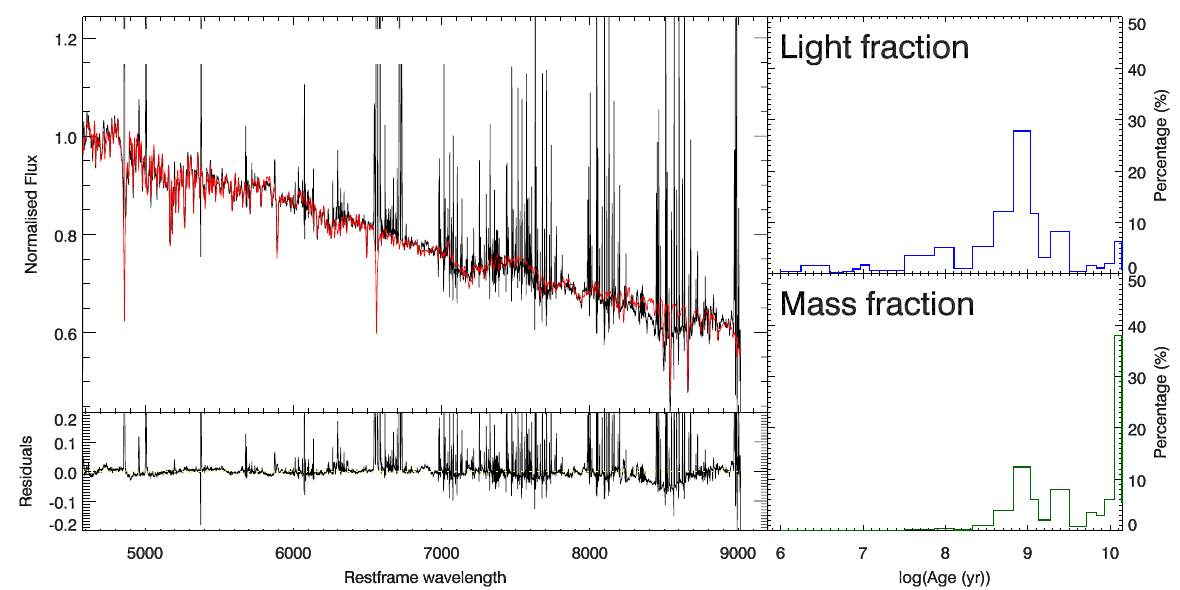}
 \includegraphics[width=\columnwidth]{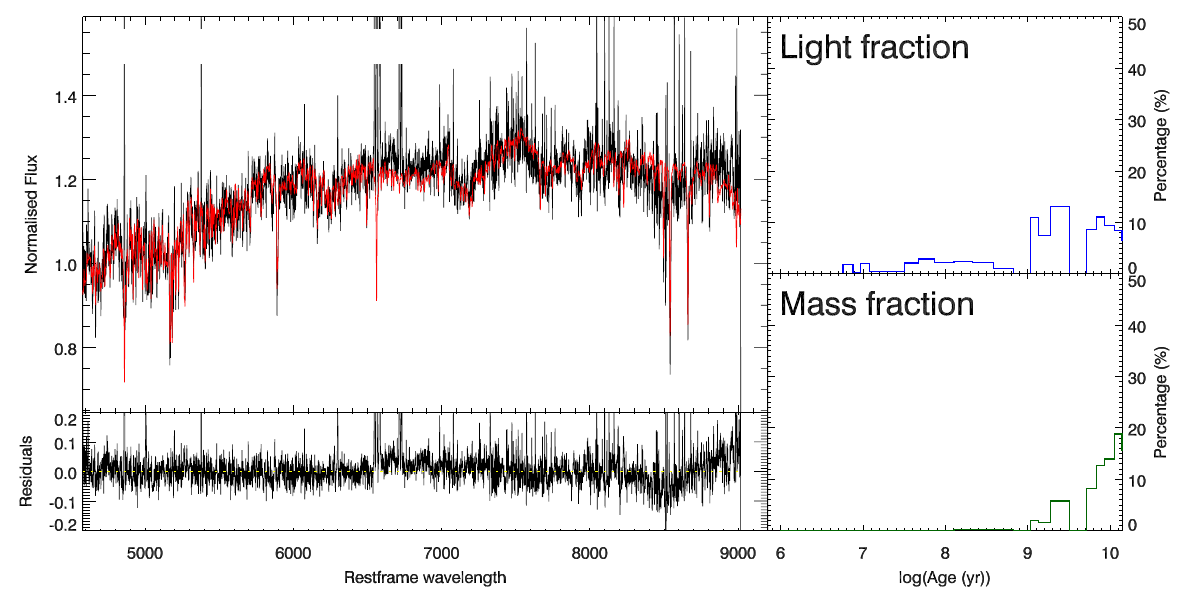}
	\includegraphics[width=\columnwidth]{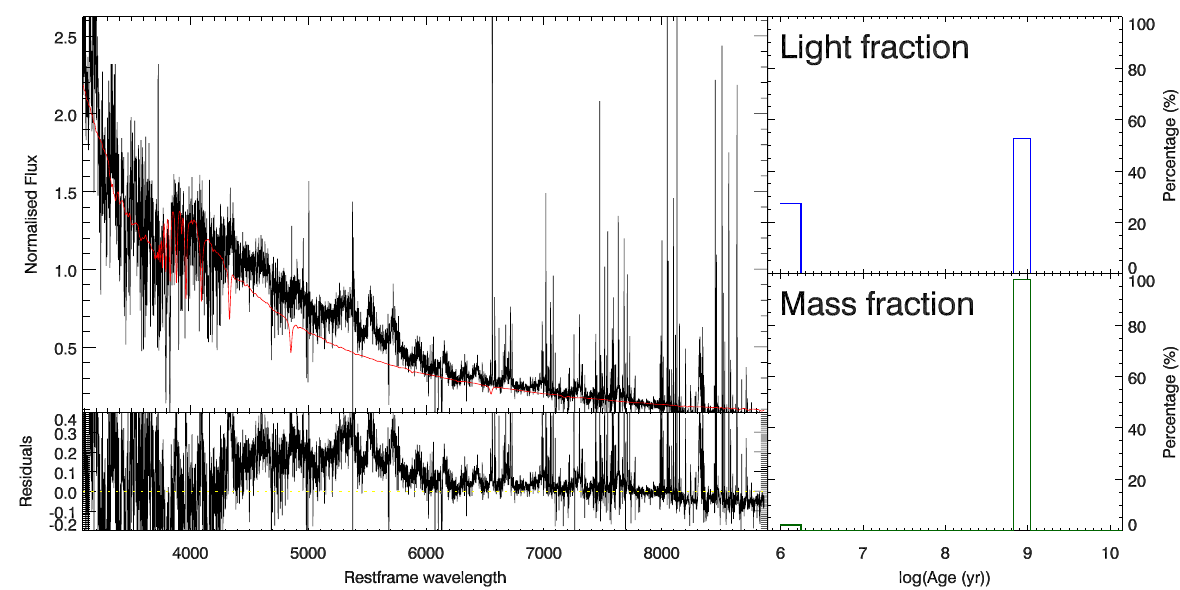}
	    \caption{Stellar population fits of different representative spectra. Top panel: combined spectrum of the host, middle panel: integrated spectrum of the core, bottom panel: the GRB site using the X-shooter observations from 14 months after the GRB instead of the SN contaminated MUSE spectra. In each panel, the top left box shows the observed spectrum (black) with the fitted spectrum over plotted in red, the lower left panel the residuals of the fit. The boxes on the right show the light (top) and mass (bottom) fraction for SPs at different ages.}
    \label{Fig:starlightall}
\end{figure}

In Fig.~\ref{Fig:starlightall} we show the spectra of different regions in the host together with the result of the stellar population (SP) fit performed, as described in section~\ref{starlight}. For the integrated spectrum of the galaxy the residuals of the fit are very good with $<$1$\%$. This stellar population fit returns an average extinction of $A_V=0.34$ mag. Both the mass and the light fraction show a dominant population at around 900 Myr with additional contributions from older stars, but also a significant population with ages below 100 Myr and even some under 10 Myr, indicating recent SF activity. 

\begin{figure*}[h!]
	\includegraphics[width=6cm]{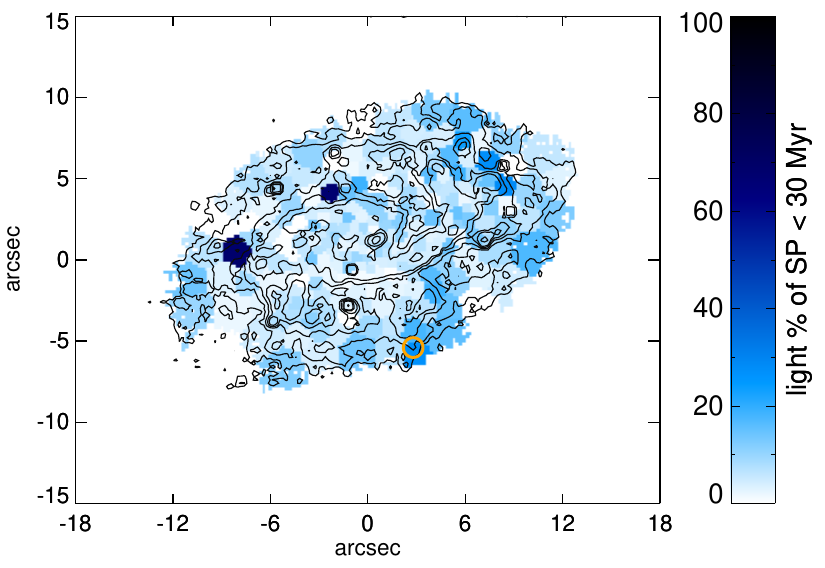}
	\includegraphics[width=6cm]{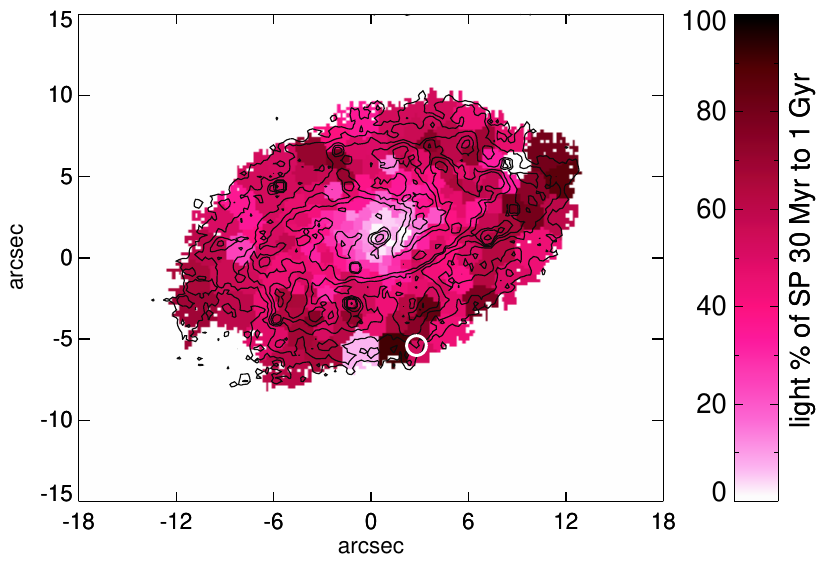}
	\includegraphics[width=6cm]{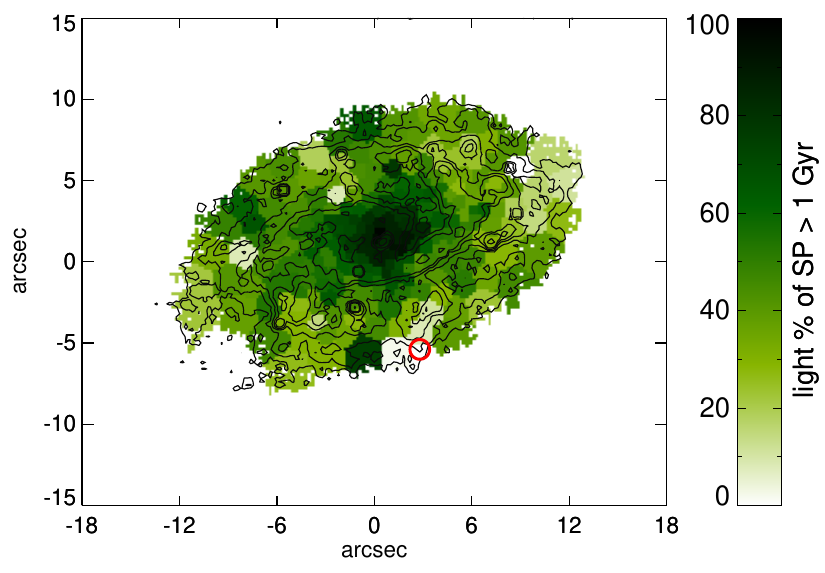}
	    \caption{Maps of the light percentage of the stellar population, divided into three age bins: The light percentage of the population $<$30 Myr, the intermediate-age population of 30 Myr -- 1 Gyr and the old population of $>$ 1 Gyr. The white region in all three plots is a Voronoi bin affected by a foreground star and was therefore excluded from the analysis.}
    \label{fig:starlightages}
\end{figure*}

 The spectrum from the core reveals a population dominated by old stars, but still with a contribution of a few percent to the light by younger stars in the 100 Myr range. This region has also a larger than average extinction of $A_V=0.89$ mag. To study the GRB region, we use the late time X-shooter spectrum from $\sim$14 months after the GRB where the contamination from the SN is negligible. We do detect some broad features, mainly at 5000--6000 \AA{} which could be Fe-lines from the SN or residuals from the overlap between UVB and VIS arm, and which we hence exclude before fitting the stellar continuum. The H~{\sc ii} region next to the GRB site, in the original cube, has a very blue continuum from the younger stars and a lower extinction of $A_V=0.26$ mag. 
 Using the data derived from the stellar population fit, we then make maps in Voronoi binning corresponding to the age distribution in three different age ranges (Fig. \ref{fig:starlightages})

The SP fit at the GRB site shows an older population of $\sim$1\,Gyr contributing almost all of the stellar mass, and another peak with a very young population of only a few Myr contributing to the light but not the mass of the region. The GRB region has a contribution of $\sim$ 30\% from a SP $<$3\,Myr, but 60\% of the light comes from a SP of $\sim$1\,Gyr, possibly from the last major SF event of the galaxy. Two H~{\sc ii} regions show a light fraction for the youngest SP bin of $>60\%$. These regions as well as the GRB region show a small fraction of an intermediate population and a negligible contribution from a SP of $>$1\,Gyr. The other two young regions, however, are located at the opposite side of the galaxy compared to the GRB and within the SF ``ring'', while the GRB region lies at a larger distance. This confirms that, indeed, there has been a recent onset of SF at the GRB site, which has produced the massive GRB progenitor. We will have a closer look at the other three young regions in Sect. \ref{sect:discussion}.

Another, less direct, method to determine the age of the dominant stellar population is to measure the equivalent width (EW) of H$\alpha$, hence the line luminosity compared to the continuum. Some of the other bright star-forming regions in the SF ring of the host show rather high EWs between --60 and --80\,\AA{}, which corresponds to a population age of $\sim$8--10\,Myr according to STARBURST99 models at half solar metallicity \footnote{https://www.stsci.edu/science/starburst99/docs/default.htm} \citep{Leitherer99}. The EW at the GRB site from the MUSE cube is very low with \mbox{$\sim$--14\,\AA}, however, this is only a lower limit due to the presence of the SN. For the GRB site we therefore measure the EW from the late time X-shooter spectrum, also used for the SP analysis described above. Despite the absence of the SN continuum, the EW remains low with only $\sim$ --18\,\AA{}, not indicating a particularly young SP age. This probably results from the significant light contribution of an older stellar population at the site mentioned above.

\section{Kinematic analysis}\label{sect:kinematics}
The kinematics of different states of the gas can give clues to possible star-formation triggers such as gas inflows or past mergers or interactions. The rich dataset for this host allows us to analyse the velocity and dispersion maps of the ionized gas by probing H$\alpha$, the molecular gas via CO emission and the stellar kinematics using NaD absorption lines as well as investigate the environment for possible interacting neighbour galaxies.

\subsection{The galactic environment of the host}
The large field-of-view (FOV) of MUSE allows to search for nearby galaxies at the same redshift. In the HST image there are several elongated objects and a round, low-surface brightness structure with a bright center at the North of the host. There is also a prominent object consisting of several bright knots $\sim$\,6\,arcsec East-South-East of the host, which at first glance could be a satellite of the host galaxy. We checked for emission lines in all those objects and conclude that none of them are at the redshift of the GRB (see Fig. \ref{fig:companion}). All objects described above belong to two groups of galaxies at $z=0.61$ and $z=0.62$, including the bright knots to the East-South-East, which hence is clearly not a satellite galaxy. The fact that most of the galaxies in those groups are edge on is an observational bias since edge-on galaxies are more easily detectable than face-on objects due to their higher apparent surface brightness. 

Our observations in \ion{H}{i} revealed another object at a similar redshift as the GRB \citep{deUgarte24}, a large H~{\sc i} disk from another spiral galaxy at a projected physical distance of 183\,kpc to the North-West with a velocity difference of only $\sim$30\,km\,s$^{-1}$ (see Fig.~\ref{fig:largefield}). As comparison, the LMC and SMC are located at 50 and 62\,kpc, the Andromeda galaxy is at 770\,kpc from the Milky Way. The companion has an H~{\sc i} mass similar to the host ($\mathrm{log~M}=$9.45\,M$_\odot$ and 9.49\,M$_\odot$ for the host and companion, respectively) and shows a somewhat more regular H~{\sc i} disk than the host (see Fig.~\ref{fig:VLA}). We do not detect any H~{\sc i} gas between the two galaxies, which, if present, would likely be too faint to be detected in our data. As calculated in \cite{deUgarte24}, assuming a transverse velocity between those two galaxies of $\sim$ 200\,km\,s$^{-1}$, the last encounter of the two galaxies happened 900\,Myr ago, making an influence on the current SF of the host unlikely, however, we have no information no the actual relative velocity between the two objects.

The distant companion galaxy is also a spiral galaxy, albeit looking more irregular and with some bright star-forming regions. Unfortunately, no high resolution imaging of that galaxy is available to further study its morphology. Our PMAS data do not have a high enough S/N to allow for a detailed analysis and we only obtain resolved maps of H$\alpha$ and [O~{\sc iii}] (see Fig.~\ref{figapp3}). H$\beta$ is too weak and [N~{\sc ii}] is almost not detected (the value listed in Tab. \ref{tab:integratedspecs} is rather an upper limit than a detection). The outer spiral arms are very weak in H$\alpha$ and we note that the [O~{\sc iii}] emission in stronger in the S-E part, while H$\alpha$ is more uniform. The metallicity of the galaxy is similar to the host of GRB 171205A (12+log(O/H)=8.47), but it has a considerably lower SFR of only 0.11 M$_\odot$y$^{-1}$ and also a very low H$\alpha$ EW, hence not indicating a very young stellar population. The two galaxies are clearly part of a group, however, their evolution was likely going different paths.

\subsection{Ionized gas kinematics}
\begin{figure}[h!]
	\includegraphics[width=\columnwidth]{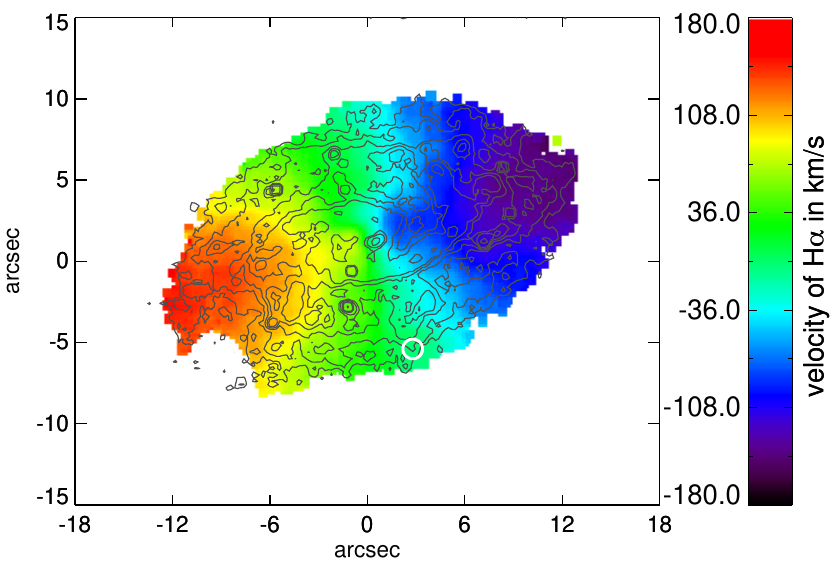}
	\includegraphics[width=\columnwidth]{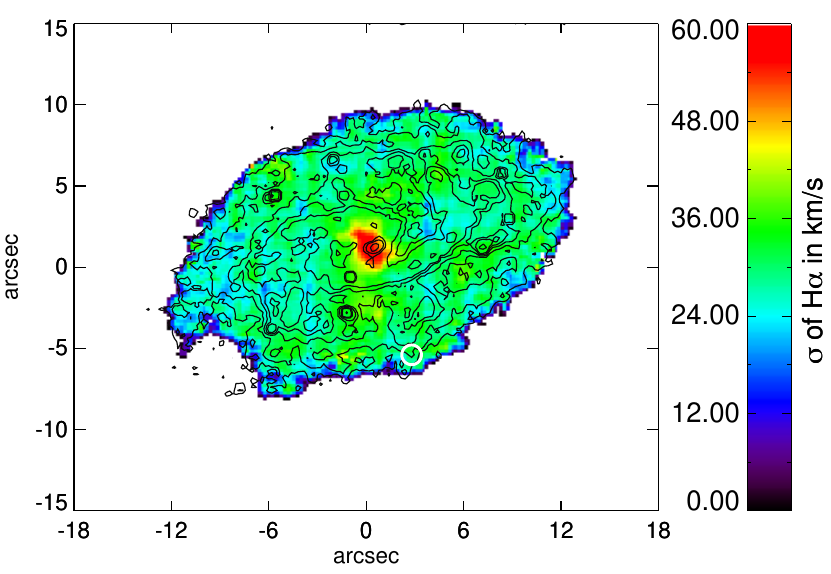}
    \caption{Velocity and dispersion derived from a single Gaussian fit to the H$\alpha$ line. The dispersion was corrected for the instrumental resolution of R$\sim$3000.}
    \label{fig:velfield}
\end{figure}

The ionized gas kinematics is traced using the H$\alpha$ emission lines in the MUSE cube. In Fig.~\ref{fig:velfield} we show the velocity and dispersion obtained from fitting a single Gaussian to H$\alpha$. Zero velocity is determined as the velocity at the brightest spaxel at the center of the host. The dispersion map is corrected for the instrumental resolution of MUSE, R$\sim$3000 at $\lambda$6600\,\AA{}. At first sight, the velocity field appears like a regular rotating disk. The dispersion map only shows a larger dispersion in the center due to the larger amount of material in the bulge and otherwise has values around 20--30\, km\,s$^{-1}$ with slightly higher velocities in the arms and somewhat lower in the interarm regions.

\begin{figure}[h!]
	\includegraphics[width=\columnwidth]{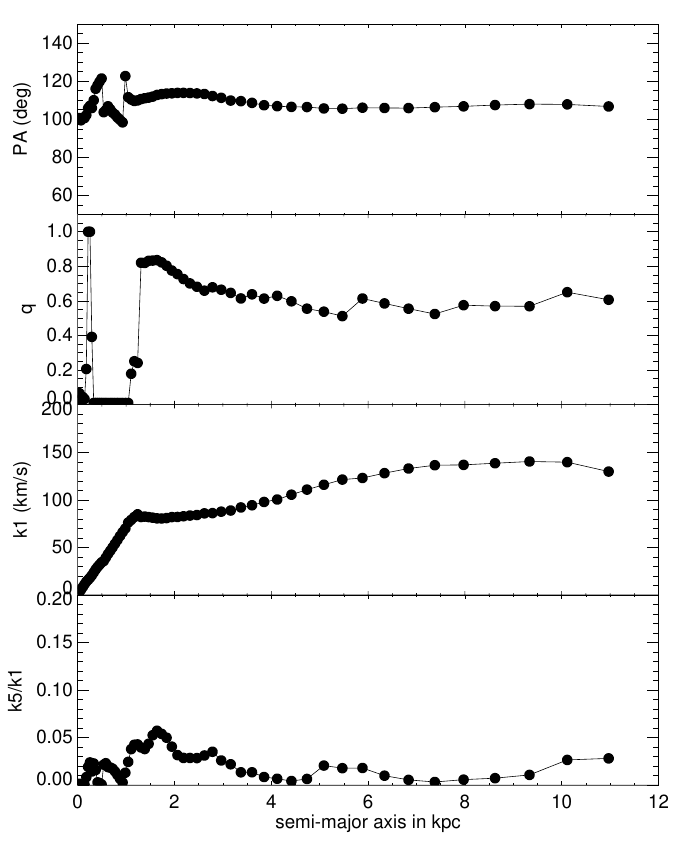}
    \caption{Top: Parameters obtained from the kinemetry analysis \citep{kinemetry06}. The PA of the fitted ellipses, their ellipticity $q$, $k1$ is the first kinematic moment (equivalent to the rotation curve) and $k5/k1$ the deviation from simple rotation. 
    }
    \label{fig:kinemetry}
\end{figure}

To further analyse the kinematics, we fit kinemetry models \citep{kinemetry06} to the velocity field. This technique extracts the velocity profiles along ellipses around the center of the galaxy (since we assume disks are intrinsically round and appear as ellipses due to the inclination), which can be described by harmonic terms in Fourier analysis. The method looks for the best fitting ellipses to minimise those terms, which again then yield the parameters for these ellipses as a function of distance from the center of rotation. Fig.~\ref{fig:kinemetry} shows the kinematic position angle (PA), the flattening of the ellipse $q$ and the kinematic moments $k$1 and $k$1/$k$5 as a function of deprojected distance. The PA traces the position of maximum velocity, hence the PA of the best fit ellipses, $q$ is related to the inclination by $q=\sin(i)$ and $k$1 as the first kinematic moment corresponding to the rotation curve of the galaxy. $k$5 traces the higher-order deviations from simple rotation and hence deviations from simple rotation. The kinematic center of the galaxy ($v=0$\,km\,s$^{-1}$) is the brightest spaxel of H$\alpha$ in the center, and corresponds to a redshift of $z=0.03714$. The kinemetry code \cite{kinemetry06} is then run for all ellipses with a minimum covering fraction of velocities of 0.7. No limits are given on $q$ or PA. 

Fig. \ref{fig:kinemetry} shows that the galaxy has a rotation curve (parameter $k$1) with a steep, linear, component out to $\sim$ 1.5\,kpc after which it follows a smooth behaviour out to the maximum distance where we can fit the  H$\alpha$ line. This is indeed the textbook behaviour for a bar (which behaves as a solid body) and disk galaxy (rising and then flat rotation curve). In the same inner 1.5\,kpc the ellipticity changes somewhat erratically with a sudden jump at 1.5\,kpc and there are some deviations in the PA. The deviations $k$5/$k$1 from simple rotation is in general small but shows some higher values in the central 2\,kpc, likely associated to the influence of the bar. At the end of the bar beyond $\sim$ 2\,kpc the velocity field becomes much more regular, although some enhanced $k$5/$k$1 is present out to 4\,kpc, possibly due to some remaining influence of the bar. A further increase in $k$5/$k$1 at the outermost part of the galaxy is likely a simple issue of S/N and coverage of the best fitting ellipse with the velocity data from H$\alpha$. For the same reason, the analysis does not allow to claim that the rotation curve actually starts to decline in the outermost part.

\subsection{Molecular gas kinematics}
\begin{figure*}[h!]
		\includegraphics[width=8.7cm]{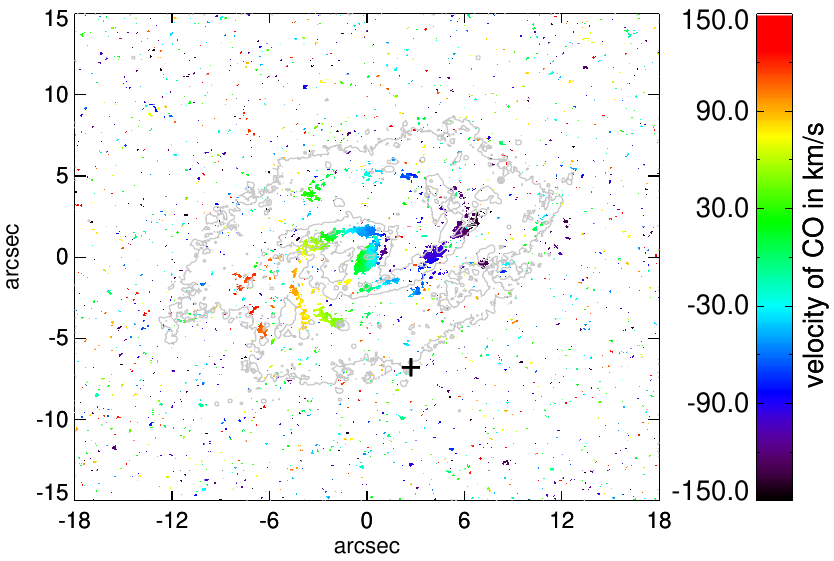}
			\includegraphics[width=8.7cm]{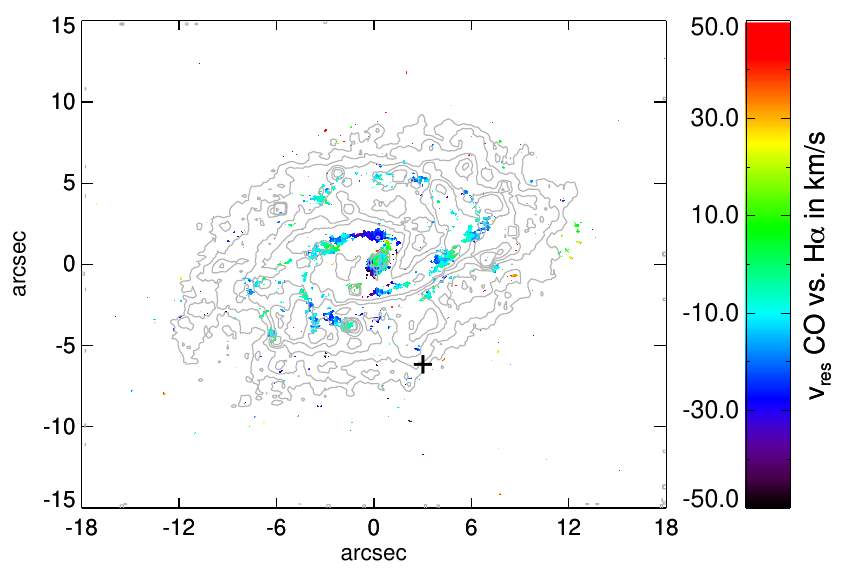}
    \caption{CO velocity map (left) and residuals compared to the  H$\alpha$ velocity field (right). Contours are from the late-time HST F475W image. The GRB site is marked by a +.}
    \label{plot:residualsCO}
\end{figure*}

The molecular gas traced by CO(1--0) emission has been presented in \cite{deUgarte24}. Due to the much sparser velocity field probed by the CO emission we cannot do a full kinemetry analysis for the molecular gas, instead, we compare the H$\alpha$ emission to the CO velocity map (see Fig.~ \ref{plot:residualsCO}). 

The velocity differences are $<$\,15 km\,s$^{-1}$ except the regions at the end of the bar/onset of the inner spiral arms, which show residuals up to $\sim$ --50\,km\,s$^{-1}$. Disk galaxies have often higher rotation velocities for CO gas than for H$\alpha$ (see e.g. \citealt{Levy18}) with typical residuals of $\sim$ 25 km\,s$^{-1}$, associated to extraplanar diffuse gas or a thick disk. At redshifts $>1$ there are both examples of similar CO and H$\alpha$ rotation curves and some with large discrepancies \citep{Uebler18, Girard19}. Dispersions are often found to be larger for H$\alpha$ like we observe in the host of GRB\,171205A. In general, the gas dispersion in a galaxy seems to increase with redshift \citep{Uebler19}. 

The largest residuals are measured at the Northern end of the bar, the same region that shows indications of shocked material (see Sect.~\ref{sect:shocks}). The negative residuals could be an indication of an inflow of gas, as it is commonly observed in barred galaxies \citep[see e.g.][and references therein]{Yu22}. Observations of galaxies in the local Universe have revealed details on gas transport and kinematics, especially in central regions and around the bar \citep[see e.g. a study on M83][]{DellaBruna22}. Comparing the line width of H$\alpha$ (see Fig.\ref{fig:velfield} and \citealt{deUgarte24}), there is a similar pattern with a high dispersion in the center and lower width in the spiral arms, however, both values are $\sim$\,15\,km\,s$^{-1}$ higher for H$\alpha$. There is no detection of CO at the GRB site itself, preventing further analysis of the gas content or  kinematics of molecular gas in this part of the galaxy.
5
 Only about 15 GRB hosts have to date been detected in CO using different transitions of CO \citep{Stanway15, Michalowski18, Arabsalmani18, Hatsukade20, deUgarte20}. Kinematical analysis has been limited to very rough velocity maps \citep{Hatsukade20}, but often the only information is the CO line width due to the faintness of the emission. \cite{deUgarte20} make a comparison between the width of the CO line and H$\beta$ in the host of GRB\,190114C and find a very similar value. \cite{Arabsalmani20} have CO(2-1) observations at comparable resolution for the much closer host of GRB 980425 at d$_L \sim$38\,Mpc, but not presenting a kinematic analysis of the full galaxy. Even for field galaxies, detailed 2D studies comparing ionized and molecular gas kinematics have been limited to galaxies in the local Universe such as the THINGS/SINGS/metal-THINGS survey \citep{Kennicutt03,Walter08,metalTHINGS23} and PHANGS survey \citep[e.g.][]{Leroy21, Emsellem22}, only now being extended to higher redshifts by e.g the CALIFA/EDGES survey \citep[e.g.][]{Levy18}.

\subsection{Neutral gas kinematics}
{We rereduced the JVLA data already described in \cite{Arabsalmani22} and analyze both the host and the companion in H~{\sc i}, the latter of which is not mentioned in \cite{Arabsalmani22}. Figure~\ref{fig:VLA} shows the moment 0 (flux) and moment 1 (velocity) maps for both galaxies. As mentioned in \cite{deUgarte24}, the companion has a higher total H~{\sc i} flux and also its spatial distribution is larger. Both galaxies have a velocity field of a rotating disk with a gap in the center of the disk and a slightly asymetric distribution with more HI gas in one half of the galaxy. \cite{Arabsalmani22} found an extra component to the south of the GRB position (below the Western lobe) in the VLA data, which we cannot recover in our re-reduction of the same data. Instead, there seems to be a cloud offset South of the Eastern lobe.

\begin{figure*}[ht!]
		\includegraphics[width=8.5cm]{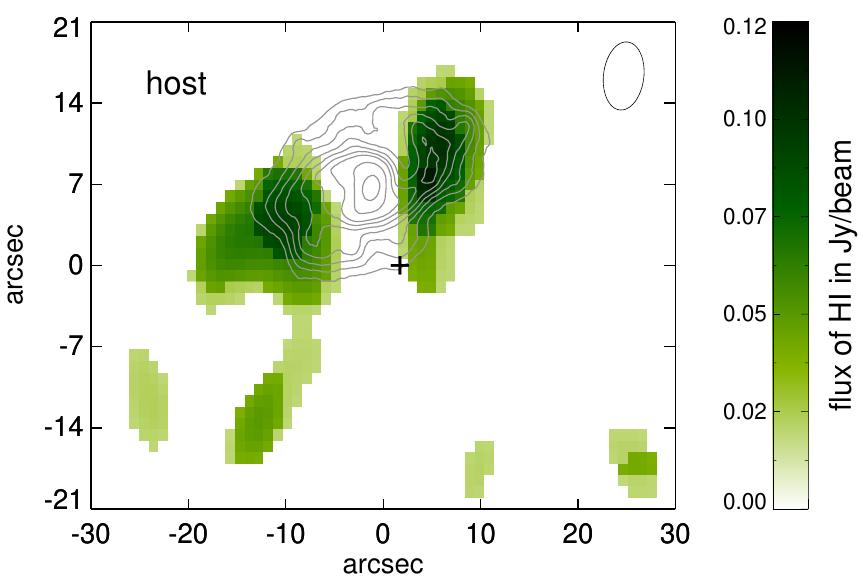}
  \includegraphics[width=8.5cm]{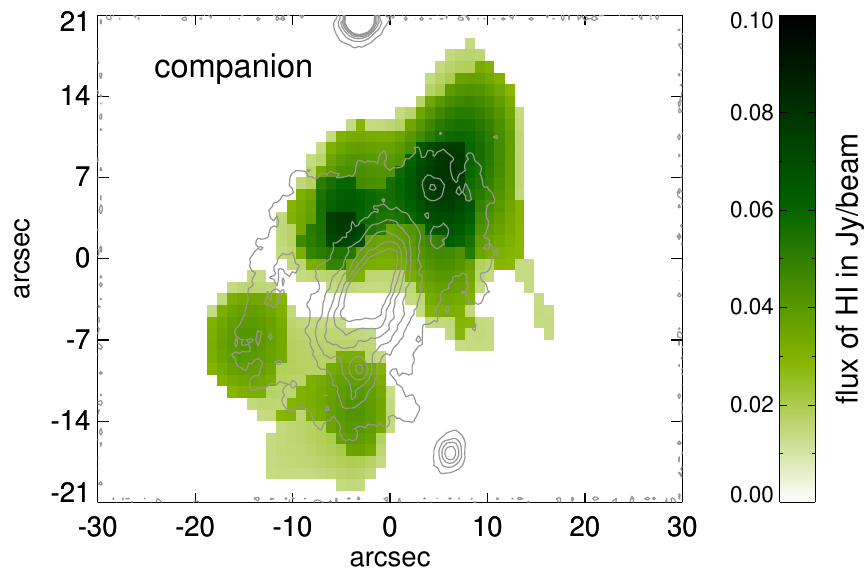}\\
   \includegraphics[width=8.5cm]{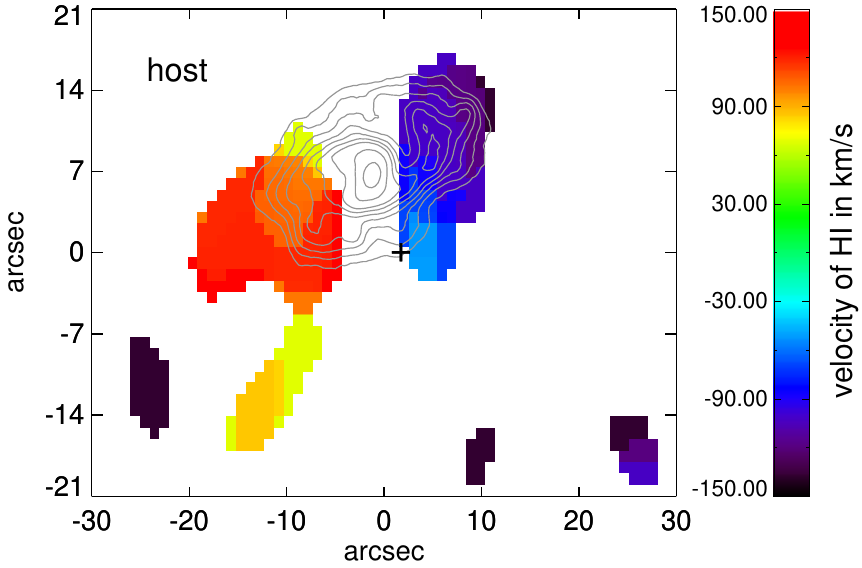}
   \includegraphics[width=8.5cm]{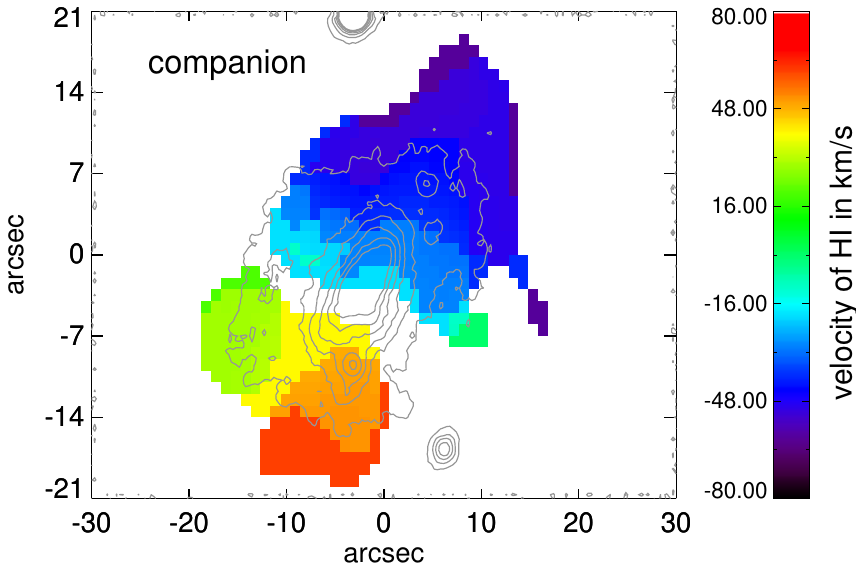}\\
    \caption{Left: Moment 0 and 1 maps of the host galaxy observed in HI with VLA. The ellipse in the top figure indicates the beam size. Right: Same for the distant companion galaxy. Overlaid both objects are r'-band contours from PanSTARRS. Note that the FOV is larger than for the MUSE plots. }
    \label{fig:VLA}
\end{figure*}

We use the high resolution JVLA data to analyse the kinematics of the H~{\sc i} gas and compare it to the ionized and molecular gas (see Fig.~\ref{fig:VLAfit}). The model velocity field used here is:
\(v = v_0\, \tanh(r/h) \, \sin(i)\, \cos(\phi-\mathrm{PA_k}) \), where $v$ is the line-of-sight velocity, $r$ is the radius with respect to the centre of the galaxy, $\phi$ is the angle with respect to the positive y-axis, $\textrm{PA}_\textrm{k}$ is the kinematic position angle of the galaxy, $i$ is the inclination of the galaxy, and $v_0$ and $h$ are constants constrained by the true rotation velocity. This model has been used before to create mock galaxy velocity fields in \cite{Stark18} and is similar to the two-parameter arctan function described in \cite{Courteau97} that has been used to describe galaxy rotation curves. A warp can be introduced to this model by changing the $\textrm{PA}_\textrm{k}$ with radius at a specific rate. This acts as an additional parameter in the model.

We fit the inclination, PAk, $v_0$, $h$ and the warp using the spatial and velocity data obtained from the JVLA observations. After a best-fit model is found, we calculate the normalised root mean squared error (NRMSE) to estimate how well the model fits the data. The NRMSE for the GRB 171205A host and companion are 0.073 and 0.078, respectively. While this fit does not exclude past interactions, it shows that the observed gas velocity can be described by the model used here. It also does not support the conclusions reached by \cite{Arabsalmani22} of a recent interaction with a smaller satellite, of which there is no trace in the optical data.

\begin{figure*}[ht!]
\centering
   \includegraphics[width=16cm]{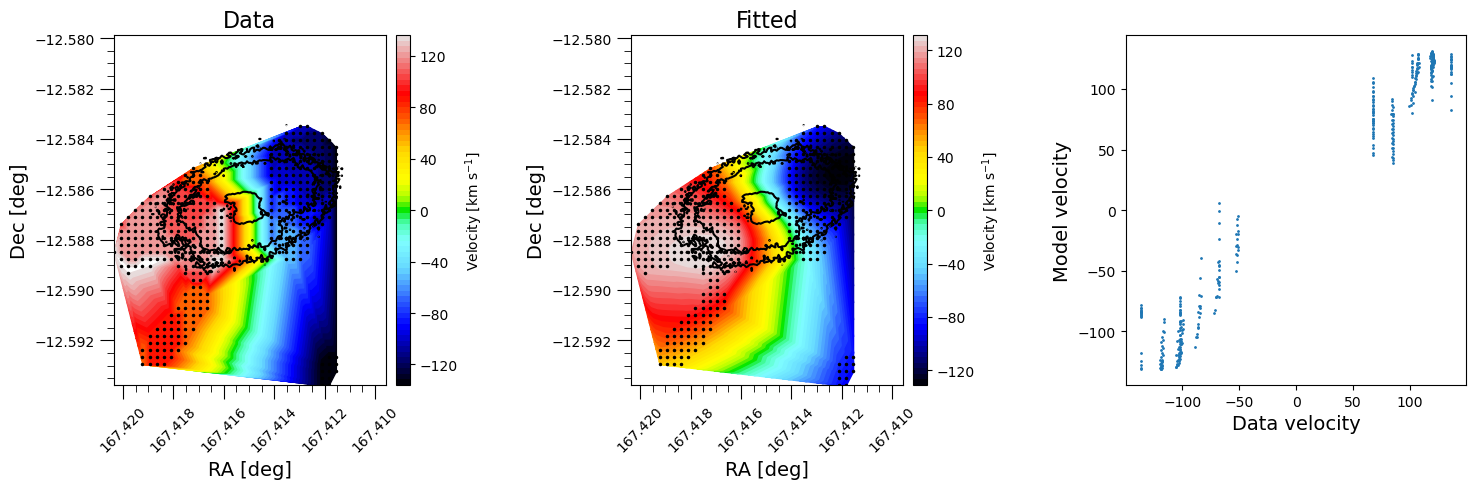}\\
   \includegraphics[width=16cm]{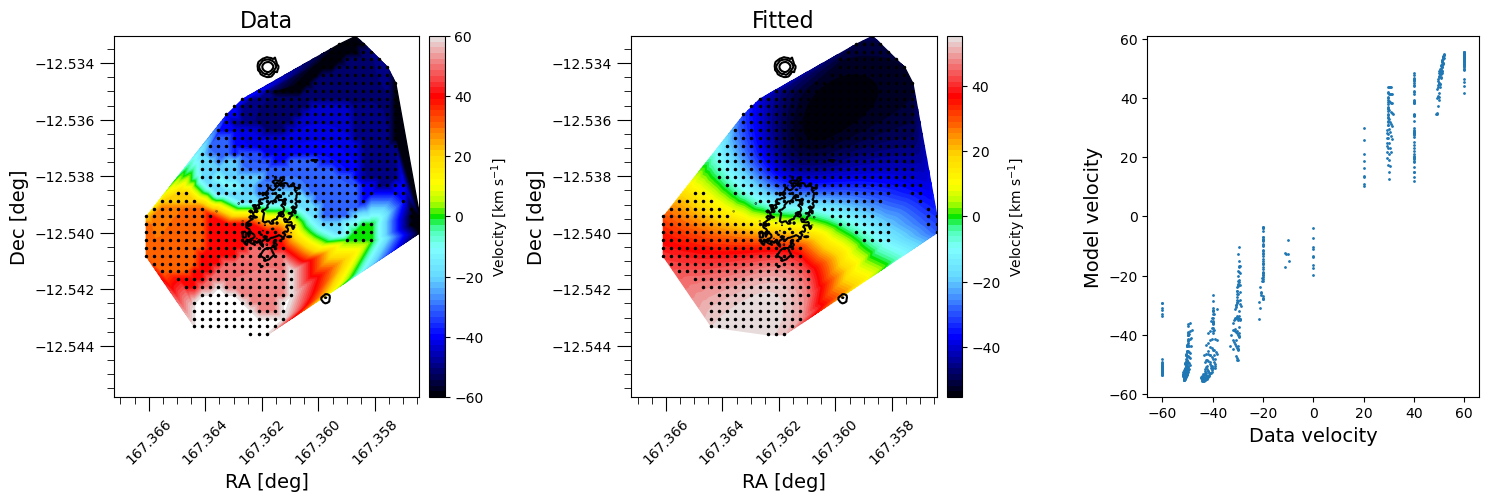}\\
    \caption{Velocity fits to the host (top) and companion (bottom) as described in the text. Dotted values are regions of measured flux, overlaid are contours from PanSTARRS images as in Fig.\ref{fig:VLA}. }
    \label{fig:VLAfit}
\end{figure*}

\section{Discussion}\label{sect:discussion}

\subsection{The GRB site compared to the host properties}\label{sect:discussion_local}
To check for the peculiarity (or lack of thereof) of the GRB H~{\sc ii} region, we compare a number of properties derived in Sect.~\ref{sect:analysis} to 1) the average properties of all spaxels in the galaxy and 2) the average properties of the ISM in a ring of $\pm$1\,kpc compared to the galactocentric distance of the GRB of $\sim$\,8kpc. We determine the mean and standard deviation for each property in all spaxels with sufficient S/N in (a) the entire host (see also Tab.~\ref{tab:integratedprops}) and (b) in a bin of $\pm$1\,kpc around the deprojected distance of the GRB. We subsequently determine the deviation of the GRB site as (prop$_{GRB}$-prop$_{gal/1kpc}$)/standdev$_{gal/1kpc}$ (see Fig.~\ref{fig:deviation}). SFR/spaxel, ionisation and H$\alpha$ EW at the GRB site are rather similar to the global and local values, the latter is due to the fact that the GRB site shows a rather low EW for an actively star-forming region. The extinction is considerably lower as it is consistent with zero at the GRB site. The metallicity is $>$1$\sigma$ lower at the GRB site, the fact that the deviation is less compared to the global metallicity is due to the larger standard deviation taking all spaxels in the galaxy. The sSFR does not have a large deviation in $\sigma$ owing to the large standard deviation for this value in the galaxy, on an absolute scale, the sSSFR in the GRB H~{\sc ii} region and site are twice the average in the galaxy. We also note that the global properties derived from the integrated galaxy spectrum match rather well with the average values of all spaxels.

We furthermore determine the position of the GRB site in the BPT diagram, the different spaxels in the galaxy as well as other long GRB hosts from the literature (see Fig.~\ref{fig:BPT}). BPT diagrams serve to distinguish regions excited by hot stars and those excited by AGN activity. In the most commonly used version, plotting [N~{\sc ii}]/H$\alpha$ vs. [O~{\sc iii}]/H$\beta$ do not show any indication for AGN excitation using the demarcation line of \cite{Kewley07}, but parts of the galaxy, including the GRB region, are in the intermediate region \cite{Kauffmann03}. The other two BPT diagrams using [S~{\sc ii}]/H$\alpha$ and [O~{\sc i}]/H$\alpha$ do show parts of the galaxy, especially the outer regions, to fall into the AGN part. We also checked for a possible contribution from diffuse ionised gas (DIG), which can contribute several tens of \% to the emission \citep[see e.g.][]{Poetrodjojo19, DellaBruna22}.DIG is often found in interarm regions and is contributing a larger fraction in the outskirts of a galaxy. To remove a possible contribution from DIG we only plot spaxels with a H$\alpha$ EW of $<$--6\AA{}. Applying this cut, we get a DIG fraction of 1.6$\%$, which is low for a spiral galaxy, but possibly our S/N anyway removes most of the possible DIG contribution before any further analysis. Figure~\ref{fig:shocksEW} shows that DIG might contribute a small part in the outskirts but almost nothing in the center of the galaxy. Hence we exclude DIG contributing significantly to the emission anywhere in the host. The values for individual spaxels show a trend from very low ionization regions in the center to values closer to the values for GRB hosts in the outskirts, including the GRB site. The GRB site has a relatively low ionization and high metallicity compared to those for other GRB hosts or sites, which are distributed more to the upper left of the diagram \citep{Kruehler15}. Only for the BPT diagram with [O~{\sc i}]/H$\alpha$ the values of the GRB site and galaxy fall in the same region as other long GRB hosts, however, we only have [O~{\sc i}] clearly detected in four other GRB hosts. Since the high ionization regions outside the clearly photoionized regions in the [SII]/H$\alpha$ and [OI]/H$\alpha$ BPT diagrams are not due to extended amounts of DIG nor do we see any sign for AGN activity, together with the fact that most spaxels are in the outskirts of the host, we think that this high ionization is rather due to shocks.

\begin{table*}[h!]
\caption{Emission lines of GRB host, GRB site and companion galaxy. The GRB host and HII regions were measured from integrated spectra extracted from the MUSE cube, values at the GRB site were obtained from the late time X-shooter spectrum, the spectrum of the companion was obtained from separate PMAS/CAHA observations of the galaxy. The values in ``GRB site'' refer to the values from the X-shooter spectrum. In the MUSE data, the [O~{\sc ii}]$\lambda$3727,29 is out of range. Fluxes in units of $10^{-16}$ erg s$^{-1}$ cm$^{-2}$ and are corrected for Galactic and intrinsic extinction using the Balmer decrement.} 
\label{tab:integratedspecs}      
\centering    
\small
\begin{tabular}{ccccccc|c}       
\hline\hline                
Line&$\lambda_\mathrm{rest}$&GRB host & Core+Bar        & GRB site       & Region 15       & Region 17       & Companion     \\
\hline\hline 
O~{\sc ii}         & 3726.032 & ---          &  ---            & 15.49$\pm$2.49  &---            &  ---            &      ---      \\
O~{\sc ii}         & 3728.815 & ---          &  ---            & 14.67$\pm$1.51  &---            &  ---            &      ---      \\
H$\beta$   & 4861.333 & 328.16$\pm$14.98 & 19.04$\pm$0.25 & 2.88$\pm$0.44  & 5.39$\pm$0.09 & 3.77$\pm$0.05 & 25.0$\pm$13.1 \\
O~{\sc iii}        & 4958.911 & 77.54$\pm$15.3 & ---             & 1.55$\pm$0.38   & 0.81$\pm$0.13 & 0.54$\pm$0.06 & 17.5$\pm$7.9   \\
O~{\sc iii}        & 5006.843 & 157.6$\pm$10.1 & 3.90$\pm$0.96 & 5.50$\pm$0.67   & 2.21$\pm$0.11 & 2.13$\pm$0.06 & 34.8$\pm$7.1  \\
He~{\sc i}         & 5875.624 & 35.4$\pm$1.9 & 1.73$\pm$0.11 & ---             &0.51$\pm$0.04 & 0.38$\pm$0.03 & ---            \\
O~{\sc i}          & 6300.304 & 34.68$\pm$4.04 & 1.31$\pm$0.07 & ---            &0.57$\pm$0.06 & 0.40$\pm$0.02 & ---            \\
N~{\sc ii}         & 6548.050 & 100.55$\pm$20.93 & 6.98$\pm$0.28 & 0.47$\pm$0.26  &1.74$\pm$0.06 & 1.07$\pm$0.04 & 3.39$\pm$2.9   \\
H$\alpha$  & 6562.819 & 906.93$\pm$21.41& 52.75$\pm$0.29 & 7.97$\pm$0.35   &14.9$\pm$0.06 & 10.44$\pm$0.05 & 69.1$\pm$5.5  \\
N~{\sc ii}         & 6583.460 & 333.90$\pm$21.89& 23.84$\pm$0.30 & 1.65$\pm$0.29  &5.91$\pm$0.06 & 3.69$\pm$0.05 & 50.22$\pm$9.8  \\
S~{\sc ii}         & 6716.440 & 204.56$\pm$21.95 & 8.58$\pm$0.28 & 1.86$\pm$0.36  &3.05$\pm$0.06 & 1.95$\pm$0.04 & 13.99$\pm$4.1   \\
S~{\sc ii}         & 6730.810 & 146.18$\pm$21.48 & 7.14$\pm$0.29 & 1.38$\pm$0.39 &2.19$\pm$0.06 & 1.42$\pm$0.04 & 19.53$\pm$4.5  \\
Ar~{\sc iii}       & 7135.790 & 6.66$\pm$1.29  & 0.82$\pm$0.11 &    1.17$\pm$0.20        & 2.25$\pm$0.54 & 0.22$\pm$0.02 & ---           \\
\hline\hline
\end{tabular}
\end{table*}

\begin{table*}[h!]
\caption{Properties for the integrated spectrum of the GRB host, the average of all spaxels in the host, the GRB site and the companion galaxy. The properties for the two young regions in Tab. \ref{tab:integratedspecs} are already listed in Tab.\ref{tab:HIIregions}.
}    
\label{tab:integratedprops}      
\centering    
\small
\begin{tabular}{lccc|c}       
\hline\hline                
Property&GRB host & Host spaxels avg. & GRB site    & Companion     \\
\hline
12+log(O/H) \scriptsize{(O3N2)} & 8.50$\pm$0.04  & 8.48$\pm$0.06 & 8.33$\pm$0.10& 8.54$\pm$0.16  \\
SFR  \scriptsize{[M$_\odot$y$^{-1}$]} & 2.28$\pm$0.05  & 0.0002$\pm$0.0002 & 0.020$\pm$0.001  & 0.17$\pm$0.01\\
sSFR \scriptsize{[M$_\odot$y$^{-1}$LL*]} &  1.16$\pm$0.03  &6.1$^{+7.9}_{-6.1}$&  26.4$\pm$1.01  & 0.74$\pm$0.06  \\
EW H$\alpha$ \scriptsize{[\AA{}]} & --27$\pm$3   &  --17$\pm$13     &$<$--18 &  --8$\pm$2 \\
E(B--V) \scriptsize{[mag]}& 0.26$\pm$ 0.21  &  0.20$\pm$0.17   & 0.39$\pm$0.4  &  0.03$\pm$0.03  \\
log U  & --2.57$\pm$0.23  &   --2.92$\pm$0.25    & --2.85$\pm$0.54 &   --2.77$\pm$0.40  \\
log N/O  &  --0.89$\pm$0.23 &  --0.99$\pm$0.17     &--1.23$\pm$0.52  &   --0.64$\pm$0.57 \\
\hline\hline 

\end{tabular}
\end{table*}

\begin{figure}[ht!]
\centering
	\includegraphics[width=\columnwidth]{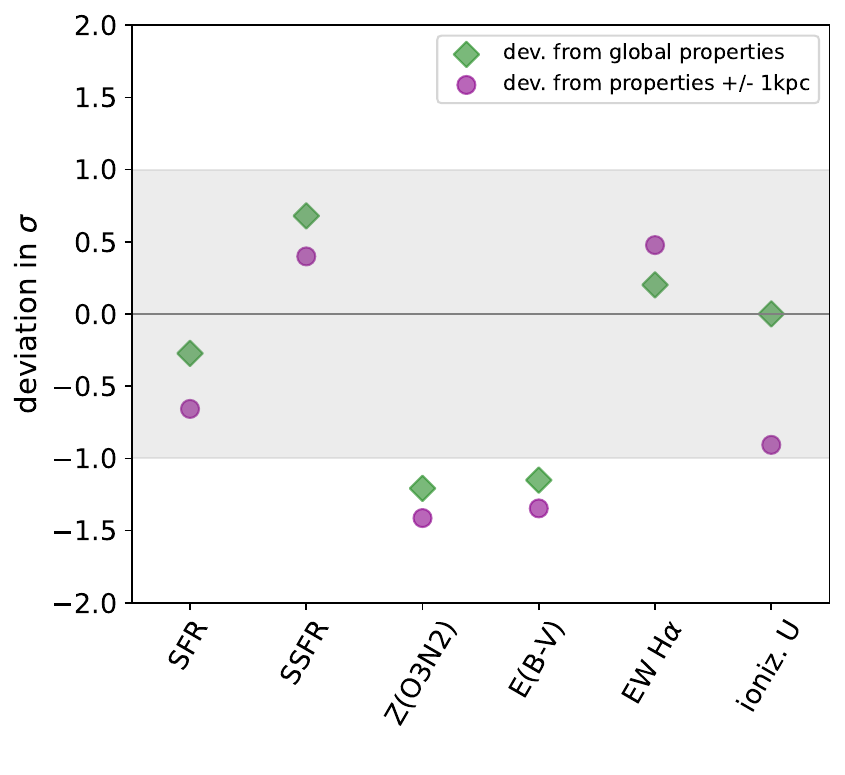}
    \caption{Deviation in standard deviation $\sigma$ of different properties in the GRB H~{\sc ii} region compared to the mean value in the host (green diamonds) and compared to the average value at  $\pm$1\,kpc of the deprojected distance of the GRB (8\,kpc, magenta dots). The SFR is determined as SFR/spaxel to account for the fact that the spectrum of the GRB HII region was integrated over several spaxels.}
    \label{fig:deviation}
\end{figure}

There are two other regions in the host that have a very young stellar population, region 15 and 17  (for the H~{\sc ii} region numbering and location, please refer to Fig.~\ref{fig:voronoi_segmap}) at the opposite side of the galaxy, with a higher fraction of a young population than the GRB site itself. Both regions have a moderate, but not very high SSFR and a metallicity at the host average. The lowest metallicity regions (20, 26 and 39, region 39 is next to the GRB site) do not show any particular pattern nor a very young stellar population and a lower SSFR than the GRB site. Several regions at the same side of galaxy as the GRB (region 1, 4, 5 and 8) have both a high EW and SSFRs ($>$3 M$_\odot$y$^{-1}$(L/L*)$^{-1}$) as do region 3 and 7, at the opposite, Northern, side of the galaxy, in fact, region 3 has the highest values both in SSFR (4.87 \,M$_\odot$y$^{-1}$(L/L*)$^{-1}$) and EW (--77\AA{}) and a metallicity on the lower side. The SP ages derived from STARBURST99 models, however, indicate SP ages of 7\,Myr for the highest EW regions. Possibly several of these regions mentioned above might have also been able, in principle, to produce a GRB progenitor, but due to the scarcity of GRBs (GRBs are about 1/10,000 times less frequent than core-collapse SNe, see e.g. \citealt{Spinelli23}) and particularly rare in high-mass galaxies \citep{Taggart21} it is unlikely to detect another GRB in the same galaxy. This scarcity might also point to some very peculiar stellar properties, that all have to occur in the same star, to actually make a star explode as a GRB, with many more parameters needing to be fulfilled than only the ISM properties from which the progenitor originates.

\subsection{The GRB site and host in comparison to other GRB hosts}

%r_50 of the host: 6.18 arcsec or 4.54 kpc
%deprojected distance GRB: 5.9 kpc
%measured distance: 7.12 arcsec or 5.23 kpc
As a grand-design spiral, the host of GRB\,171205A is a very rare type of GRB host. While it is difficult to determine the exact morphology of GRB hosts at higher redshifts, out of 47 long GRB hosts at $z<0.5$ there are only 10 confirmed spiral galaxies (including the host of GRB\,171205A), hence $\sim$20$\%$. However, two of them are disputed as being the actual host galaxy (GRB 190829A, a very large spiral host which might have a dwarf behind being the actual host, and GRB 051109B, where the host association is unclear), two are SN-less long GRBs (GRB 060505, a dwarf spiral, \citealt{Thoene14}), and GRB 111005A, a nearly edge-on large spiral, \citealt{Michalowski18}) and for GRB 011121 fits to the images suggest a disk and bulge, but definite imaging is missing. This is somewhat surprising since 40\% of low mass galaxies (log M$<$9.5 M$_\odot$) and 70\% of high mass galaxies (log M$>$9.5 M$_\odot$) at $z<1$ are spiral galaxies or classified as ``disky'', while the fraction of spirals goes to almost zero at $z\sim4$ (see the recent JWST study by \citealt{Jacobs23}). In Fig.~\ref{fig:hostcomparison} we plot the GRB hosts mentioned above in relative physical size together with the dwarf irregular host of GRB\,100416D and the smallest host detected, the host of GRB 060218 with an r$_\mathrm{50}$ of only 0.36$^{+0.00}_{-0.01}$\,kpc \citep{Lyman17}. The typical GRB host has an r$_\mathrm{50}$ and r$_\mathrm{80}$ of 1.7$\pm$0.2 and 3.1$\pm$0.4 kpc respectively \citep{Lyman17}, while the r$_\mathrm{50}$ for the host of GRB 171205A is 4.54\,kpc, more than twice the median for long GRB hosts.

Long GRBs usually have a small offset from their host, in contrast to short GRBs. Since most long GRB hosts are dwarfs, the absolute median offset is only 1.0--1.3~kpc depending on the sample used \citep{Blanchard16,Lyman17}. The absolute distance for the GRB site of GRB 171205A is 5.23\,kpc, almost 5 times larger than the median offset of previous samples. However, due to the different physical sizes of GRB hosts, it is more instructive to use the offset normalized to r$_{50}$. The site of GRB 171205A has a normalized offset of 1.15, which puts it among the 20\% largest offsets for long GRB hosts \citep{Blanchard16, Lyman17} (see Fig. \ref{fig:cumulative}), on the upper end of values for the offset but still well within a standard GRB host.

\begin{figure*}[ht!]
\centering
	\includegraphics[width=\textwidth]{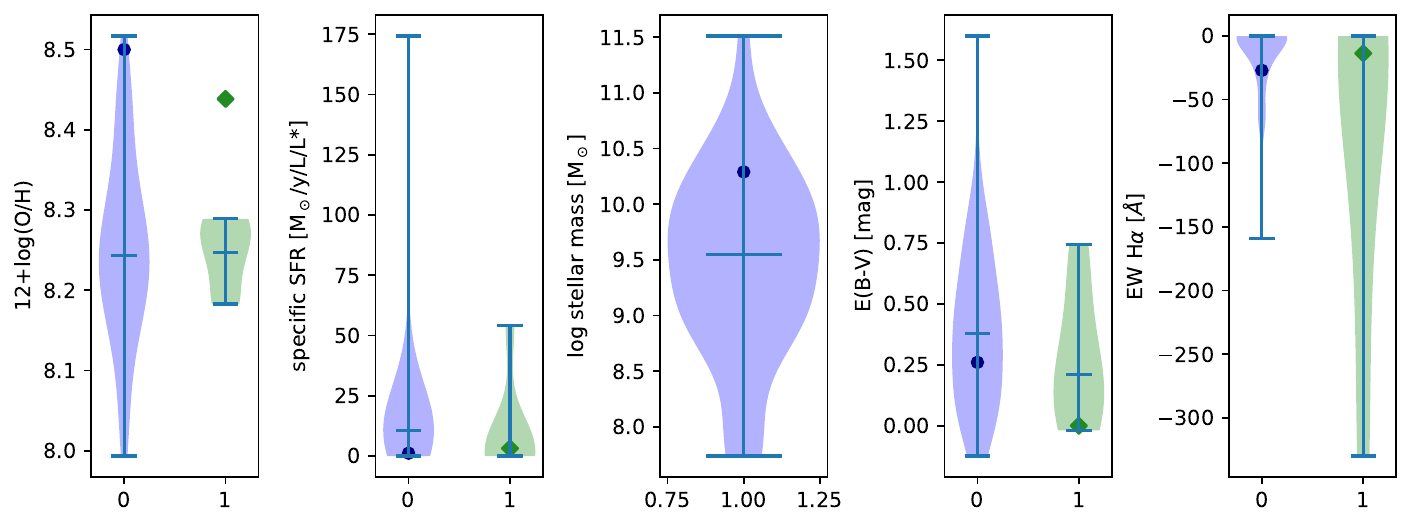}
    \caption{Violin plots of different properties of global/integrated GRB host spectra (blue plots) and the same properties at the resolved GRB sites (green plots). The dots are the global values for the host of GRB 171205A, diamonds are the values at the GRB site of GRB 171205A. Emission lines for the integrated host spectra were obtained from the samples of \cite{Kruehler15, Han10}, with additional single GRBs from \cite{DellaValle06, Christensen08, Kelly13, Schulze14, Thoene14, Izzo17, Heintz18, Cano17, deUgarte18, Melandri19, deUgarte20}. Stellar masses were taken from the GHostS database (\protect\url{http://www.grbhosts.org}) and \cite{Palmerio19}. Rest-frame B-band magnitudes to derive the SSFR were taken from the TOUGH sample \citep{Hjorth12} and \cite{Kruehler11}. H$\alpha$ EW were determined from public spectra stored in the GRBspec database \citep{deUgarteGRBspec}. The values for the GRB sites were taken from \cite{Christensen08, Thoene14, Izzo17, Cano17, deUgarte18, Melandri19}. 
    }
    \label{fig:violinplot}
\end{figure*}

To compare the host of GRB\,171205A with the general population of long GRB hosts we collect properties of 54 long GRB hosts from the literature up to $z \sim 2.4$ with an average redshift of 0.93 using global or integrated spectra of the host as well as 7 low redshift ($z<0.65$) resolved GRB sites (IFU and longslit). We derive metallicities using the O3N2 parameter \citep[using the calibration in][]{MarinoZ} (47 hosts), the specific SFR (39 hosts), stellar mass (38 hosts), extinction (47 hosts) and H$\alpha$ EWs (16 hosts) of both the integrated spectra and the sample of resolved GRB locations (except for the stellar mass where we do not have measurements for the GRB site only), visualized as violin plots in Fig.~\ref{fig:violinplot} (for references to the different literature samples used see the figure caption). 

Both the host and GRB site of GRB\,171205A show a high metallicity compared to other GRB hosts/sites, the GRB site would even fall in the upper metallicity range for the global sample. Note that we derived all metallicities consistently using the O3N2 parameter in the calibration from \cite{MarinoZ}, which gives lower values than some of the calibrators used in e.g. \citet{Kruehler15} and we do not obtain any super solar values for the same GRB hosts in their study. The location in the BPT diagram at the lower end of the distribution for GRB hosts, implying a low ionization and high metallicity, is in line with this observation. Regarding the SSFR, the GRB site is somewhat above the median while the value of the integrated galaxy value falls below the median of the sample, not surprising given that the host is not a starburst galaxy. The mass of the galaxy is in the upper third of the distribution but not among the most massive hosts, which, however, are all at redshifts z$>$1, in line with the findings of \cite{Palmerio19} that GRB host masses increase with redshift. Previous studies have found a metallicity threshold for GRB hosts \citep{PerleySHOALS2, Vergani17, Schulze14}, however at rather high values of 0.5--0.9\,Z$_\odot$, which is not in tension with our findings. The extinction of the host is average and the GRB site has one of the lowest values of the sample, also not surprising given its location in a small, blue star-forming region in the outskirts of the galaxy.

\begin{figure*}[h!]
	\includegraphics[width=\textwidth]{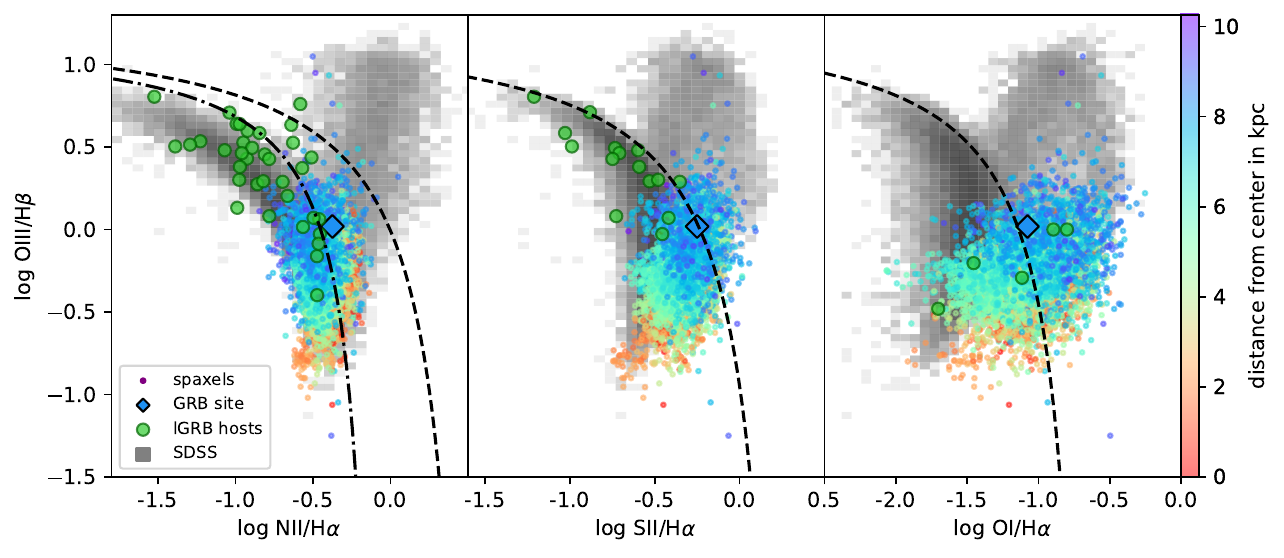}
    \caption{BPT diagrams using [N~{\sc ii}]/H$\alpha$, [S~{\sc ii}]/H$\alpha$ and [O~{\sc i}]/H$\alpha$ for individual spaxels compared to the values for other long GRB hosts (green dots) and star forming galaxies in the SDSS (grey grid). The spaxels are color coded by their deprojected distance from the galaxy center as shown in the color bar. The value of the GRB site (color coded by its distance) is shown as a diamond. Emission lines for the GRB sample have been obtained from the same data used in Fig.~\ref{fig:violinplot}. We only plot spaxels with EW$<$--6\AA{} to remove diffuse interstellar gas (DIG) as explained in the text.
    \label{fig:BPT}}
\end{figure*}

\subsection{Kinematics of neutral vs. molecular vs. ionized gas and possible star formation triggers}\label{sect:discussion_alldata}

The velocity fields of ionized (traced by H$\alpha$), and molecular (traced by CO(1--0)) gas are surprisingly smooth and show no sign of any major interaction in the recent past. Interaction is known as one of the triggers of star-formation, in particular outside of denser cluster regions. The only major companion at the same redshift is 183\,kpc from the host and any possible interaction would have happened at timescales much beyond what is needed for a recent burst of star-formation.

In contrast, the H~{\sc i} gas distribution is more asymmetric. \cite{Arabsalmani22} publish a high resolution map of H~{\sc i} of the VLA data which shows a butterfly like shape and an extra component in the S-W. While the main double-lobe of H~{\sc i} gas is likely the normal H~{\sc i} disk of the host, the extra component was interpreted as the H~{\sc i} disk of a recently interacting satellite. Our reanalysis of the same dataset did not result in any extraplanar emission in the Western lobe, instead, we observe some extra component south of the Eastern lobe, but also at low significance. Thus we issue caution here in the overinterpretation of any extraplanar feature. The kinematics of both the component plotted in \cite{Arabsalmani22} and the different one we found in our analysis are consistent with being a part of the rotating H~{\sc i} disk. We do not find any evidence for other emission in the MUSE cube for any of these proposed extraplanar components, neither in continuum nor H$\alpha$.

Star-formation in the host is concentrated into a ring around 5--7\,kpc from the center, somewhat closer to the center than the GRB site, corresponding to an undisturbed inside-out star formation process in this galaxy. However, in the S-E part of the galaxy, where the GRB exploded, there is enhanced star-formation  both in- and outside of the main ring (see Fig.~\ref{fig:deprojectedhost}). This could suggest that the H~{\sc i} inflow/small merger triggered some additional SF in this region of the galaxy. There are a few regions with high SSFR, low metallicity and low SP age at this side of the galaxy, region 4, 5 and 8 discussed in Sect. \ref{sect:discussion_local} as well as the GRB region. The most extreme H~{\sc ii} regions, however, are at the opposite side of the galaxy, where, if the merger has been rather recent, this seems unlikely as the trigger of high SF in that opposite part of the galaxy. 

Our comparison between neutral, molecular and ionized gas do not show any obvious indication for this proposed merger to be the needed SF trigger. \cite{Michalowski12,Michalowski15} have found large amounts of H~{\sc i} gas close to the GRB site in several GRB hosts and propose inflow of H~{\sc i} gas as possible SF trigger. \cite{Arabsalmani19} found a ring-like structure in the host of GRB\,980425 in H~{\sc i} and interpret this as remaining from a previous interaction with a dwarf galaxy. Inflow or interaction only visible in H~{\sc i} gas might also be at place here, but we still lack any evidence in other wavelengths or properties to confirm this theory.

\section{Conclusions}
In this paper we presented the most extensive resolved multi-wavelength study of a long GRB host to date, GRB 171205A at a redshift of $z=0.036$ or 163\,Mpc using MUSE/VLT IFU spectroscopy, HST UV and VIS imaging as well as ALMA CO(1-0) and H~{\sc i} 21cm observations. These are the main conclusions from this extensive dataset: 
\begin{itemize}
    \item The host is a grand-design spiral galaxy with a short bar, a rare type for long GRB hosts, in particular at low redshift ($<$20\%), which are primarily irregular galaxies. In contrast, almost half of low redshift galaxies have a disk morphology. 
    \item The host shows a smooth, negative, metallicity gradient of --0.015 dex~kpc$^{-1}$. The total SFR is not very high with $\sim$1\,M$_{\odot}$~yr$^{-1}$.
    \item The GRB was located in a small star-forming region in an outer spiral arm at a deprojected distance of $\sim$\,8\,kpc.
   The metallicity at the GRB site is around half solar and the absolute SFR is low, likewise, the region is only among the 45$\%$ brightest pixels in the host in UV light. Given the size of the H~{\sc ii} region, however, the luminosity weighted SSFR is one of the highest in the host (3.5\,M$_\odot$y$^{-1}$(L\,L*)$^{-1}$) and high compared to other GRB sites. SP modeling gives a significant contribution from a very young stellar population. The region seems dust-free with zero extinction.
   \item The GRB site is not a uniquely special site in the host. Several other H~{\sc ii} regions show a similar combination of low metallicity, high SSFR and young age. The H~{\sc ii} region in which the GRB occurred, shows the 6th highest SSFR and the 4th lowest metallicity, a comparison of the age is complicated by the presence of the underlying GRB-SN.
   \item The velocity field of the ionized and molecular gas is very regular with only small deviations in the center, connected to the bar. The end of the bar shows some further evidence of shocked regions from optical emission line ratios. The width of CO is $\sim$15 km~s$^{-1}$ lower than the one of the H$\alpha$, but the velocity fields of CO and H$\alpha$ match well. 
    \item The H~{\sc i} emission is different from ionized and molecular gas with an excess to the South-East of the galaxy, which could be connected to an inflow of gas. This has also been observed in the past for several other GRB hosts, but not for the hosts of other types of supernovae.
    
\end{itemize}

Our study shows that long GRBs can occur in many different types of (star-forming) galaxies, including a rather standard, yet star-forming, galaxy for its redshift, and that the conditions at the actual site are more important than general host properties. Previous resolved studies of GRB hosts have also shown some, but generally not extreme differences between the GRB site and the global host properties \citep{Christensen08, Kruehler17, Thoene14, Izzo17}. The sample is still rather small, but it is likely that the larger and more unusual the host is, the larger are the deviations of the GRB site from the host properties. This would speak in favour of GRBs requiring a set of very limited conditions which can occur in different types of galaxies, as long as some star-forming region has just the right conditions to form a GRB. On the other hand, the region hosting a GRB progenitor, especially in larger hosts, might not be a unique region. Considering the scarcity of stars turning into a GRB at the end of their lives, this is difficult to verify for a given host galaxy.

We also note that this GRB was a so-called ``low-luminosity'' GRB (ll-GRB), which can only be observed at low redshifts due to their faint afterglow, hence the results for cosmological GRBs might be different. On the other hand, properties such as metallicity and SFR were less extreme in this host and its GRB site than found for other GRB hosts. Several studies have found the GRB production to be stifled above a certain metallicity, ranging between 0.5 and solar metallicity \citep{Schulze15, PerleySHOALS2, Palmerio19}. Only one paper has investigated and found a non-correlation between metallicity and GRB $\gamma$-ray emission \citep{Levesque10c}, however, this was based on a sample of only 16 GRBs before 2010. The fact that this GRB host and site had a relatively high metallicity and that both the GRB, afterglow and SN were relatively weak (which also made the detection of the cocoon emission possible) could have had a physical reason, e.g. that higher metallicity stars lead to an almost-choked jet. Further studies on this might be warranted.

Despite the extensive dataset across the electromagnetic spectrum, the origin for the SF hosting this GRB is still unclear. Violent SF or a SF trigger through major interactions does not seem to be needed to produce a GRB progenitor and there have been very few cases so far for any kind of recent interaction triggering new SF. Several hosts, however, do show an asymmetric, offset, extraplanar presence of H~{\sc i} gas, often close to the location of the GRB. Further, highly detailed and panchromatic studies, will be needed to get adequate statistics on kinematics of different gas components as well as abundances and other properties at the GRB site, and extension to higher redshifts is also highly warranted, albeit observationally challenging.

\begin{acknowledgements}
Based on observations collected at the European Organisation for Astronomical Research in the Southern Hemisphere under ESO programme 0100.D-0649. 

JFAF acknowledges support from the Spanish Ministerio de Ciencia, Innovaci\'on y Universidades through the grant PRE2018-086507.

M.J.M.~acknowledges the support of the National Science Centre, Poland through the SONATA BIS grant 2018/30/E/ST9/00208 and the Polish National Agency for Academic Exchange Bekker grant BPN/BEK/2022/1/00110.

AJL was supported by the European Research Council (ERC) under the European Union’s Horizon 2020 research and innovation programme (grant agreement No.~725246).

L.C. is supported by DFF/Independent Research Fund Denmark, grant-ID 2032--00071.

J. L. thanks Aaron Lawson and Ananthan Karunakaran for useful discussions regarding the VLA data.

This research has made use of the GHostS database (www.grbhosts.org), which is partly funded by Spitzer/NASA grant RSA Agreement No. 1287913.

This research is based on observations made with the NASA/ESA Hubble Space Telescope obtained from the Space Telescope Science Institute, which is operated by the Association of Universities for Research in Astronomy, Inc., under NASA contract NAS 5–26555. These observations are associated with program 15349.

\end{acknowledgements}

%\begin{thebibliography}{200}
\bibliographystyle{aa}
\bibliography{3D_171205A_astroph}

%\end{thebibliography}

%%%%%%%%%%%%%%%%% APPENDICES %%%%%%%%%%%%%%%%%%%%%

\newpage

\begin{appendix}
\section{Additional figures}
\begin{figure*}
\centering
    \includegraphics[width=16cm]{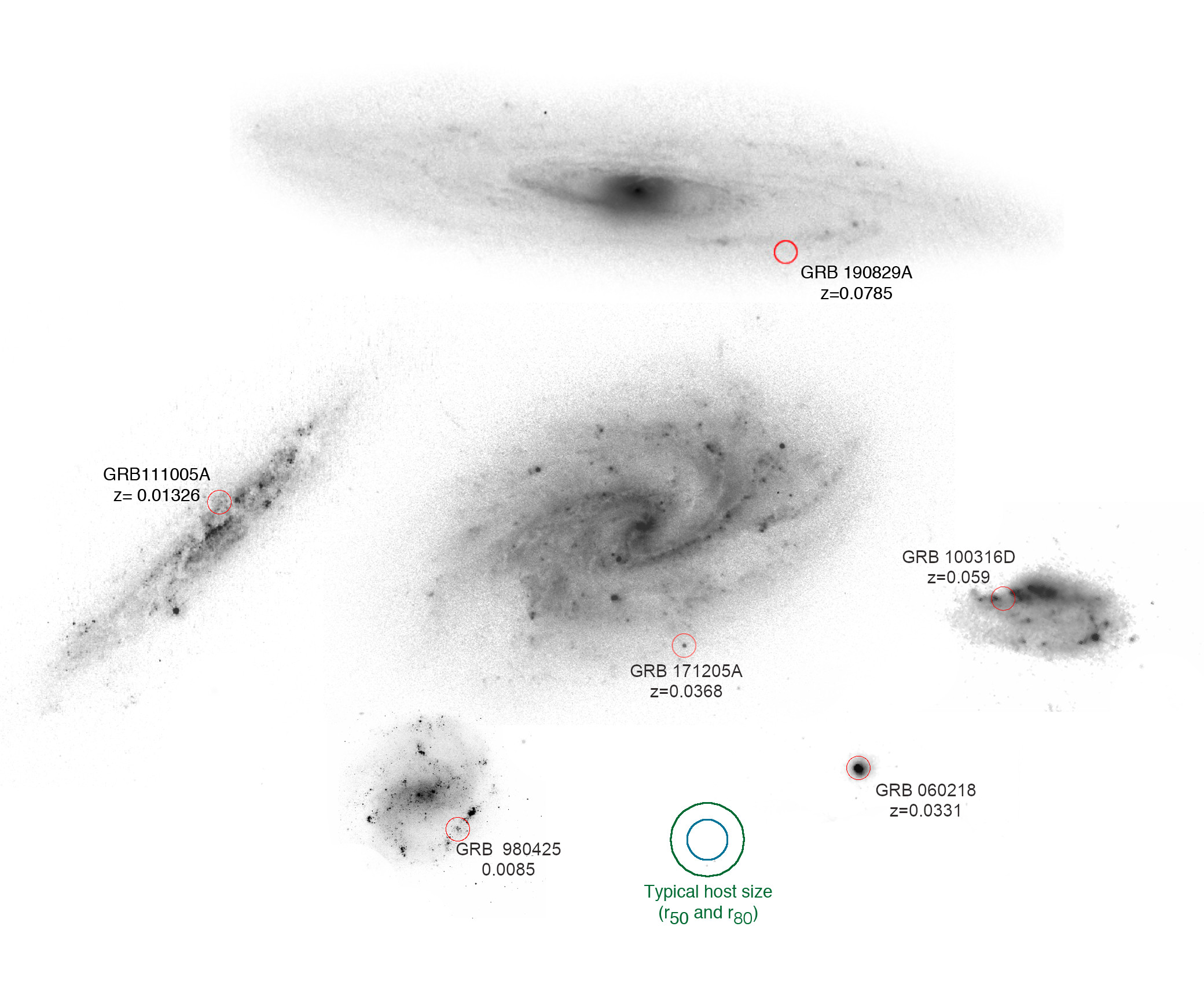}	
	    \caption{Comparison of the physical size of different low redshift long GRB hosts including the host of GRB\,171205A. The red circles have a radius of 1 kpc and are placed at the best known location of each  burst. Also indicated is the typical r$_{50}$= 1.7 kpc of long GRB hosts \citep{Lyman17}}
    \label{fig:hostcomparison}
\end{figure*}

\begin{figure*}
    \includegraphics[width=\columnwidth]{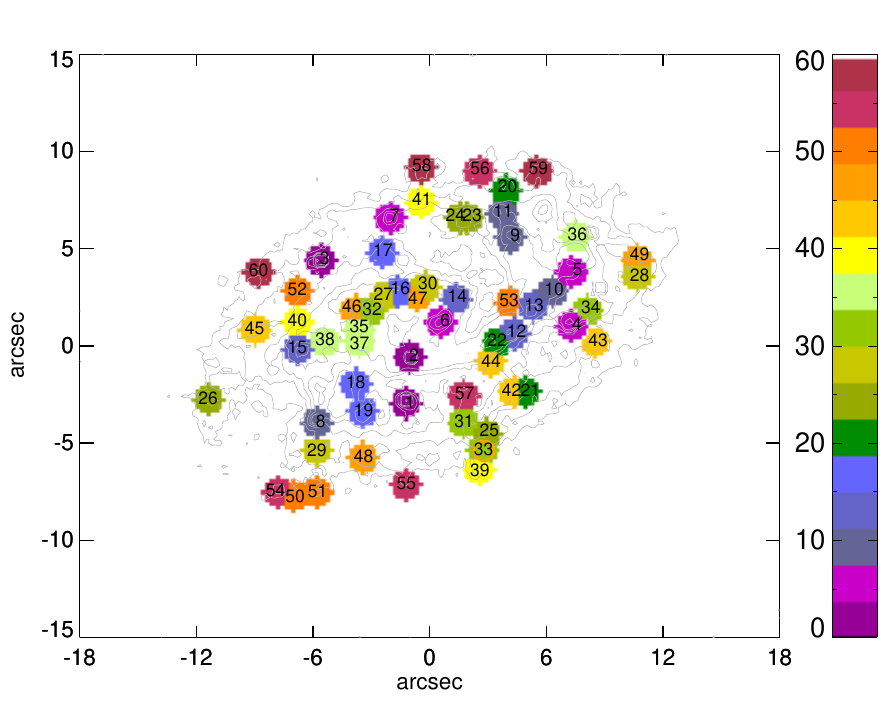}\, \, \, \, \, 
	\includegraphics[width=\columnwidth]{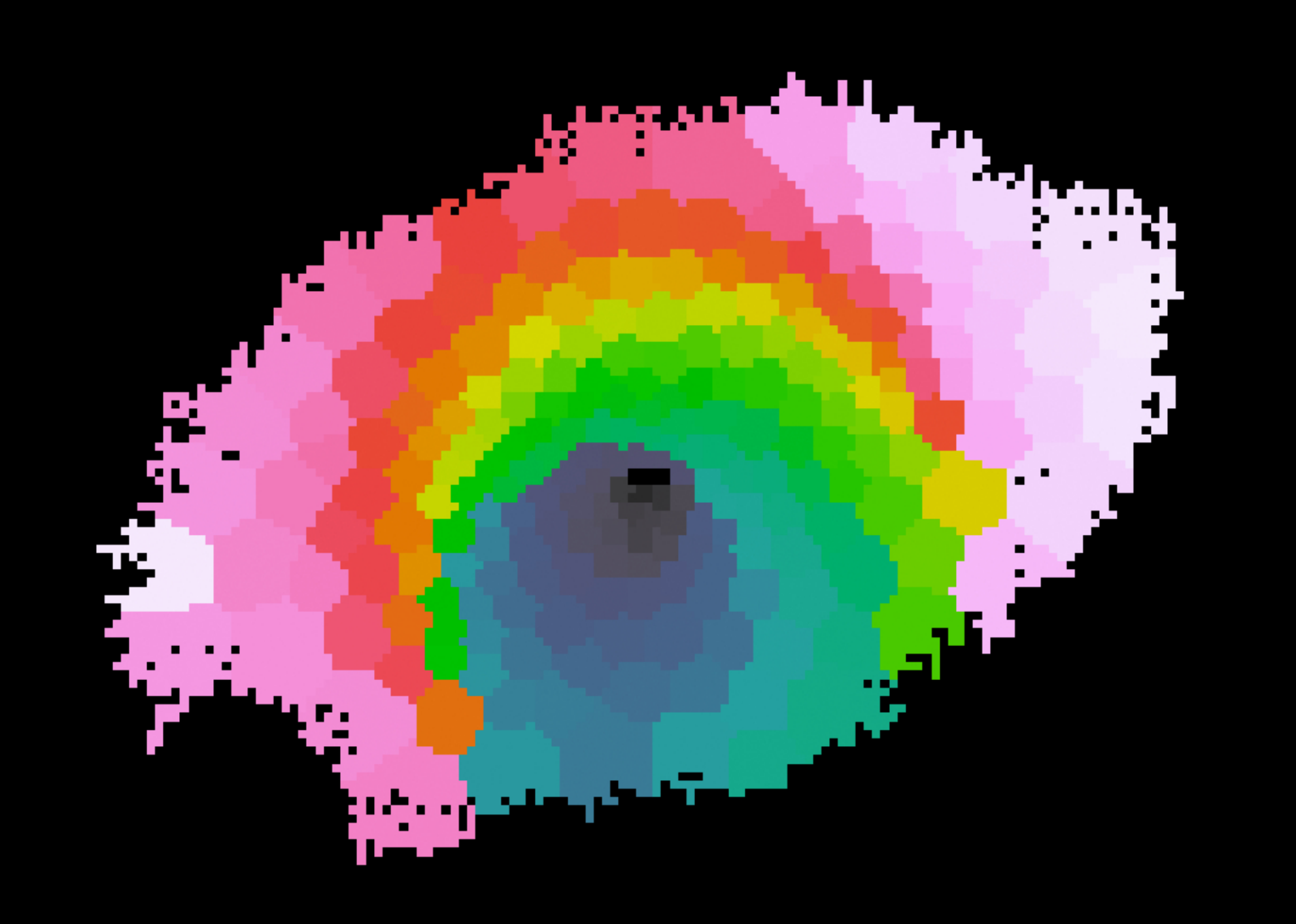}	
	    \caption{Left: H~{\sc ii} regions with extracted integrated spectra used in the analysis throughout the paper. Overlaid are continuum contours from HST. We number the H~{\sc ii} regions for easier reference in the text. Right: Voronoi tessellation of the MUSE continuum data constructed to give a signal to noise ratio of 40 per bin.}
    \label{fig:voronoi_segmap}
\end{figure*}

\begin{figure*}
\includegraphics[width=8.1cm]{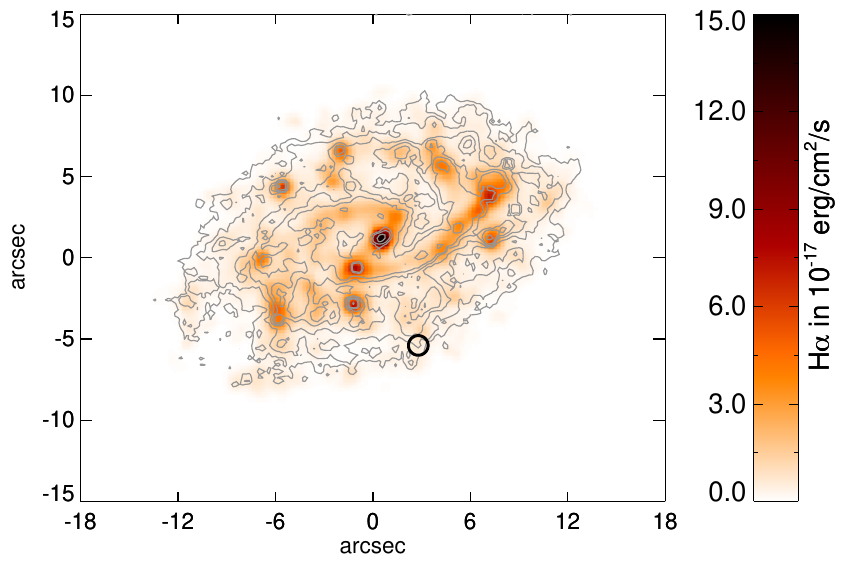}
\includegraphics[width=8.1cm]{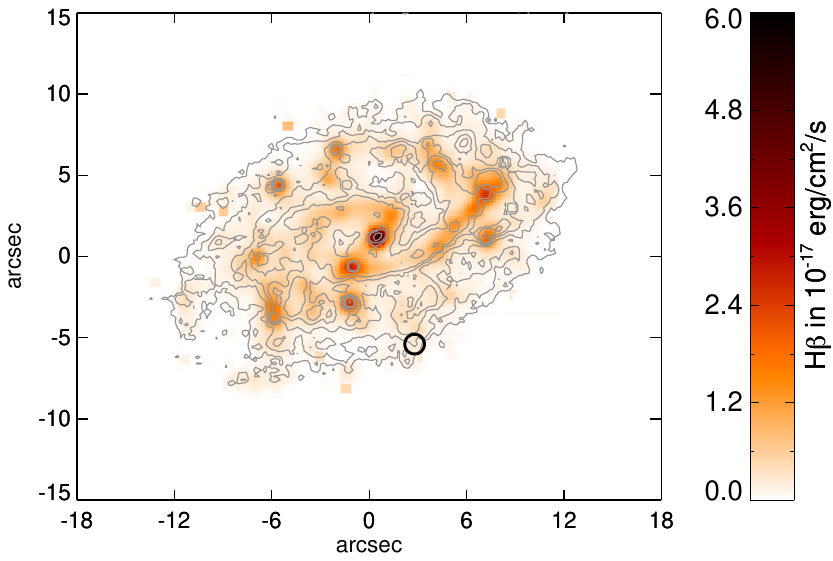}\\
\includegraphics[width=8.1cm]{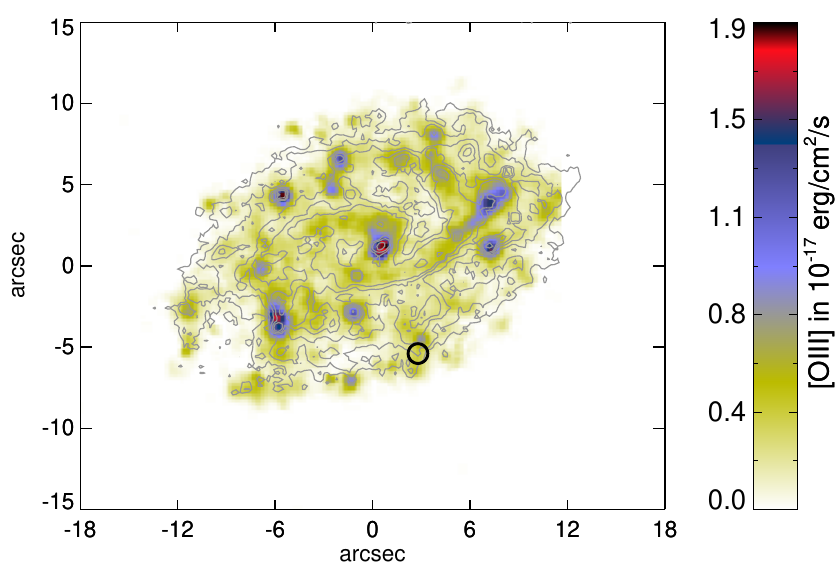}
\includegraphics[width=8.1cm]{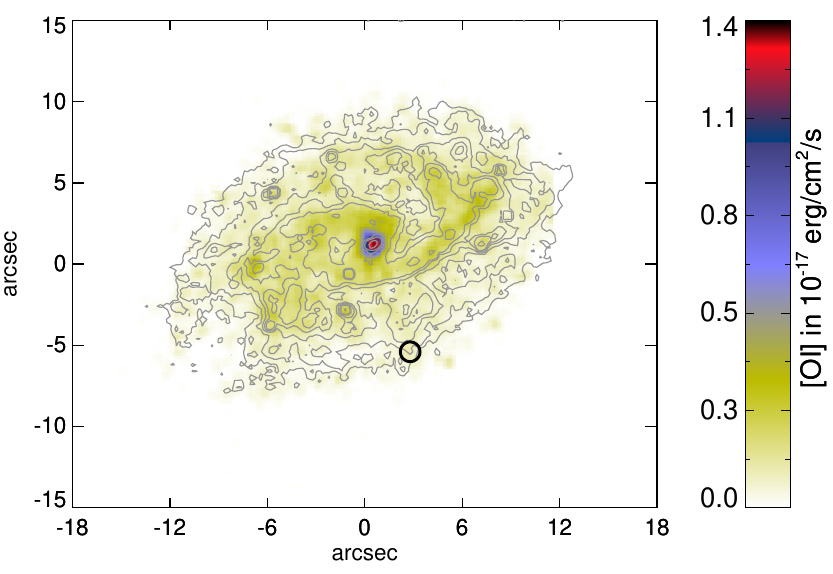}\\
	\includegraphics[width=8.1cm]{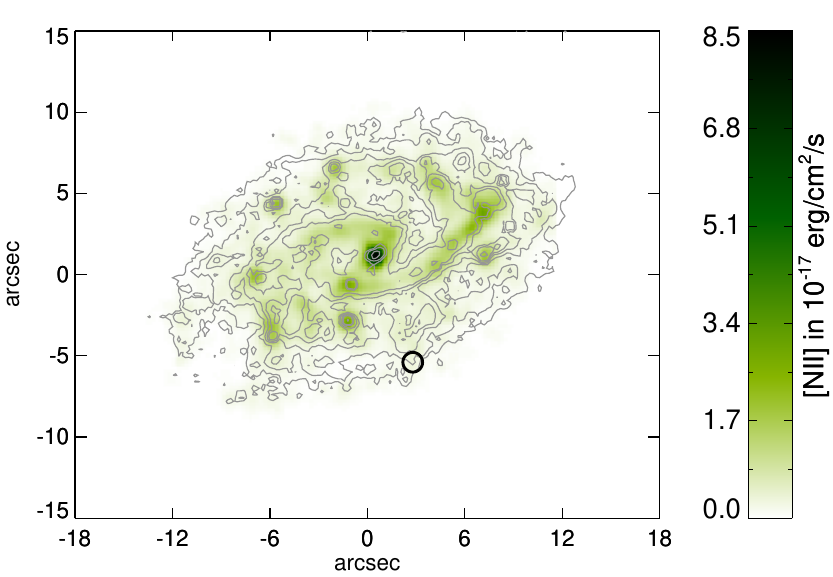}
	\includegraphics[width=8.1cm]{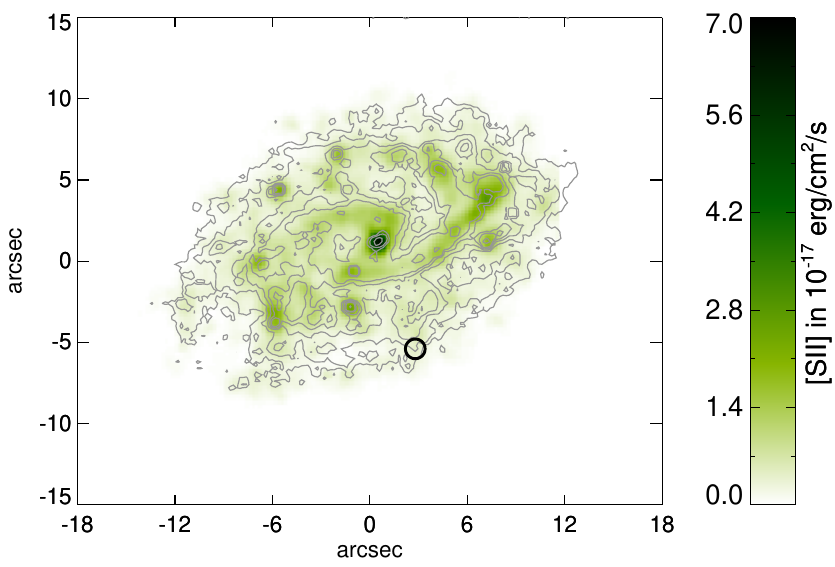}\\
	\includegraphics[width=8.1cm]{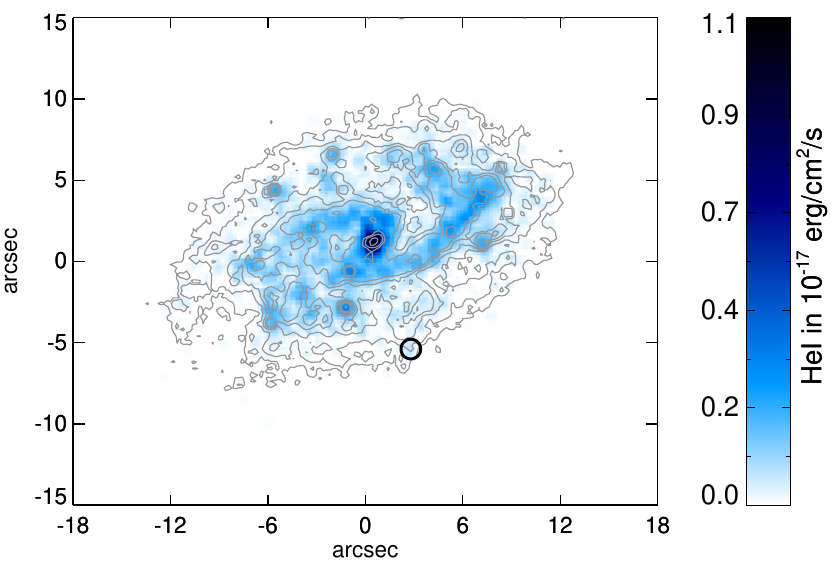}
    \includegraphics[width=8.1cm]{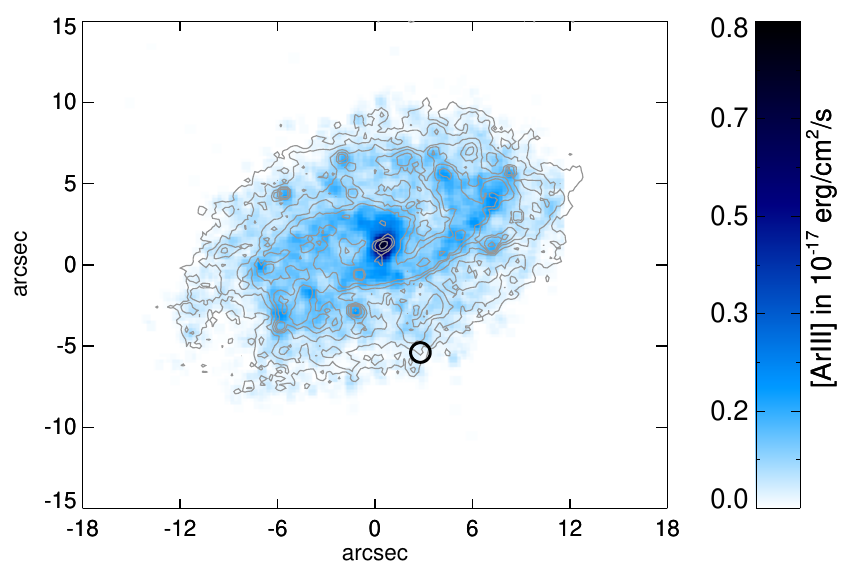}\\
    \caption{From top left to bottom left: Maps of H$\alpha$, H$\beta$, [O~{\sc iii}] $\lambda$ 5008, [O~{\sc i}] $\lambda$ 6300, [N~{\sc ii}] $\lambda$ 6585, the sum of the [S~{\sc ii}] $\lambda\lambda$ 6717,6732 doublet,  He~{\sc i} $\lambda$ 5877 and Ar~{\sc iii} $\lambda$ 7135, all fluxes are 10$^{-17}$ erg\,cm$^{-2}$\,\AA$^{-1}$ and corrected for Galactic and host intrinsic extinction. The position of the GRB is indicated with a black circle.
    \label{figapp1}}
\end{figure*}

\begin{figure}
	\includegraphics[width=8.2cm]{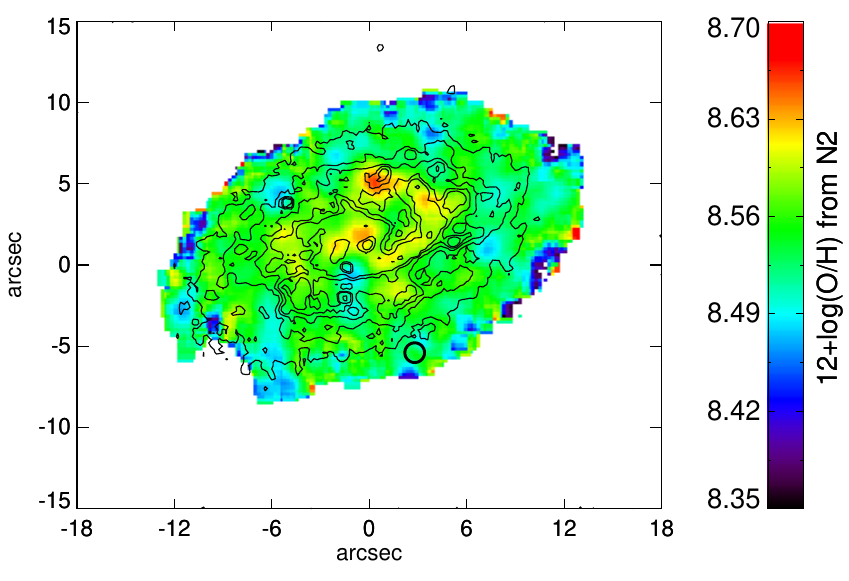}
	   \caption{Map of the metallicity determined by the N2 parameter according to \cite{MarinoZ}.
   \label{figapp2}}
\end{figure}

\begin{figure}[ht!]
	\includegraphics[width=\columnwidth]{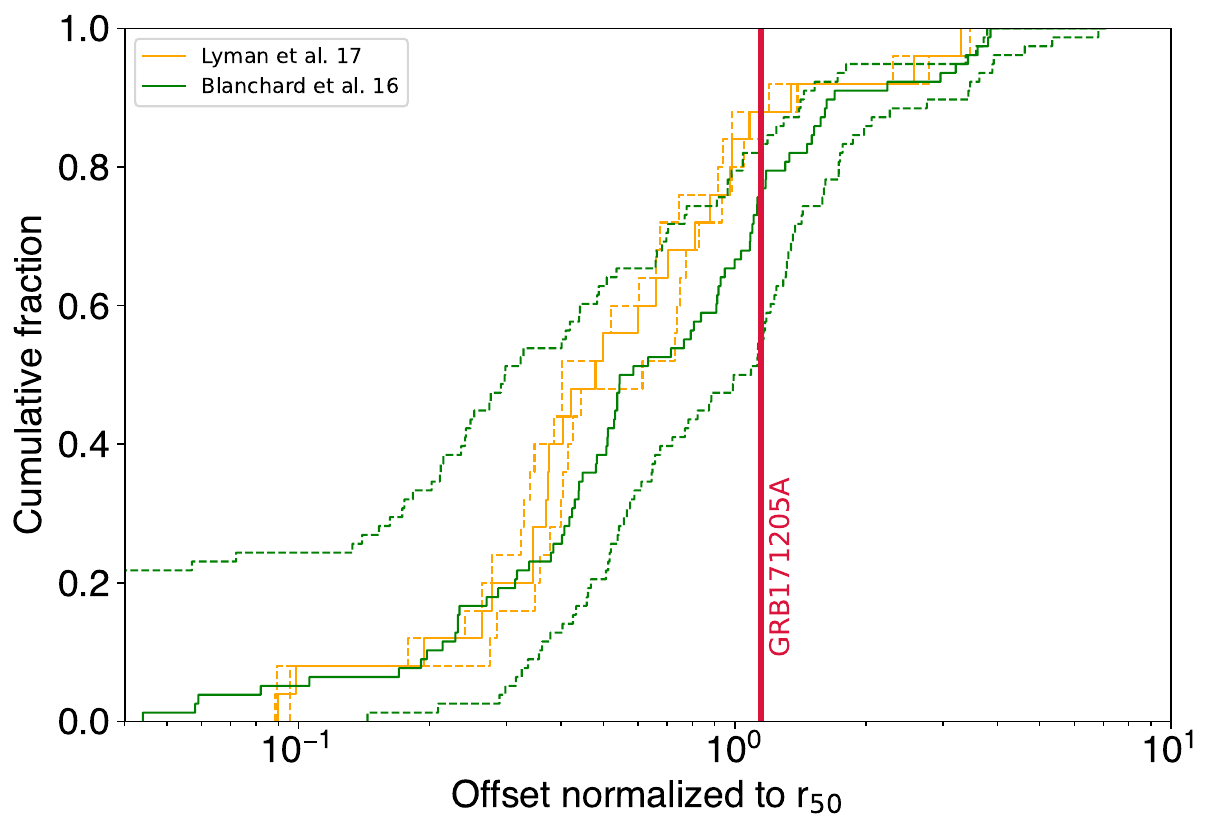}\\
 \includegraphics[width=\columnwidth]{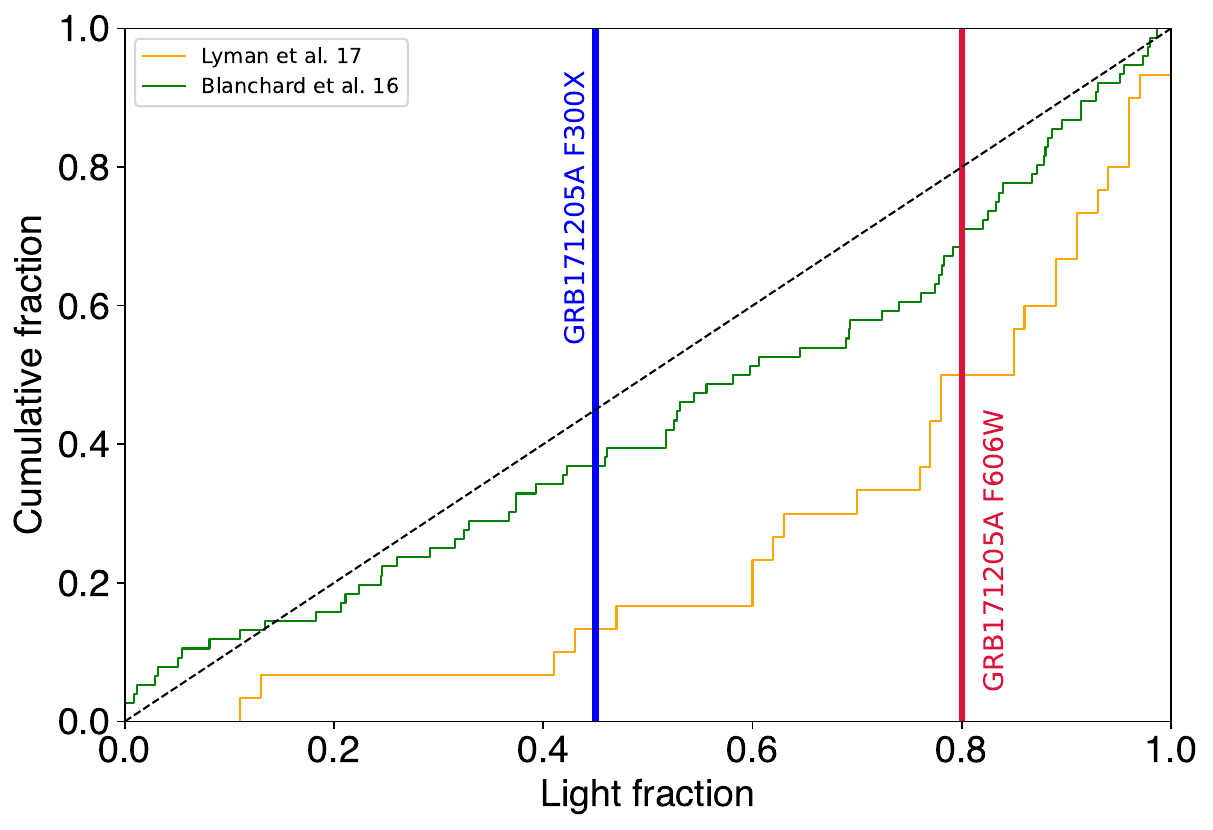}
    \caption{Top: Cumulative offsets normalized to r$_{50}$ for two samples of long GRB hosts galaxies observed with HST \citep{Lyman17,Blanchard16}. The offset for GRB\,171205A is the measured offset, not the deprojected offset used elsewhere in the paper since the comparison samples are mostly irregular galaxies where deprojection is hampered. Bottom: Cumulative distribution of the fraction of brightest pixel at the GRB site in different GRB host galaxies from the same sample as in the plot above. The dashed line indicates the GRB site following the general light distribution. The two lines indicate the light fraction for the site of GRB\,171205A in the F300X filter (UV) and F606W filter ($\sim$R-band).}
    \label{fig:cumulative}
\end{figure}

\begin{figure}
\includegraphics[width=8.7cm]{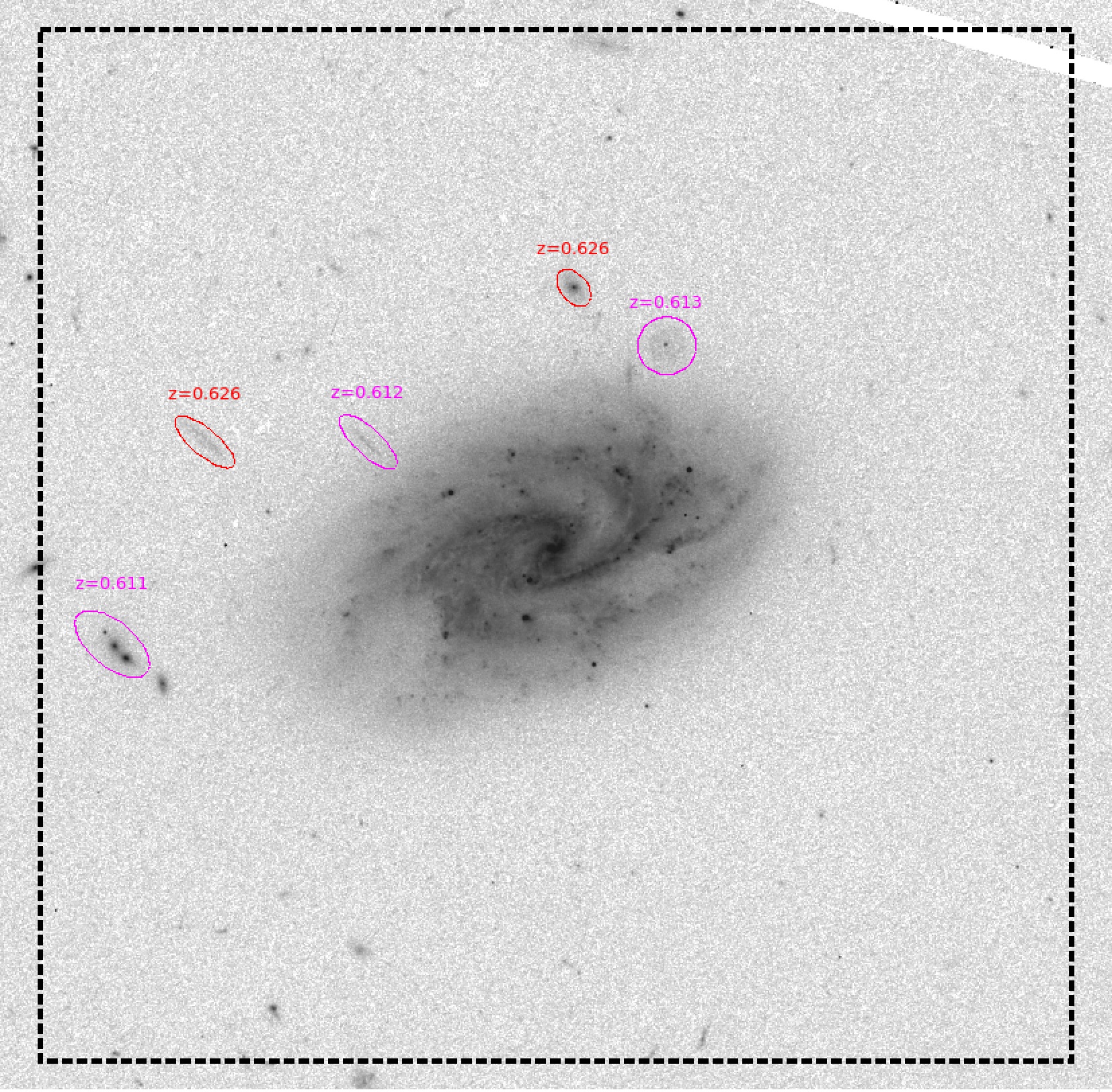}
\caption{Galaxies in the FOV of MUSE around the host of GRB 171205A with their corresponding redshifts. None of the sources are physically related to the GRB host. 
    \label{fig:companion}}
\end{figure}

\begin{figure}
\includegraphics[width=8.7cm]{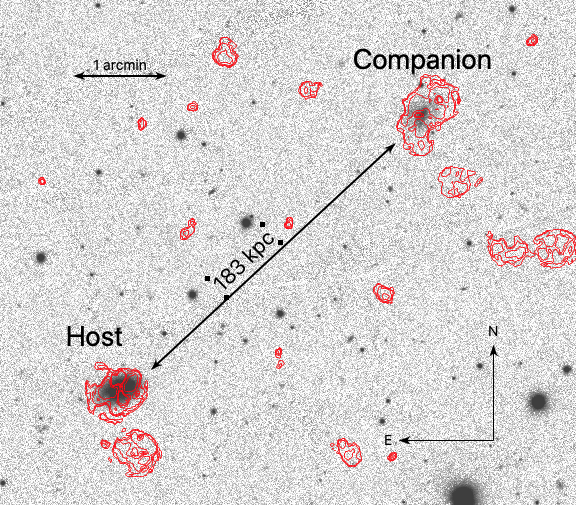}
\caption{Larger view of the field including the host galaxy and the companion mentioned throughout the paper. Red contours are from VLA. Both objects are at a very similar redshift and hence part of a group.
    \label{fig:largefield}}
\end{figure}

\begin{figure*}
\centering
\includegraphics[width=15cm]{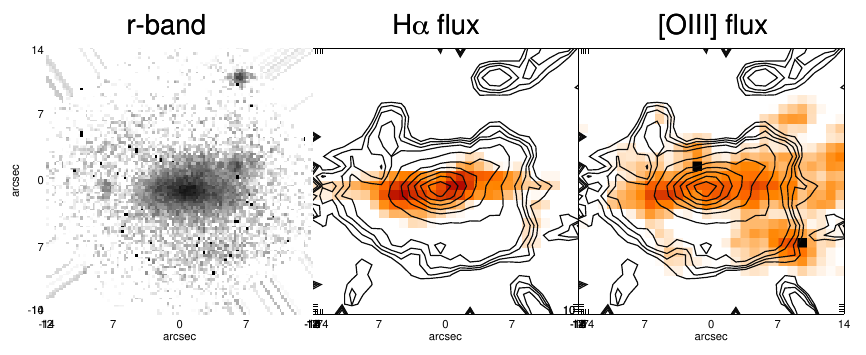}
\caption{R-band image and emission line maps of H$\alpha$ and [O~{\sc iii}] $\lambda$ 5008 for LEDA 951348, the galaxy companion of the host of GRB 171205A at 186\,kpc.
    \label{figapp3}}
\end{figure*}

\newpage

\begin{table*}

\caption{Properties of the integrated H~{\sc ii} regions, region 33 contains the GRB}            
\label{tab:HIIregions}      
\centering   
\scriptsize
\begin{tabular}{ccccccccccc}       
\hline\hline     
 Number & dist. & offset x & offset y & 12+log(O/H) & 12+log(O/H) & SFR & SSFR & EW & E(B--V) \\
  & kpc & arcsec & arcsec & (N2) & (O3N2) & 10$^{-4}$M$_\odot$/y/spaxel & M$_\odot$/y/L/L* & \AA{} & [mag] \\ \hline\hline 
           1 &   6.05 &  -8.00 & -21.00 &   8.56 &   8.52 &  10.16 & 141.20 & -65.17 &   0.22\\
           2 &   2.70 &  -7.00 &  -9.00 &   8.54 &   8.57 &  12.46 & 173.14 & -42.17 &   0.29\\
           3 &   9.09 & -30.00 &  16.00 &   8.54 &   8.45 &   7.76 & 107.76 & -77.17 &   0.22\\
           4 &   7.64 &  34.89 &  -1.11 &   8.54 &   8.47 &   8.10 & 107.90 & -69.88 &   0.22\\
           5 &   7.47 &  35.00 &  13.00 &   8.57 &   8.50 &  11.81 & 164.08 & -52.57 &   0.36\\
           6 &   0.22 &   1.00 &   0.00 &   8.65 &   8.56 &  19.14 & 265.98 & -15.95 &   0.66\\
           7 &   9.18 & -12.00 &  27.00 &   8.55 &   8.47 &   7.07 &  98.29 & -60.14 &   0.30\\
           8 &   8.74 & -31.00 & -26.00 &   8.56 &   8.46 &   7.27 & 100.98 & -67.52 &   0.24\\
           9 &   6.69 &  19.04 &  21.85 &   8.58 &   8.52 &   7.19 &  95.87 & -32.76 &   0.28\\
          10 &   6.24 &  30.00 &   8.00 &   8.60 &   8.52 &   8.58 & 109.48 & -30.89 &   0.38\\
          11 &   8.21 &  16.93 &  28.13 &   8.57 &   8.49 &   4.82 &  61.55 & -37.43 &   0.22\\
          12 &   4.71 &  20.40 &  -2.81 &   8.59 &   8.54 &   7.09 &  86.41 & -28.95 &   0.33\\
          13 &   5.17 &  24.89 &   3.89 &   8.60 &   8.52 &   7.15 &  91.22 & -24.05 &   0.35\\
          14 &   1.88 &   5.11 &   6.20 &   8.61 &   8.54 &   8.51 & 110.98 & -17.35 &   0.48\\
          15 &   7.46 & -36.00 &  -7.08 &   8.61 &   8.51 &   6.11 &  83.15 & -23.66 &   0.46\\
          16 &   3.64 &  -9.53 &   8.25 &   8.63 &   8.53 &   4.57 &  46.60 & -15.36 &   0.32\\
          17 &   6.90 & -14.00 &  18.00 &   8.59 &   8.48 &   4.89 &  67.96 & -26.43 &   0.40\\
          18 &   5.59 & -21.00 & -15.92 &   8.62 &   8.52 &   5.08 &  69.20 & -18.51 &   0.41\\
          19 &   6.99 & -19.00 & -23.08 &   8.60 &   8.51 &   4.17 &  56.74 & -21.75 &   0.30\\
          20 &   9.95 &  18.05 &  34.36 &   8.53 &   8.38 &   2.17 &  27.01 & -31.30 &   0.13\\
          21 &   8.46 &  23.92 & -18.00 &   8.58 &   8.48 &   2.67 &  27.30 & -31.18 &   0.22\\
          22 &   4.02 &  15.34 &  -5.09 &   8.61 &   8.53 &   4.90 &  48.58 & -18.22 &   0.31\\
          23 &   7.79 &   8.92 &  27.00 &   8.59 &   8.46 &   3.43 &  35.02 & -21.49 &   0.27\\
          24 &   7.88 &   4.62 &  27.00 &   8.60 &   8.47 &   4.27 &  35.09 & -21.72 &   0.38\\
          25 &   9.74 &  13.34 & -28.39 &   8.55 &   8.42 &   2.12 &  22.86 & -28.81 &   0.10\\
          26 &  12.46 & -59.00 & -20.00 &   8.54 &   8.40 &   1.75 &  24.30 & -38.31 &   0.13\\
          27 &   4.07 & -13.88 &   6.78 &   8.63 &   8.53 &   5.06 &  45.91 & -14.85 &   0.41\\
          28 &  10.78 &  52.00 &  11.56 &   8.57 &   8.45 &   2.47 &  30.08 & -32.79 &   0.22\\
          29 &  10.38 & -31.00 & -33.44 &   8.55 &   8.44 &   2.58 &  31.41 & -37.34 &   0.22\\
          30 &   3.06 &  -2.45 &   9.55 &   8.65 &   8.52 &   4.81 &  51.85 & -11.37 &   0.45\\
          31 &   8.36 &   6.85 & -26.04 &   8.58 &   8.46 &   2.19 &  28.56 & -23.76 &   0.18\\
          32 &   3.95 & -16.79 &   2.85 &   8.64 &   8.53 &   4.77 &  44.61 & -13.66 &   0.43\\
          33 &  10.99 &  11.89 & -33.26 &   8.51 &   8.40 &   1.38 &  13.70 & -14.97 &   0.00\\
          34 &   8.29 &  39.45 &   3.45 &   8.57 &   8.49 &   3.08 &  34.96 & -33.17 &   0.14\\
          35 &   4.23 & -20.10 &  -1.59 &   8.63 &   8.51 &   4.32 &  35.55 & -12.79 &   0.45\\
          36* &   8.68 &  36.00 &  22.08 &   8.59 &   8.49 &   3.90 &  53.04 & -24.41 &   0.24\\
          37 &   4.18 & -20.00 &  -5.92 &   8.63 &   8.53 &   2.86 &  29.17 & -12.31 &   0.31\\
          38 &   5.98 & -28.85 &  -4.96 &   8.64 &   8.52 &   3.27 &  43.62 & -13.70 &   0.37\\
          39 &  12.43 &  10.93 & -38.50 &   8.49 &   8.39 &   1.02 &  12.10 & -25.15 &   0.00\\
          40 &   7.78 & -36.00 &   0.08 &   8.62 &   8.49 &   2.61 &  35.49 & -17.43 &   0.30\\
          41 &   9.59 &  -4.00 &  31.00 &   8.58 &   8.48 &   1.95 &  27.09 & -21.64 &   0.25\\
          42 &   7.66 &  19.08 & -18.00 &   8.60 &   8.49 &   2.16 &  22.08 & -20.99 &   0.18\\
          43 &   9.49 &  41.33 &  -5.28 &   8.58 &   8.48 &   1.64 &  19.99 & -26.94 &   0.14\\
          44 &   4.91 &  13.83 & -10.48 &   8.61 &   8.52 &   2.75 &  32.69 & -15.28 &   0.27\\
          45 &  10.00 & -47.00 &  -2.00 &   8.60 &   8.46 &   1.55 &  21.48 & -18.28 &   0.18\\
          46 &   5.15 & -22.00 &   3.52 &   8.64 &   8.53 &   3.08 &  27.04 & -11.76 &   0.33\\
          47 &   2.26 &  -5.04 &   5.67 &   8.68 &   8.51 &   3.28 &  22.31 &  -8.00 &   0.39\\
          48 &  10.15 & -19.00 & -35.00 &   8.58 &   8.45 &   1.13 &  15.71 & -16.09 &   0.21\\
          49 &  10.95 &  52.00 &  16.92 &   8.58 &   8.45 &   1.45 &  14.84 & -24.87 &   0.11\\
          50 &  13.66 & -36.59 & -45.24 &   8.51 &   8.43 &   0.90 &   9.42 & -40.12 &   0.00\\
          51 &  12.99 & -30.78 & -43.93 &   8.54 &   8.43 &   0.87 &  11.35 & -34.45 &   0.01\\
          52 &   8.77 & -36.00 &   8.08 &   8.62 &   8.46 &   1.69 &  23.01 & -15.34 &   0.29\\
          53 &   3.84 &  18.42 &   5.22 &   8.66 &   8.53 &   1.94 &  21.99 & -10.49 &   0.25\\
          54 &  13.69 & -41.62 & -43.85 &   8.52 &   8.44 &   0.83 &   9.41 & -39.13 &   0.00\\
          55 &  12.24 &  -8.00 & -42.00 &   8.53 &   8.42 &   0.67 &   9.34 & -27.24 &   0.00\\
          56 &  11.27 &  11.00 &  39.00 &   8.58 &   8.44 &   0.71 &   9.84 & -19.19 &   0.05\\
          57 &   6.28 &   7.00 & -18.92 &   8.62 &   8.50 &   1.83 &  24.93 & -13.61 &   0.25\\
          58 &  12.26 &  -4.00 &  39.92 &   8.57 &   8.43 &   0.53 &   7.16 & -21.07 &   0.00\\
          59 &  11.47 &  26.00 &  39.00 &   8.58 &   8.46 &   0.65 &   9.02 & -15.32 &   0.05\\
          60 &  11.64 & -46.00 &  12.92 &   --- &    --- &   --- &   --- &   ---&  ---\\
\hline\hline

\end{tabular}
\tablefoot{* This region marginally contains a foreground star and was therefore omitted in the analysis}
\end{table*}

\begin{table*}
\caption{Properties of the integrated H~{\sc ii} regions - continued, region 33 contains the GRB}    
\label{table:hii}      
\centering    
\scriptsize
\begin{tabular}{ccccccccccc}       
\hline\hline                
 Number & log$\frac{\mathrm{[OIII]}}{\mathrm{[OI]}}$ & log U & log$\frac{\mathrm{[OIII]}}{\mathrm{H\beta}}$ & log$\frac{\mathrm{[NII]}}{\mathrm{H\alpha}}$ & log$\frac{\mathrm{[SII]}}{\mathrm{H\alpha}}$ & HeI/H$\alpha$ & log$\frac{\mathrm{[OI]}}{\mathrm{H\alpha}}$ & log$\frac{\mathrm{[OIII]}}{\mathrm{H\alpha}}$ & log$\frac{\mathrm{[NII]}}{\mathrm{[SII]}}$ &  N/O \\ \hline\hline 
           1 &  -0.59 &  -2.52 &  -0.32 &  -0.40 &  -0.44 &   0.17 &  -1.35 &  -0.76 &   0.04 &  -0.81\\
           2 &  -0.32 &  -2.29 &  -0.61 &  -0.44 &  -0.53 &   0.17 &  -1.37 &  -1.04 &   0.09 &  -0.74\\
           3 &  -0.85 &  -2.68 &  -0.06 &  -0.44 &  -0.40 &   0.17 &  -1.35 &  -0.50 &  -0.04 &  -0.91\\
           4 &  -0.80 &  -2.61 &  -0.12 &  -0.43 &  -0.43 &   0.17 &  -1.36 &  -0.56 &  -0.00 &  -0.86\\
           5 &  -0.58 &  -2.65 &  -0.22 &  -0.37 &  -0.36 &   0.18 &  -1.24 &  -0.66 &  -0.01 &  -0.87\\
           6 &  -0.17 &  -2.65 &  -0.31 &  -0.20 &  -0.29 &   0.34 &  -0.91 &  -0.74 &   0.09 &  -0.75\\
           7 &  -0.75 &  -2.62 &  -0.14 &  -0.41 &  -0.42 &   0.17 &  -1.32 &  -0.57 &   0.01 &  -0.85\\
           8 &  -0.78 &  -2.75 &  -0.03 &  -0.40 &  -0.34 &   0.17 &  -1.25 &  -0.47 &  -0.06 &  -0.93\\
           9 &  -0.40 &  -2.69 &  -0.30 &  -0.35 &  -0.31 &   0.22 &  -1.14 &  -0.74 &  -0.04 &  -0.91\\
          10 &  -0.41 &  -2.69 &  -0.25 &  -0.32 &  -0.31 &   0.24 &  -1.10 &  -0.68 &  -0.00 &  -0.86\\
          11 &  -0.59 &  -2.72 &  -0.18 &  -0.37 &  -0.32 &   0.19 &  -1.20 &  -0.62 &  -0.04 &  -0.91\\
          12 &  -0.31 &  -2.62 &  -0.35 &  -0.33 &  -0.34 &   0.24 &  -1.09 &  -0.78 &   0.01 &  -0.85\\
          13 &  -0.38 &  -2.70 &  -0.24 &  -0.31 &  -0.31 &   0.29 &  -1.05 &  -0.68 &  -0.00 &  -0.86\\
          14 &  -0.20 &  -2.64 &  -0.32 &  -0.29 &  -0.32 &   0.31 &  -0.95 &  -0.75 &   0.03 &  -0.82\\
          15 &  -0.33 &  -2.74 &  -0.19 &  -0.28 &  -0.28 &   0.28 &  -0.95 &  -0.63 &  -0.00 &  -0.86\\
          16 &  -0.27 &  -2.76 &  -0.24 &  -0.24 &  -0.24 &   0.33 &  -0.94 &  -0.67 &   0.00 &  -0.85\\
          17 &  -0.53 &  -2.76 &  -0.07 &  -0.34 &  -0.31 &   0.26 &  -1.04 &  -0.51 &  -0.03 &  -0.89\\
          18 &  -0.35 &  -2.75 &  -0.20 &  -0.26 &  -0.27 &   0.36 &  -0.98 &  -0.63 &   0.01 &  -0.85\\
          19 &  -0.40 &  -2.80 &  -0.18 &  -0.31 &  -0.25 &   0.28 &  -1.01 &  -0.62 &  -0.05 &  -0.93\\
          20 &  -1.00 &  -2.80 &   0.25 &  -0.46 &  -0.38 &   0.23 &  -1.19 &  -0.19 &  -0.07 &  -0.95\\
          21 &  -0.55 &  -2.82 &  -0.10 &  -0.35 &  -0.27 &   0.22 &  -1.09 &  -0.54 &  -0.08 &  -0.96\\
          22 &  -0.22 &  -2.72 &  -0.30 &  -0.29 &  -0.27 &   0.32 &  -0.95 &  -0.74 &  -0.02 &  -0.88\\
          23 &  -0.53 &  -2.93 &   0.00 &  -0.32 &  -0.21 &   0.29 &  -0.96 &  -0.44 &  -0.12 &  -1.01\\
          24 &  -0.51 &  -2.93 &  -0.01 &  -0.32 &  -0.20 &   0.34 &  -0.95 &  -0.45 &  -0.12 &  -1.01\\
          25 &  -0.77 &  -2.93 &   0.12 &  -0.42 &  -0.26 &   0.25 &  -1.09 &  -0.32 &  -0.16 &  -1.06\\
          26 &  -0.87 &  -2.94 &   0.18 &  -0.44 &  -0.27 &   0.23 &  -1.12 &  -0.26 &  -0.17 &  -1.07\\
          27 &  -0.21 &  -2.79 &  -0.22 &  -0.26 &  -0.24 &   0.38 &  -0.87 &  -0.66 &  -0.02 &  -0.89\\
          28 &  -0.61 &  -2.89 &  -0.00 &  -0.37 &  -0.25 &   0.27 &  -1.05 &  -0.44 &  -0.12 &  -1.01\\
          29 &  -0.67 &  -2.91 &   0.03 &  -0.41 &  -0.25 &   0.21 &  -1.07 &  -0.41 &  -0.16 &  -1.07\\
          30 &  -0.25 &  -2.87 &  -0.15 &  -0.20 &  -0.18 &   0.45 &  -0.83 &  -0.58 &  -0.02 &  -0.89\\
          31 &  -0.52 &  -2.86 &  -0.02 &  -0.35 &  -0.26 &   0.26 &  -0.98 &  -0.46 &  -0.09 &  -0.97\\
          32 &  -0.16 &  -2.79 &  -0.21 &  -0.22 &  -0.23 &   0.45 &  -0.81 &  -0.65 &   0.01 &  -0.85\\
          33 &  -0.54 &  -2.90 &   0.10 &  -0.51 &  -0.21 &   0.38 &  -0.80 &  -0.26 &  -0.31 &  -1.24\\
          34 &  -0.56 &  -2.82 &  -0.15 &  -0.37 &  -0.26 &   0.22 &  -1.15 &  -0.59 &  -0.10 &  -0.99\\
          35 &  -0.26 &  -2.85 &  -0.15 &  -0.23 &  -0.20 &   0.38 &  -0.84 &  -0.58 &  -0.03 &  -0.90\\
          36* &  -0.39 &  -2.85 &  -0.15 &  -0.34 &  -0.23 &   0.28 &  -0.98 &  -0.59 &  -0.10 &  -0.99\\
          37 &  -0.12 &  -2.82 &  -0.23 &  -0.24 &  -0.21 &   0.40 &  -0.79 &  -0.66 &  -0.03 &  -0.89\\
          38 &  -0.24 &  -2.86 &  -0.18 &  -0.22 &  -0.19 &   0.40 &  -0.86 &  -0.62 &  -0.03 &  -0.90\\
          39 &  -0.79 &  -2.81 &   0.15 &  -0.54 &  -0.30 &   0.24 &  -1.01 &  -0.22 &  -0.24 &  -1.16\\
          40 &  -0.36 &  -2.91 &  -0.05 &  -0.26 &  -0.19 &   0.31 &  -0.85 &  -0.49 &  -0.07 &  -0.94\\
          41 &  -0.39 &  -2.91 &  -0.11 &  -0.36 &  -0.20 &   0.26 &  -0.94 &  -0.55 &  -0.16 &  -1.06\\
          42 &  -0.47 &  -2.89 &  -0.12 &  -0.31 &  -0.21 &   0.28 &  -1.03 &  -0.56 &  -0.11 &  -1.00\\
          43 &  -0.44 &  -2.89 &  -0.12 &  -0.36 &  -0.22 &   0.24 &  -1.00 &  -0.56 &  -0.15 &  -1.04\\
          44 &  -0.22 &  -2.82 &  -0.23 &  -0.28 &  -0.22 &   0.37 &  -0.89 &  -0.67 &  -0.06 &  -0.94\\
          45 &  -0.48 &  -2.99 &   0.02 &  -0.30 &  -0.16 &   0.34 &  -0.90 &  -0.42 &  -0.14 &  -1.04\\
          46 &  -0.22 &  -2.87 &  -0.19 &  -0.22 &  -0.18 &   0.45 &  -0.84 &  -0.62 &  -0.04 &  -0.91\\
          47 &  -0.23 &  -3.02 &  -0.03 &  -0.14 &  -0.08 &   0.50 &  -0.69 &  -0.47 &  -0.05 &  -0.92\\
          48 &  -0.52 &  -2.95 &   0.06 &  -0.35 &  -0.21 &   0.37 &  -0.90 &  -0.37 &  -0.14 &  -1.03\\
          49 &  -0.54 &  -2.95 &   0.04 &  -0.34 &  -0.21 &   0.25 &  -0.94 &  -0.40 &  -0.14 &  -1.03\\
          50 &  -0.86 &  -2.69 &  -0.03 &  -0.51 &  -0.34 &   0.16 &  -1.25 &  -0.39 &  -0.17 &  -1.07\\
          51 &  -0.57 &  -2.91 &   0.02 &  -0.45 &  -0.26 &   0.20 &  -0.99 &  -0.42 &  -0.19 &  -1.10\\
          52 &  -0.43 &  -3.02 &   0.06 &  -0.27 &  -0.15 &   0.37 &  -0.80 &  -0.37 &  -0.13 &  -1.02\\
          53 &  -0.12 &  -2.92 &  -0.19 &  -0.19 &  -0.13 &   0.42 &  -0.74 &  -0.63 &  -0.06 &  -0.94\\
          54 &  -0.83 &  -2.62 &  -0.06 &  -0.48 &  -0.35 &   0.16 &  -1.22 &  -0.39 &  -0.13 &  -1.03\\
          55 &  -0.81 &  -2.74 &   0.07 &  -0.46 &  -0.29 &   0.15 &  -1.09 &  -0.28 &  -0.16 &  -1.06\\
          56 &  -0.51 &  -2.93 &   0.07 &  -0.36 &  -0.23 &   0.35 &  -0.87 &  -0.37 &  -0.13 &  -1.02\\
          57 &  -0.22 &  -2.94 &  -0.10 &  -0.26 &  -0.16 &   0.35 &  -0.76 &  -0.54 &  -0.10 &  -0.98\\
          58 &  -0.53 &  -2.99 &   0.10 &  -0.38 &  -0.20 &   0.30 &  -0.87 &  -0.34 &  -0.17 &  -1.08\\
          59 &  -0.28 &  -3.07 &  -0.03 &  -0.35 &  -0.11 &   0.35 &  -0.74 &  -0.47 &  -0.24 &  -1.17\\
          60 &   --- &    --- &   ---&    --- &   --- &   --- &   --- &    --- &   --- &    ---\\
\hline\hline
\end{tabular}
\tablefoot{* This region marginally contains a foreground star and was therefore omitted in the analysis}
\end{table*}

\end{appendix}

\end{document}